\documentclass[runningheads]{llncs}
\usepackage[T1]{fontenc}
\usepackage[english]{babel}

\usepackage{xspace}
\usepackage{amssymb}
\usepackage{amsmath}
\usepackage{hyperref}
\usepackage[capitalise]{cleveref}
\usepackage{graphicx}
\usepackage{todonotes}

\usepackage{ebproof}
\usepackage{stmaryrd}
\usepackage{cmll}
\usepackage{tikz-cd}
\usepackage{ulem}
\usepackage{lineno}
\usepackage{tcolorbox}

\usepackage{amsthm}
\usepackage{mathtools}
\usepackage{thmtools}
\usepackage{thm-restate}

\usepackage{macros}

\author{
    Sandra Alves\inst{1,2}
    \and 
    Delia Kesner\inst{3}
    \and 
    Miguel Ramos\inst{1,3,4,}\thanks{Corresponding author.}
    }

\institute{
    DCC/FCUP - Faculty of Sciences of the University of Porto 
  \and
  CRACS/INESC-TEC - Center of Advanced Computing Systems
  \and
  Universit\'e Paris Cit\'e, IRIF, CNRS
  \and
  LIACC - Artificial Intelligence and Computer Science Laboratory
}

\title{Extending the Quantitative Pattern-Matching Paradigm\thanks{Supported by: FCT, within project LA/P/0063/2020, and grant 2021.04731.BD; Base Funding UIDB/00027/2020 for LIACC through FCT/MCTES (PIDDAC).}}

\begin{document}

\maketitle

\begin{abstract}
  We show how (well-established) type systems based on non-idempotent
  intersection types can be extended to characterize termination properties of
  functional programming languages with pattern matching features. To model such
  programming languages, we use a (weak and closed) $\lam$-calculus integrating
  a pattern matching mechanism on algebraic data types (ADTs). Remarkably, we
  also show that this language not only encodes Plotkin's CBV and CBN
  $\lam$-calculus as well as other subsuming frameworks, such as the
  bang-calculus, but can also be used to interpret the semantics of effectful
  languages with exceptions. After a thorough study of the untyped language, we
  introduce a type system based on intersection types, and we show through
  purely logical methods that the set of terminating terms of the language
  corresponds exactly to that of well-typed terms. Moreover, by considering
  \emph{non-idempotent} intersection types, this characterization turns out to
  be quantitative, \ie the size of the type derivation of a term $t$ gives an
  upper bound for the number of evaluation steps from $t$ to its normal form.
\end{abstract}

\section{Introduction}

Pattern matching is a very useful and powerful mechanism in programming, as
witnessed by functional programming languages. Indeed, Haskell and OCaml use
pattern matching mechanisms to interact with \emph{algebraic data types} (ADTs),
notably by providing a mechanism to  deconstruct complex data structures in a
concise and readable manner. ADTs are types built by combining other types, such
as \emph{product types}  and \emph{sum types}. Product types are represented by
tuples and sum types by tagged unions. Pattern matching also enhances
expressiveness, making it easier to manipulate complex data structures
intuitively. Due to the significant role that pattern matching plays in
functional programming languages, it is not only  important to study its
standard semantics, but also its quantitative semantics. In particular, this
work focuses on developing a model that allows the study of \emph{quantitative}
properties of such languages. Semantics for pattern matching have been studied
in the
literature~\cite{CirsteaK01,Kahl04,JayK09,BucciarelliKR15,AccattoliB17,AlvesDFK18,Barenbaum18,Alves2019,BucciarelliKR21},
both for formalizing programming
languages~\cite{CirsteaK01,Kahl04,JayK09,AccattoliB17,AlvesDFK18} and theorem
provers~\cite{BucciarelliKR15,Barenbaum18,Alves2019,BucciarelliKR21}. However,
they are either too simple (do not capture
ADTs)~\cite{BucciarelliKR15,AlvesDFK18,BucciarelliKR21}, lack (quantitative)
semantics~\cite{CirsteaK01,Kahl04,AlvesDFK18,JayK09}, or do not deal with
types~\cite{AccattoliB17}.

\para{Pattern Matching.} (Purely) functional programming languages can be
modeled by the $\lam$-calculus. More specifically, they are modeled by the
\emph{weak} $\lam$-calculus (\ie no evaluation inside functions), and terms to
be evaluated are \emph{closed} (\ie no occurrences of free variables).
Therefore, it is natural to focus on pattern matching mechanisms in the
framework of weak and closed $\lam$-calculi. In this work, we consider an
extension of the $\lam$-calculus based on \emph{generalized $\lam$-abstractions}
(see \eg~\cite{Kesner97,KesnerPT96}), which are $\lam$-abstractions of the form
$\lam p.t$, where $t$ is a term and $p$ is a pattern specifying the expected
structure of its argument; and \emph{case expressions} of the form
$\case{u}{(\branch{p_1}{t_1}, \ldots, \branch{p_n}{t_n})}$, where $t_1, \ldots,
    t_n, u$ are terms and $p_1, \ldots, p_n$ are patterns, further generalizing
these $\lam$-abstractions. These constructions can be encoded into more
elementary syntax, but we prefer to work with a more expressive high-level
language. Case expressions with a single branch could be understood as
generalized $\lam$-abstractions applied to an argument. For example,
$\case{t}{(\branch{\const{\name{one}}{(x)}}{\case{u}{(\branch{\const{\name{one}}{(y)}}{xy})}})}$
can be alternatively expressed as $(\lam \const{\name{one}}{(x)}. (\lam
    \const{\name{one}}{(y)}.xy))\ t\ u$. However, while in the former case $u$
appears inside a branch, and thus can only be evaluated after $t$, in the latter
case this order of evaluation is no longer imposed (at least) \textit{a priori}.
However, these two constructions do not behave in the exact same way. Indeed,
while (generalized) $\lam$-abstractions can in principle postpone the evaluation
of their arguments (see~\cref{sec:calculus}), case expressions cannot, since
they need to evaluate their arguments until a data matches some pattern in order
to decide the branch to pick.

In this work, patterns are either variables (for which matching always
succeeds) or \emph{tagged products}, which are products tagged by a unique name.
As an example, given a pair (binary product) of the form $(x, y)$, we can build
the \emph{tagged} product $\const{\texttt{pair}}{(x,y)}$ by preceding $(x,y)$
with the (unique) name $\texttt{pair}$. By requiring all patterns in case
expressions to be unique tagged products, case expressions allow matching over
tagged unions. Indeed, a function such as
$\lam x.\case{x}{(\branch{\name{pair}(x,y)}{y}, \branch{\name{triple}(x,y,z)}{x})}$ is able to handle both data such as \emph{pairs} \emph{or} \emph{triples} precisely because of the case expression inside its body.
In sum, by adding generalized $\lam$-abstractions and case expressions to the
weak and closed $\lam$-calculus, we obtain a simple, yet expressive
formal language that can be used to study the semantics of programming languages
with pattern matching.

\para{Intersection Types.} Type systems based on intersection types can
not only guarantee termination (a typable term is terminating), but also
characterize it (a terminating term is typable)~\cite{CoppoD78}.
\emph{Intersection types} extend simple types with an intersection constructor
$\cap$. Intuitively, a program $t$ is typable with the intersection of types
$\tau \cap \sig$ if $t$ is typable with both $\tau$ and $\sig$
independently~\cite{CDC78}. Intersection types were first introduced as a model,
for capturing computational properties of the $\lam$-calculus in a broader
sense. For example, termination of different evaluation strategies can be
characterized by typability in some appropriate intersection type
system~\cite{BucciarelliKV17,BonoD20}, that is, a program $t$ is
terminating in a precise sense if and only if $t$ is typable in an appropriate
type system. Moreover, by considering non-idempotent
intersections~\cite{KfouryW99,NeergaardM04}, none of the power of intersection
types is lost, but there are substantial improvements.
For example, typability in a type system using non-idempotent intersections does
not only characterize a \emph{qualitative} property such as termination, but
also provides \emph{quantitative} information about termination, such as
\emph{upper bounds} for the number of evaluation steps needed to reach a normal
form~\cite{Gardner94,Kfoury00,deCarvalho2007,deCarvalho2018}. Very roughly, for
type systems based on non-idempotent intersection types, every evaluation step
strictly decreases the size of the type derivation tree associated to a
well-typed program~\cite{BucciarelliKV17}. This shift of perspective, from
idempotent to non-idempotent intersection types, goes beyond lowering the
logical complexity of the proof: the quantitative information provided by typing
derivations in the non-idempotent setting unveils crucial quantitative relations
between typing (statics) and evaluation (dynamics) of programs. For example, one
particular consequence is that type inhabitation is undecidable for idempotent
types, but becomes decidable for non-idempotent ones~\cite{BucciarelliKR18}.
Moreover, there is a tight correspondence between non-idempotent intersection
types and the multiplicative connective $\oc$ (read ``of course'') of linear
logic~\cite{Girard87}. Indeed, let $A$ be a set of types. Then, $\oc A$ denotes
a \emph{multiset} of types of $A$. Since associativity, commutativity and
idempotency are granted if intersections are denoted by sets, therefore, it is
natural to represent non-idempotent intersections by multisets, as adopted in
this work.

\para{Related Work.} Over the last couple of decades, a lot of work has been put
towards studying extensions of the $\lam$-calculus with pattern matching
mechanisms, either from an operational point of
view~\cite{HL91,SekarR93,CirsteaK01,Kahl04,JayK09,Kha90} or from a logical
one~\cite{KesnerPT96,CerritoaK99,Zeilberger08,Krishnaswami09}. More
recently~\cite{BucciarelliKR15,BucciarelliKR21}, a notion of observability for a
calculus with pattern matching was introduced, in which the $\lam$-calculus is
extended with generalized $\lam$-abstractions, but patterns are restricted to
pairs. Inspired by the pioneering works
of~\cite{Gardner94,Kfoury00,deCarvalho2007,deCarvalho2018} and
later~\cite{AccattoliGK20}, type systems providing upper-bounds and exact
measures for the length of evaluation sequences and the size of normal forms
were introduced in~\cite{Alves2019} for a language virtually identical to that
in~\cite{BucciarelliKR15,BucciarelliKR21}. In~\cite{AccattoliB17}, it was shown
that matching steps can be assigned zero cost, as they are linear in the number
of $\beta$-steps and the size of the initial term. The language considered
in~\cite{AccattoliB17} extends the $\lam$-calculus with case expressions and
tagged products, but not with generalized $\lam$-abstractions. Moreover, no type
system is studied in that work, and no proper treatment of \emph{stuck} case
expressions is proposed. As a consequence, it is not possible to clearly
distinguish programs that get stuck from those that halt because they have
\emph{successfully} finished computing. Another recent work involving the
$\lam$-calculus with pattern matching is~\cite{Barenbaum18}, where
non-idempotent intersection types are used to characterize weak normalization
for a $\lam$-calculus based on the Calculus of Inductive Construction, as
introduced in~\cite{GregoireL02}. Roughly, the latter is an extension of the
$\lam$-calculus with a fixpoint operator, case expressions, and tagged products.
However, it is worth noting that their type system does \emph{not} enforce
\emph{type safety}, \ie many programs are syntactically well-formed but have no
meaning at all. For instance, it is possible to write down a program applying
data to an argument, which is clearly effectively meaningless. Type safety is
then a desirable property that can be enforced by a type system in order to
restrict the set of programs to those that do not evaluate to errors. This idea
was  explored by Milner in~\cite{Milner78}, which coined it with the slogan
``well-typed programs cannot go wrong''. Indeed, in~\cite{Barenbaum18},
well-typed program \emph{can} ``go wrong''. This does not mean that programs
containing errors cannot be semantically sound. It simply means that, in order
to ensure \emph{type safety}, errors need to be introduced in a controlled
manner, \eg by using exceptions~\cite{Moggi91}.

All the aforementioned works use \emph{open} $\lam$-calculi, where terms are
allowed to have free occurrences of variables. However, as it is customary in
$\lam$-calculi modeling programming languages, programs are modeled by closed
terms. Syntactically, the language considered in~\cite{AccattoliB17} is the
closest to the one considered in this work. Still, there are two crucial
differences. Besides case expressions, we also include generalized
$\lam$-abstractions, and we allow patterns to be nested.

\begin{itemize}
    \item[$\bullet$] Including generalized $\lam$-abstractions allows
        single-branched case expressions to be written as applications using
        generalized $\lam$-abstractions. Currently, since we are going to adopt
        a weak call-by-name strategy and the calculus enjoys the one-step
        diamond property (\cref{lem:diamond}), this redundant syntax is more
        expressive but completely inconsequential. However, if we consider a
        call-by-need (CBNeed) strategy with a garbage collecting step (which we
        leave as future work), it is possible to optimize single-branched case
        expressions by rewriting them as applications using generalized
        $\lam$-abstractions. On one hand, case expressions must \emph{always}
        reduce their arguments until a data matches some pattern in order to
        select \emph{one} of its branches. On the other hand, applications may
        (potentially) postpone the evaluation of their arguments for longer,
        since evaluating the abstraction's body does not necessarily require the
        argument to be fully evaluated.

    \item[$\bullet$]
        Nested patterns, such as $\const{\name{succ}}{(\const{\name{succ}}{(x)})}$, do
        not result in any additional computational expressivity, but could be more
        convenient. Indeed, nested patterns can be encoded by generalized
        $\lam$-abstractions. For instance, the term $\lam \const{\name{succ}}{(x)}.((\lam
            \const{\name{succ}}{(y)}.y)x)$ is an encoding of $\lam
            \const{\name{succ}}{(\const{\name{succ}}{(x)})}.x$ that does \emph{not} use
        nested patterns.
\end{itemize}

In contrast to previous contributions in the domain, here we also study
\emph{stuck case expressions}. These case expressions are those whose
evaluation is stopped prematurely due to an inability to match any of the
branches. Consider the following case expression:

\[ \begin{array}{c}
        t =
        \case{\const{\name{pair}}{(r,u)}}{(\branch{\const{\name{one}}{(x)}}{x},
            \branch{\const{\name{triple}{(x,y,z)}}}{y})}
    \end{array} \]

Evaluation is \emph{stuck} for $t$ because $\name{pair}(r,u)$ neither
matches $\name{one}(x)$ nor $\name{triple}{(x,y,z)}$. That is, $t$ is a
\emph{stuck case expression}. It is also a normal form, but clearly meaningless.
In practice, stuck case expressions like the previous one should be untypable,
but we believe that they should also be studied and understood from a purely
operationally (\ie untyped) point-of-view \ie by providing a characterization of
normal forms. Finally, the crucial difference between~\cite{Barenbaum18} and
this work is that our type system does \emph{not} type errors: all
typable programs terminate without errors
(see~\cref{lem:typability-implies-cfness}, where \emph{clash} is our notion of
error). Indeed, our type system characterizes \emph{error free}
termination (see~\cref{thm:one}).

\para{Contributions and Overview.} This work extends the pattern calculi
in~\cite{BucciarelliKR15,BucciarelliKR21,Alves2019} with tagged products and
case expressions, which capture ADT's. The calculus that is obtained, called the
$\calc$-calculus, models functional programming languages more closely. The main
contributions of this work are threefold. First, we provide a detailed
operational treatment of the $\calc$-calculus in~\cref{sec:calculus}, for which
we propose an evaluation strategy $\redd$ based on an extension of the
well-known \emph{weak head reduction} strategy. We also give a purely
syntactical characterization of \emph{clash-free} (\ie error-free) normal forms.
Then, we take a slight detour in~\cref{sec:encodings} to show a
surprising and unexpected consequence of our work: that the operational
semantics of our calculus not only subsumes CBN and CBV~\cite{Plotkin75}, but
can also encode another subsuming frameworks, such as the
bang-calculus~\cite{BucciarelliKRV23}.

Finally, we define a non-idempotent intersection type system in~\cref{sec:types}
that naturally extends type system $\mathcal{U}$ from~\cite{Alves2019}.  We
provide a purely \emph{logical} and \emph{quantitative} characterization of
termination, \ie we show that a program $t$ is typable if, and only if, its
evaluation terminates (in a \emph{clash-free} normal form). Moreover, the
characterization result is \emph{quantitative}: we show that the size of type
derivations provide upper bounds for the number of evaluation steps.
In~\cref{sec:conclusions} we conclude and suggest some future work.
\cref{sec:appendix} includes some auxiliary lemmas needed in order to show the
main results, as well as all the proofs and in full detail.

\section{The Pattern-Matching Calculus}
\label{sec:calculus}

Before introducing the syntax of the pattern-matching $\calc$-calculus, we
clarify some basic notation related to \emph{tagged products}.

\para{Tagged Products.} Let $\ttc$ be a \defn{tag} taken out of an enumerable
set of tags $\setofconsts$. A \defn{constructor} $\ttc^n$ is built by assigning
a unique \emph{arity} $n \in \nat$ to a \emph{tag} $\ttc$. Since each tag has a
\emph{unique} arity, constructors are identified uniquely by their tags.
Constructors cannot be applied partially: a constructor $\ttc^n$ \emph{must} be
applied to $n$ arguments. Accordingly, the application $\ttc^n\ a_1 \cdots\ a_n$
will be written $\const{\ttc^n}{(a_1, \ldots, a_n)}$ and called a \defn{tagged
    product}, where the sequence of arguments $a_1 \cdots a_n$ is represented as a
\emph{tuple} (or \emph{product} or \emph{vector}) of arguments $(a_1, \ldots,
    a_n)$ of length $n$. It is possible to take advantage of the tuple notation and
simply write $(a_1, \ldots, a_n)$ as $\vect{a_i}_n$ (meaning that $1 \leq i \leq
    n$) in order to keep notation light. Tagged products represent \defn{data} and
will thus be referred to as such throughout the rest of this work. Moreover, the
arity of constructors will always be left implicit, and all data is assumed to
be well-formed (constructors are always applied to tuples of arguments of the
appropriate length).

\para{The Syntax of the $\calc$-Calculus.} Given a infinite but countable set of variables $x, y, z, \ldots \in \setofvars$, we introduce the following grammars:
{\small\[ \begin{array}{rrcl}
        \textbf{(Patterns)}      & p, q          & ::= & x \mid \const{\ttc}{\vect{p_i}_n}
        \\
        \textbf{(Branches)}      & b             & ::= & \branch{\const{\ttc}{\vect{p_i}_n}}{t}
        \\
        \textbf{(Terms)}         & t, u, r, v, s & ::= & x \mid \lam p.t \mid t u \mid t \match{p}{u} \mid \const{\ttc}{\vect{t_i}_n} \mid \case{t}{(b_1, \ldots, b_n)}
        \\
        \textbf{(List Contexts)} & \MC           & ::= & \chole \mid \MC \match{p}{t}
    \end{array} \]}

Let $\terms$ denote the set of terms of the $\calc$-calculus. In order to keep
notation light, symbols $\ptuple$, $\btuple$, and $\ttuple$ are going to be used
to denote \defn{tuples of patterns}, \defn{tuples of branches}, and \defn{tuples
    of terms}, respectively. The pattern $x$ is called a \defn{variable pattern},
and $\const{\ttc}{\ptuple}$ is a \defn{data pattern}. The term $\lam p.t$ is a
\defn{generalized abstraction}, $t \match{p}{u}$ is a \defn{matching closure}
(where $\match{p}{u}$ is a called an \defn{explicit matching operation}),
$\const{\ttc}{\ttuple}$ is a \defn{data term}, and $\case{t}{\btuple}$ is a
\defn{case expression}. Remark in particular that the explicit matching operator
is  a  new constructor in the language, and it is not on the meta-level. We
write $\I$ as a special term representing the identity function $\lam x.x$.

\para{Free and Bound Variables.} The \defn{set of variables} of a pattern $p$ is
denoted by $\var{p}$ and defined as expected. It will be useful to write
$\hat{p}$ to denote \emph{data patterns}, so that
$\case{t}{\vect{\branch{\const{\ttc_i}{\ptuple_i}}{u_i}}_n}$ can be written as
$\case{t}{\vect{\branch{\hat{p_i}}{u_i}}_n}$, where $\hat{p_i} =
    \const{\ttc_i}{\ptuple_i}$. The \defn{sets of free and bound variables} of
branches, terms and list contexts are also defined as expected. In particular:

{\small \[ \begin{array}{rcl}
        \fv{\lam p.t}                                   & \defeq & \fv{t} \setminus \var{p}
        \\
        \fv{t \match{p}{u}}                             & \defeq & (\fv{t} \setminus \var{p}) \cup \fv{u}
        \\
        \fv{\case{t}{\vect{\branch{\hat{p}_i}{u_i}}_n}} & \defeq & \fv{t} \bigcup_{i \in \interval{1}{n}} (\fv{u_i} \setminus \var{\hat{p}_i})
        \\
        \bv{\MC \match{p}{t}}                           & \defeq & \bv{\MC} \cup \var{p}
    \end{array} \]}

Terms are $\alpha$-equivalent if they only differ on the names of their bound
variables. In this work, terms are going to be considered equal up-to
$\alpha$-equivalence. In particular, substitutions $t \subs{x}{u}$ are capture
avoiding, \ie $x \not\in \bv{t}$.

\para{List Contexts.} Terms can be surrounded by list contexts, which are
sequences of explicit matching operations. Indeed, given a list
context $\MC$ and a term $t$, $\MC \lhole{t}$ denotes the term obtained by
\defn{plugging the term $t$ inside context $\MC$}, \ie by replacing the unique
occurrence of $\chole$ in $\MC$ with $t$ (possibly allowing the capture of free
variables of $t$). Thus, \eg if  $\MC =
    \chole\match{x_1}{u_1}\match{x_2}{u_2}$, then $\MC \lhole{x_1}$ is the term
$x_1\match{x_1}{u_1}\match{x_2}{u_2}$. List contexts are important as they allow
matching operations to be postponed, thus exposing redexes that
might otherwise stay hidden due to premature normal forms (see~\cref{thm:one}). We use two special predicates to distinguish terms possibly affected by a list of explicit matching operators: $\isabs{t}$ iff $t = \MC
    \lhole{\lam p.u}$; $\iscase{t}$ iff $t = \MC \lhole{\case{u}{\btuple}}$; and
$\isconst[\ttc]{t}$ iff $t = \MC \lhole{\const{\ttc}{\ttuple}}$, in which case
we may also write $\isconst{t}$ whenever $\ttc$ is irrelevant.

\para{Linear Patterns.} Patterns are assumed to be \defn{linear}, \ie each variable occurs at most once, and so are tuples of patterns, \ie no variable is shared between patterns in a tuple. Most modern programming languages with pattern matching implement linear patterns since this does not affect the expressivity the language, while avoiding to check for equality of terms.

\para{Case Expressions.} As mentioned in the introduction, case expressions
capture the pattern-matching mechanism over tagged unions. Note that each branch
of a case expression is built out of a tagged product and a term. Therefore, we require all tagged products to have a different top-level tag,
thus branches are disjoint and matching is thus unambiguous. As such, case
expressions can indeed be understood as \emph{deconstructors} for \defn{tagged
    unions}. As an example, the following case expression is a function whose
behavior depends on whether its argument is a \emph{pair} or a \emph{triple}, like in $\lam x.\case{x}{(\branch{\const{\name{pair}}{(x,y)}}{y}, \branch{\const{\name{triple}}{(x,y,z)}}{x})}$.

\subsection{Picking an Evaluation Strategy}

We define our evaluation strategy as a having a lazy behavior. Indeed, with
respect to generalized abstractions built from \emph{variable patterns},
applications behave like in CBN, \ie arguments are never evaluated before they
are plugged into a term (see reduction rule (\ruleE) bellow); however,
generalized abstractions built from \emph{data patterns} and case expressions
behave differently: arguments need to be (partially) evaluated to some data in
order to satisfy the matching conditions (see reduction rules (\ruleC) and
(\ruleM) bellow).

The \defn{weak head reduction relation}  $\red{\HC}$ is defined as the closure over \defn{weak head contexts}
{\small \[ \begin{array}{rcl}
            \HC & ::= & \chole \mid \HC\ t \mid \HC \match{p}{u} \mid t
            \match{\hat{p}}{\HC} \mid \case{\HC}{\btuple}\end{array}
    \]}
of the following four \defn{reduction rules}:
{\small \[ \begin{array}{rcl}
        \MC \lhole{\lam p.t} u                                                                         & \onered[\ruleBeta] & \MC \lhole{t \match{p}{u}} \text{ if $\bv{\MC} \cap \fv{u} = \eset$}
        \\
        \case{\MC \lhole{\ttc \vect{u_i}_n}}{(\ldots, \branch{\const{\ttc}{\vect{p_i}_n}}{t}, \ldots)} & \onered[\ruleC]    & \MC \lhole{t \match{p_1}{u_1} \cdots \match{p_n}{u_n}} \text{ if $\bv{\MC} \cap \fv{t} = \eset$}
        \\
        t \match{\const{\ttc}{\vect{p_i}_n}}{\MC \lhole{\const{\ttc}{\vect{u_i}_n}}}                   & \onered[\ruleM]    & \MC \lhole{t \match{p_1}{u_1} \cdots \match{p_n}{u_n}} \text{ if $\bv{\MC} \cap \fv{t} = \eset$}
        \\
        t \match{x}{u}                                                                                 & \onered[\ruleE]    & t \subs{x}{u}\end{array} \]}

Rule (\ruleBeta) is the usual starting rule of $\lam$-calculi working \emph{at a
    distance}. Indeed, since matchings and substitutions can be delayed, all
terms are formally plugged into a (possibly empty) list of contexts. Thus,
in applications, abstractions and arguments might be separated by a list
context. Rules (\ruleC) and (\ruleM) solve (top-level) matches between data
patterns and data terms, respectively. It is worth noticing that the side conditions in these three rules are necessary in order to avoid the capture of free variables. And rule (\ruleE) produces a
\emph{capture avoiding} substitution as a result of solving a
matching between a variable pattern and a term. Also, note that it is safe to
assume that all meta-level substitutions are capture avoiding due to
$\alpha$-conversion.

We write $t \red{\HC(s)} t'$ if $t \red{\HC} t'$ for $t = \HC \lhole{u}$, $t' =
    \HC \lhole{u'}$, and $u \onered[s] u'$, where $s$ is a reduction rule.
Similarly, we write $t \red{\HC(\overline{s})} t'$ if $u \onered[s'] u'$ and
$s'$ is some reduction rule different from $s$.

Let $\red{\R}$ be some reduction relation. The \defn{reflexive-transitive
    closure} of $\red{\R}$ is written $\plusred{\R}$. In particular, if
$\red{\R} = \red{\HC}$, then $t \plusred{\HC}^{(b,c,m,e)} u$ denotes that $t \plusred{\HC} u$ using $b$ \ruleBeta-steps, $c$
$\ruleC$-steps, $m$ $\ruleM$-steps, and $e$ \ruleE-steps.

\begin{example}
    \label{ex1}
    Let $\ttc_0, \ttc_1, \ttc_2 \in \setofconsts$. Going back to the last example of the previous section, we have the following:
    {\small\[ \begin{array}{rcl}
             &                      & (\lam x.\case{x}{(\branch{\const{\name{pair}}{(x,y)}}{y}, \branch{\const{\name{triple}}{(x,y,z)}}{x})})\ \const{\name{triple}}{(\ttc_0, \ttc_1, \ttc_2)}
            \\
             & \red{\HC(\ruleBeta)} & \case{x}{(\branch{\const{\name{pair}}{(x,y)}}{y}, \branch{\const{\name{triple}}{(x,y,z)}}{x})} \match{x}{\const{\name{triple}}{(\ttc_0,\ttc_1,\ttc_2)}}
            \\
             & \red{\HC(\ruleE)}    & \case{\const{\name{triple}}{(\ttc_0,\ttc_1,\ttc_2)}}{(\branch{\const{\name{pair}}{(x,y)}}{y}, \branch{\const{\name{triple}}{(x,y,z)}}{x})}
            \\
             & \red{\HC(\ruleC)}    & x \match{x}{\ttc_0} \match{y}{\ttc_1} \match{z}{\ttc_2} \red{\HC(\ruleE)} x \match{x}{\ttc_0} \match{y}{\ttc_1}
            \red{\HC(\ruleE)} x \match{x}{\ttc_0}
            \red{\HC(\ruleE)} \ttc_0
        \end{array} \]}
\end{example}

The reduction relation $\red{\HC}$ is not deterministic, \ie given a term which is not in normal form, then it can be reduced in different ways. For example, going back to the previous example, after the $\ruleC$-step, we could have chosen the following reduction sequence:
{\small \[ \begin{array}{rcl}
         &                           & (\lam x.\case{x}{(\branch{\const{\name{pair}}{(x,y)}}{y}, \branch{\const{\name{triple}}{(x,y,z)}}{x})})\ \const{\name{triple}}{(\ttc_0, \ttc_1, \ttc_2)}
        \\
         & \plusred{\HC}^{(1,1,0,1)} & x \match{x}{\ttc_0} \match{y}{\ttc_1} \match{z}{\ttc_2} \red{\HC(\ruleE)} \ttc_0 \match{y}{\ttc_1} \match{z}{\ttc_2} \red{\HC(\ruleE)} \ttc_0 \match{z}{\ttc_2} \red{\HC(\ruleE)} \ttc_0
    \end{array} \]}
where instead of applying $\ruleE$-steps from right-to-left, these are applied from left-to-right.
However, it is possible to fix a particular evaluation strategy for $\red{\HC}$
(see~\cref{prop:determinism}). Despite nondeterminism, as will be shown,
$\red{\HC}$ is not only confluent (\cref{lem:confluence}), but all evaluation
sequences leading to a normal form have the same size (\cref{lem:diamond}).

Reduction relation $\red{\R}$ is said to enjoy the \defn{diamond property} if
for every term $t$, if $u \lred{\R} t \red{\R} r$, there exists a term $v$, such
that $u \red{\R} v \lred{\R} r$ (see Def.1.1.8.(v) of~\cite{Terese03}). Note,
however, that $\red{\HC}$ is not reflexive. Still, by assuming that $u \neq r$,
the following alternative version of the diamond property can be shown for
$\red{\HC}$. A reduction relation $\red{\HC}$ is said to enjoy the
\defn{one-step diamond property} if for every term $t \in \terms$, if $u
    \lred{\HC} t \red{\HC} r$ with $u \neq r$, there exists a term $v \in \terms$,
such that $u \red{\HC} v \lred{\HC} r$ (see Prop.1 of~\cite{LagoM08}).

\begin{restatable}[One-Step Diamond]{lemma}{lemdiamond}
    \label{lem:diamond}
    Reduction relation $\red{\HC}$ enjoys the one-step diamond property.
\end{restatable}

As a corollary we obtain:

\begin{restatable}[Confluence]{corollary}{lemconfluence}
    \label{lem:confluence}
    \begin{itemize}
        \item[$\bullet$] Reduction relation $\red{\HC}$ is confluent.
        \item[$\bullet$] Any two different reduction paths to normal form have the same length.
    \end{itemize}
\end{restatable}

\para{Weak Head Evaluation.} One of the main distinguishing characteristics
between reduction \emph{relations} and evaluation \emph{strategies} is that,
while the former are (usually) \emph{not} deterministic, the latter \emph{are}
(usually) deterministic, \ie for any term $t$, if $t$ is not a normal form,
exactly one reduction rule applies. This is exactly the case for the
\defn{weak head evaluation strategy} $\redd$, which is defined by the
set of reduction rules in~\cref{fig:figure1}.

\begin{figure}[]
    \begin{tcolorbox}[colback=white]
        {\small \[ \hspace{-.2cm}\begin{array}{c}
                    \begin{prooftree}
                        \hypo{\bv{\MC} \cap \fv{u} = \eset} \infer1[(\ruleBeta)]{\MC \lhole{\lam p.t} u \redd \MC \lhole{t \match{p}{u}}}
                    \end{prooftree}
                    \sep
                    \begin{prooftree}
                        \hypo{\neg\isabs{t}}
                        \hypo{t \redd t'}
                        \infer2[(\ruleAppL)]{tu \redd t'u}
                    \end{prooftree}
                    \\[.8cm]
                    \begin{prooftree}
                        \hypo{\bv{\MC} \cap \fv{t} = \eset} \infer1[(\ruleM)]{t \match{\const{\ttc}{\vect{p_i}_n}}{\MC \lhole{\const{\ttc}{\vect{u_i}_n}}} \redd \MC \lhole{t \match{p_1}{u_1} \cdots \match{p_n}{u_n}}}
                    \end{prooftree}
                    \sep
                    \begin{prooftree}
                        \hypo{\phantom{BUUUU}} \infer1[(\ruleE)]{t \match{x}{u} \redd t \subs{x}{u}}
                    \end{prooftree}
                    \\[.5cm]
                    \begin{prooftree}
                        \hypo{\neg\isconst[\ttc]{u}}
                        \hypo{t \redd t'}
                        \infer2[(\ruleEsL)]{t \match{\const{\ttc}{\ptuple}}{u} \redd t' \match{\const{\ttc}{\ptuple}}{u}}
                    \end{prooftree}
                    \sep
                    \begin{prooftree}
                        \hypo{\neg\isconst[\ttc]{u}}
                        \hypo{t \not\redd}
                        \hypo{u \redd u'}
                        \infer3[(\ruleEsR)]{t \match{\const{\ttc}{\ptuple}}{u} \redd t \match{\const{\ttc}{\ptuple}}{u'}}
                    \end{prooftree}
                    \\[.8cm]
                    \begin{prooftree}
                        \hypo{\bv{\MC} \cap \fv{t} = \eset} \infer1[(\ruleC)]{\case{\MC \lhole{\const{\ttc}{\vect{u_i}_n}}}{\vect{\ldots, \branch{\const{\ttc}{\vect{p_i}_n}}{t}, \ldots}} \redd \MC \lhole{t \match{p_1}{u_1} \cdots \match{p_n}{u_n}}}
                    \end{prooftree}
                    \\[.5cm]
                    \begin{prooftree}
                        \hypo{t \redd t'} \hypo{\neg\isconst[\ttc_i]{t} \text{ for all } i \in \interval{1}{n}} \infer2[(\ruleCaseIn)]{\case{t}{\vect{\branch{\const{\ttc_i}{\ptuple_i}}{u_i}}_n} \redd \case{t'}{\vect{\branch{\const{\ttc_i}{\ptuple_i}}{u_i}}_n}}
                    \end{prooftree}
                \end{array} \]}
    \end{tcolorbox}
    \caption{The Weak Head Evaluation Strategy $\redd$.}
    \label{fig:figure1}
\end{figure}

As expected, $\redd$ can be taken as an evaluation strategy:

\begin{restatable}[Determinism]{proposition}{propdeterminism}
    \label{prop:determinism}
    The reduction relation $\redd$ is deterministic.
\end{restatable}

Note that the evaluation sequence in~\cref{ex1} follows, in particular, the weak head evaluation strategy defined above. Moreover, it is not difficult to show that $\redd$ has the same set of normal forms as $\red{\HC}$. Thus, since every reduction path in $\red{\HC}$ has the same length, it is safe to work with $\redd$ instead of $\red{\HC}$.

\para{Encoding Exceptions.} Here, we portray the expressive power of the $\calc$-calculus, by giving a concrete example. Indeed, the following fragment of the $\calc$-calculus is
enough to encode exceptions à la Moggi~\cite{Moggi89,Moggi91}:
{\small \[ \begin{array}{rcl}
        v, w & ::=  & x \mid \lam x.t
        \\
        t, u & ::=  & \const{\name{v}}{(v)} \mid \const{\name{e}}{(t)} \mid v w \mid t \match{x}{u} \mid \case{t}{(\branch{\const{\name{v}}{(x)}}{u}, \branch{\const{\name{e}}{(y)}}{y})} \\
             & \mid & \case{t}{(\branch{\const{\name{v}}{(x)}}{u}, \branch{\const{\name{e}}{(y)}}{\name{e}(y)})}
    \end{array} \]} where $\name{v}$ and
$\name{e}$ are unary constructors used to tag values and exceptions,
respectively. Terms $\const{\name{v}}{(v)}$ and $\const{\name{e}}{(t)}$
distinguish between values and exceptions. Case expressions of the form
$\case{t}{(\branch{\const{\name{v}}{(x)}}{u},
        \branch{\const{\name{e}}{(y)}}{{y}})}$ compose terms sequentially (\ie
they force $t$ to be evaluated before $u$), to deal with the fact that
terms either reduce to a value or to an exception with a continuation.
Note that in the latter case, the continuation must be returned. Moreover, an application of the form $t u$ can be encoded as
    {\small \[ \begin{array}{c}
                \case{t}{(\branch{\const{\name{v}}{(x)}}{\case{u}{(\branch{\const{\name{v}}{(y)}}{xy}, \branch{\const{\name{e}}{(z)}}{\name{e}(z)})}}, \branch{\const{\name{e}}{(z)}}{\name{e}(z)})}
            \end{array}\]}

The following examples illustrate three terms encoding the expected behavior of a term of the form $(\lam x.u) t$ in the presence of exceptions, depending on whether $t$ evaluates (successfully) to $\name{v}(v)$ or (exceptionally) to $\name{e}(r)$:
{\small \[ \begin{array}{rcl}
        t_1 & = & \case{\const{\name{v}}{(v)}}{(\branch{\const{\name{v}}{(x)}}{u}, \branch{\const{\name{e}}{(y)}}{y})} \plusredd u \subs{x}{v}
        \\
        t_2 & = &
        \case{\const{\name{e}}{(r)}}{(\branch{\const{\name{v}}{(x)}}{u},
            \branch{\const{\name{e}}{(y)}}{y})} \plusredd r
        \\
        t_3 & = &
        \case{\const{\name{e}}{(r)}}{(\branch{\const{\name{v}}{(x)}}{u},
            \branch{\const{\name{e}}{(y)}}{\name{e}(y)})} \plusredd \name{e}(r)
    \end{array} \]}

The first term $t_1$ encodes a successful application $(\lam x.u) (\const{\name{v}}{(v)}) \plusredd u \subs{x}{v}$; $t_2$ encodes a short-circuited application $(\lam x.u) (\const{\name{e}}{(r)}) \rightsquigarrow r$ that handles the exception $\name{e}$; and $t_3$ encodes a short-circuited application $(\lam x.u) (\const{\name{e}}{(r)}) \rightsquigarrow \name{e}(r)$ that simply propagates exception $\name{e}$ outwards.


\para{Normal Forms and Match Operations.} In order to  syntactically describe the set of irreducible forms for $\redd$ (and thus for $\red{\HC}$), it is necessary to place special care in the treatment of \defn{stuck matchings}. The latter can arise in two ways: (1) from rule (\ruleBeta) whenever the abstraction is built from a data pattern; and (2) from rule (\ruleC). Consider the two following examples:
{\small \[ \begin{array}{ll}
        (\lam \const{\name{pair}}{(x,y)}.y) (\const{\name{duo}}{(t,u)}) \red{\ruleBeta} y \match{\name{pair}(x,y)}{\const{\name{duo}}{(t,u)}} \not\red{\ruleM} & \quad (1)
        \\
        \case{\const{\name{duo}}{(t,u)}}{(\branch{\const{\name{one}}{(x)}}{x},
        \branch{\const{\name{pair}{(x,y)}}}{y})} \not\red{\ruleC}                                                                                              & \quad
        (2)
    \end{array} \]}
In both cases, evaluation is \emph{stuck} due to a
stuck matching. Clearly, these irreducible terms are not the desired
results of computations. Moreover, the same is true if
$\const{\name{duo}}{(t,u)}$ is replaced with an abstraction. These
kinds of situations will be dealt with later by introducing the
notion of \defn{clash} and the set of \defn{clash-free} normal forms
(see~\cref{lem:syntactic-char-of-clash-free-d-normal-forms}). Still,
to provide a syntactical description of irreducible terms,
these situations need to be considered. Note that stuck matchings
that result from matching a data pattern with a different
constructor (in the examples above, data $\const{\name{duo}}{(t,u)}$
is matched against patterns $\const{\name{pair}}{(x,y)}$ and
$\const{\name{one}}{(x)}$) can be described by: (1) inspecting
the matching operation $\match{\hat{p}}{t}$ in a matching closure $u
    \match{\hat{p}}{t}$, and checking
that whenever $t$ is data, it has the same constructor as $\hat{p}$;
(2) inspecting $t$ in a case expression $\case{t}{(b_1, \ldots,
        b_n)}$, and checking that whenever
$t$ is data, it has the same constructor as the data pattern of one
of the branches $b_1, \ldots, b_n$. Naturally, different data
patterns in closure matchings and sets of data patterns in vectors
of branches generate different sets of stuck matching. Thus, the set of irreducible of the
$\calc$-calculus is the union over the sets of normal forms with
respect to a particular constructor $\ttc \in \setofconsts$. To capture
that, we define three categories of terms:
\defn{neutral}, \defn{neutral data}, and \defn{normal}. The
set of \defn{neutral forms} is written $\ne$ and does not depend on
any constructor. The \defn{set of normal (resp. neutral data)
    forms} is written $\no$ (resp. $\na$) and defined as $\no \defeq
    \bigcup_{\ttc \in \setofconsts} \no[\ttc]$ (resp. $\na \defeq
    \bigcup_{\ttc \in \setofconsts} \na[\ttc]$), where
$\no[\ttc]$ (resp. $\na[\ttc]$) is the \defn{set of normal (resp.
    neutral data) forms with respect to $\ttc \in \setofconsts$}.

Let $\setofconsts' \subseteq \setofconsts$. Then, $\no[\neg \setofconsts']
    \defeq \{t \in \no[\ttc] \mid \ttc \in \setofconsts \setminus \setofconsts '\}$,
and $\HC \lhole{\no[\neg \setofconsts']} \defeq \{ \HC \lhole{t} \mid  t \in
    \no[\neg \setofconsts'] \}$. Then, $\no[\ttc]$ is generated by the following grammars:
{\small\[ \begin{array}{rrcl}
        \textbf{(Neutral)}      & \ne\      & ::= & x \mid \na\ t \mid \ne \match{\const{\ttc'}{\ptuple}}{\no[\neg \ttc']}                        \mid \case{\no[\neg \{ \ttc_1, \ldots, \ttc_n\}]}{\vect{\branch{\const{\ttc_i}{\ptuple_i}}{u_i}}_n}
        \\
        \textbf{(Neutral Data)} & \na[\ttc] & ::= & \ne \mid \const{\ttc}{\ttuple} \mid \na[\ttc] \match{\const{\ttc'}{\ptuple}}{\no[\neg \ttc']}
        \\
        \textbf{(Normal)}       & \no[\ttc] & ::= & \na[\ttc] \mid \lam p.t \mid \no[\ttc]  \match{\const{\ttc'}{\ptuple}}{\no[\neg \ttc']}
    \end{array} \]}
and $\no[\neg \ttc' ]$ is written for $\no[\neg \{\ttc'\} ]$. Intuitively, $t \in \no[\ttc]$ if and only if $t$ is an irreducible term, and either
$\neg\isconst{t}$ or $\isconst[\ttc]{t}$. The set of neutral forms $\ne$
corresponds to irreducible terms that do not produce any redexes whenever plugged
into any context $\HC$, \ie the set of terms $t \in \no[\ttc]$, such that
$\neg\isabs{t}$ and $\neg\isdata{t}$ hold. The set $\na[\ttc]$ is the set of
irreducible terms (with respect to $\ttc$) that do not produce any redexes whenever
plugged into a context of the form $\HC\ t$, \ie the set of terms $t
    \in \no[\ttc]$, such that $\neg\isabs{t}$ holds.

\begin{example}
    \label{ex2}
    The following terms are normal forms:
    {\small \[ \begin{array}{c}
            \case{\const{\name{duo}}{(\I,\I)}}{(\branch{\const{\name{pair}}{(x,y)}}{y})} \in \ne \qquad \case{\I}{(\branch{\const{\name{pair}}{(x,y)}}{y})} \in \ne
            \\
            \qquad \const{\name{pair}}{(\I \I, \I)} \in \no[\name{pair}] \subseteq \no \hspace{3cm} \const{\name{pair}}{(\I \I,\I)}\ \I \in \no
        \end{array} \]}
    In particular, note that $\const{\name{duo}}{(\I,\I)} \in \no[\name{duo}] \subseteq \no[\neg \name{pair}]$ and $\I \in \no[\neg \name{pair}]$.
\end{example}

Set $\no$ precisely captures the notion of irreducible term as follows.

\begin{restatable}[Normal Forms]{lemma}{lemsyntnfs} %
    \label{lem:charnfs}
    Let $t \in \terms$. Then, $t \in \no$ iff $t \not\redd$.
\end{restatable}

Note that the set of normal forms was defined for \emph{open} terms. This is a technical detail that allows induction to go through smoothly in the proofs. The set of normal forms for \emph{closed} term (programs) is obtained by removing $x$ from the definition of normal forms and assuming every term to be closed.

\para{Clashes.} As was discussed in the introduction and also mentioned at the beginning of this section, not all normal forms are the desired results of a computation. Indeed, some normal forms are semantically meaningless. Of course, programs containing meaningless subprograms are also meaningless. Such programs are called \emph{clashes} and are generated by the following grammars:
{\small \[ \begin{array}{rrcl}
        \textbf{(Base Clashes)} & \BCl & ::=  & \MC \lhole{\const{\ttc}{\ttuple}}\ u \mid t \match{\const{\ttc}{\ptuple}}{\MC \lhole{\lam q.u}} \mid t \match{\const{\ttc}{\ptuple}}{\MC \lhole{\const{\ttc'}{\ttuple}}} \text{ where } \ttc \neq \ttc'
        \\
                                &      & \mid & \case{\MC \lhole{\lam p.t}}{\btuple}
        \\
                                &      & \mid & \case{\MC \lhole{\const{\ttc}{\ttuple}}}{\vect{\branch{\const{\ttc_i}{\ptuple_i}}{u_i}}_n} \text{ where } \ttc \not\in \{\ttc_1, \ldots, \ttc_n\}
        \\
        \textbf{(Clashes)}      & \Cl  & ::=  & \BCl \mid \Cl\ t \mid \Cl \match{p}{u} \mid t \match{p}{\Cl} \mid \case{\Cl}{\btuple}
    \end{array} \]}

\begin{example}
    \label{ex3}
    Going back to~\cref{ex2}, term $\const{\name{pair}}{(\I \I, \I)} \not\in \Cl$ is \emph{not} a clash. But the three following terms are clashes:
    {\small \[ \begin{array}{c}
            \case{\const{\name{duo}}{(\I,\I)}}{(\branch{\const{\name{pair}}{(x,y)}}{y})} \in \Cl
            \\
            \case{\I}{(\branch{\const{\name{pair}}{(x,y)}}{y})} \in \Cl \qquad \const{\name{pair}}{(\I \I,\I)}\ \I \in \Cl
        \end{array} \]}
\end{example}

Clearly, it is easy to refine the set of normal forms by excluding clashes. The resulting set is called the \defn{set of clash-free normal forms} and is generated by the following grammars:
{\small \[ \begin{array}{rrcl}
        \textbf{(Clash-Free Ne. Forms)} & \ncf & ::= & x \mid \ncf\ t \mid \ncf \match{\const{\ttc}{\ptuple}}{\ncf} \mid \case{\ncf}{\btuple}
        \\
        \textbf{(Clash-Free No. Forms)} & \cf  & ::= & \ncf \mid \lam p.t \mid \const{\ttc}{\ttuple} \mid \cf \match{\const{\ttc}{\ptuple}}{\ncf}
    \end{array} \]}

Set $\cf$ precisely captures the notion of clash-free normal form as follows.

\begin{restatable}[Clash-Free Normal Forms]{lemma}{lemsyntacticclashfree}%
    \label{lem:syntactic-char-of-clash-free-d-normal-forms}
    Let $t \in \terms$. Then, $t \in \cf$ iff $t \not\redd$ and $t \not\in \Cl$.
\end{restatable}

Again, note that the set of clash-free normal forms was defined for \emph{open} terms. Unsurprisingly, by restricting this set to \emph{closed} terms, the set of clash-free normal forms collapses as follows.

\begin{restatable}[Closed Clash-Free Normal Forms]{lemma}{lemsyntacticclosed}%
    \label{lem:syntactic-char-of-closed-closed-d-normal-forms}
    Let $t \in \terms$ be a closed term. Then, $t \not\redd$ and $t \not\in \Cl$ iff $t = \lam p.u$ or $t = \const{\ttc}{\ttuple}$ for $\ttc \in \setofconsts$.
\end{restatable}

Programs that are
neither clashes nor in normal form can still \emph{evaluate} to clashes. As
an example, term $((\lam x.\const{\name{pair}}{(\I,\I)}) \I) \I \not\in \Cl$, but evaluates to the \emph{clash} $\const{\name{pair}}{(\I,\I)}\ \I \in \Cl$ from~\cref{ex3}.

Formally, a program $t$ is a \defn{clash-free program} iff there is no $u
    \in \Cl$ such that $t \plusredd u$. It is clearly not possible to provide a
syntactical characterization of clash-freeness. However, as mentioned in the
introduction, type systems can be used in order to provide a \emph{logical} one.
Indeed, a term $t$ is \defn{weak head terminating} if weak head evaluation terminates for
$t$ with a clash-free normal form. In~\cref{sec:types}, we purpose such a type
system, which can be seen as a natural extension of type system $\mathcal{U}$
from~\cite{Alves2019} to data (patterns) and case expressions, thus
characterizing the set of weak head terminating terms of the $\calc$-calculus.

\section{The \texorpdfstring{$\calc$-Calculus}{Pattern-Matching Calculus} as a Subsuming Framework}
\label{sec:encodings}

In this section, we take a slight detour in order to explore some of the
encoding capabilities of the $\calc$-calculus. In particular, it is shown that
our language encodes Plotkin's CBV and CBN $\lam$-calculus~\cite{Plotkin75}, and
also the bang-calculus, written $\lam^\oc$, that was proposed
in~\cite{BucciarelliKRV23} as a subsuming framework for CBN and CBV
$\lam$-calculus based on Girard's encodings~\cite{Girard1987}. While the
$\calc$-calculus is clearly capturing expressive pattern matching programs, it
is neither natural nor intuitive that is also able to subsume two different
calling paradigms such as  CBN and CBV. This is an interesting observation of
this work.

\para{Plotkin's CBN.} It is easy to see that Plotkin's CBN
$\lam$-calculus is a strict subset of our language. The set of \defn{terms} and \defn{CBN contexts} of Plotkin's CBN $\lam$-calculus are generated by the following grammars:
{\small \[ \begin{array}{c}
        \textbf{(Terms)} \       t, u     ::= x \mid \lam x.t \mid t u
        \qquad
        \textbf{(CBN Contexts)} \ \name{N} ::= \square \mid \name{N} t
    \end{array} \]}

The CBN reduction relation $\ra_\name{n}$ is defined as the closure by (pure) CBN contexts $\name{N}$ of the $\beta$-rule:
{\small \[ \begin{array}{rcl}
        (\lam x.t) u & \red{\beta} & t \subs{x}{u}
    \end{array} \]}

Plotkin's CBN $\lam$-calculus can be fully embedded into the $\calc$-calculus through the identity translation. Let $\Lambda_\name{n}$ denote the set of terms of Plotkin's CBN $\lam$-calculus, respectively and $\calc^\name{n}$ be the image of the aforementioned (trivial) translation, extended with matching closures restricted to variable patterns and terms $t \match{x}{u}$. The set of $\defn{terms}$ of the $\calc^\name{n}$-calculus is generated by the grammar:
{\small \[ \begin{array}{crcl}
        \textbf{(Terms)} & t, u & ::= & x \mid t u \mid t \match{x}{u}
    \end{array} \]}

The operational semantics of Plotkin's CBN is captured as follows.
\begin{restatable}[CBN Simulation]{lemma}{lemsimulcbn}
    For $t, u \in \Lambda_{\name{n}}$, $t \plusred{\name{n}} u$ implies $t \plusred{\HC} u$.
\end{restatable}

\para{Plotkin's CBV.} More interestingly, Plotkin's CBV $\lam$-calculus can also
be encoded into the $\calc$-calculus. The sets of \defn{values}, \defn{terms},
and \defn{CBV contexts} of Plotkin's CBV $\lam$-calculus are generated by the
following grammars:
{\small \[ \hspace{-2pt}\begin{array}{c}
        \begin{array}{c}
            \textbf{(Values)}       \ v        ::= x \mid \lam x.t
            \quad
            \textbf{(Terms)}        \ t, u     ::= v \mid t u
            \quad
            \textbf{(CBV Contexts)} \ \name{V} ::= \square \mid \name{V} t \mid v \name{V}
        \end{array}

    \end{array} \]}

The CBV reduction relation $\ra_\name{v}$ is defined as the closure by CBV contexts $\name{V}$ of the usual $\beta$-rule restricted to values: %
\[ \begin{array}{rcl}
        (\lam x.t) v & \red{\beta_\name{v}} & t \subs{x}{v}
    \end{array} \]
Plotkin's CBV $\lam$-calculus can be fully embedded into the
$\calc$-calculus through translation $\trans{\_}$. Let $\Lambda_\name{v}$
and $\mathcal{V}$ denote the sets of terms and values of Plotkin's CBV
$\lam$-calculus, respectively. Translation $\trans{\_}$ is defined as
follows in 2 columns:
{\small \[ \begin{array}{c}
        \begin{array}{rcl}
            \trans{\_}  & : & \Lambda_{\name{v}} \mapsto \terms
            \\[.1cm]
            \trans{v}   & = & \const{\name{v}}{(\transtwo{v})}
            \\
            \trans{t u} & = & (xy) \match{\const{\name{v}}{(y)}}{\trans{u}} \match{\const{\name{v}}{(x)}}{\trans{t}}
        \end{array}
        \hspace{.5cm}
        \begin{array}{rcl}
            \transtwo{\_}       & : & \mathcal{V} \mapsto \terms
            \\[.1cm]
            \transtwo{x}        & = & x
            \\
            \transtwo{\lam x.t} & = & \lam x.\trans{t}
        \end{array}
    \end{array} \]}
where $\name{v}$ is a unary constructor used to tag values. This allows us to capture the operational semantics of  Plotkin's CBV:

\begin{restatable}[CBV Simulation]{lemma}{lemsimul} %
    \label{prop:simulcbv}
    For $t, u \in \Lambda_{\name{v}}$, $t \plusred{\name{v}} u$ implies $\trans{t} \plusred{\HC} \trans{u}$.
\end{restatable}

Note that the use of the explicit matching operation
in the translation of an application is crucial in order to obtain the simulation result above. Since in Plotkin's CBV, an application of the form $tu$ can (in general) be evaluated by first evaluating either $t$ or $u$, thus translation $\trans{\_}$ must allow this as well. Let $t \ra_\name{v} t'$. Consider an alternative translation $\trans{tu} = (\lam \name{v}(x).\lam \name{v}(y).xy) \trans{t} \trans{u}$. Then, it is not difficult to see that~\cref{prop:simulcbv} does not hold anymore. Now, consider another alternative translation $\trans{tu} = \case{u}{(\branch{\name{v}(y)}{\case{t}{(\branch{\name{v}(x)}{xy})}})}$. Then, $\red{\HC}$ reduction forces $u$ to be evaluated first. Alternatively, we could have considered another alternative translation $\trans{tu} = \case{u}{(\branch{\name{v}(y)}{\case{t}{(\branch{\name{v}(x)}{xy})}})}$. However, the same problem arises if one simply realizes that the translations must also take into account the case where $u \ra_\name{v} u'$ happens first.

\para{The Bang-Calculus.} The \defn{terms}, \defn{list contexts}, and \defn{weak head surface contexts} of the $\lam^\oc$-calculus are generated by the grammars\footnote{Some formulations of the bang-calculus also use a \emph{dereliction} constructor, but this can be encoded using the identity.}:
{\small \[ \begin{array}{rrcl}
        \textbf{(Terms)}                      & t, u & ::= & x \mid \lam x.t \mid t u \mid \oc t \mid t \match{x}{u}
        \\
        \textbf{(List Contexts)}              & \LC  & ::= & \square \mid \LC \match{x}{t}
        \\[.1cm]
        \textbf{(Weak Head Surface Contexts)} & \WC  & ::= & \square \mid \WC t \mid \WC \match{x}{u} \mid t \match{x}{\WC}
    \end{array} \]}

The reduction relation $\red{\oc}$ is defined as the closure by weak head surface contexts of the following rules:
{\small \[ \begin{array}{rcl}
        \LC \lhole{\lam x.t}\ u        & \red{\name{dB}} & \LC \lhole{t \match{x}{u}} \ \text{if $\bv{\LC} \cap \fv{u} = \eset$}
        \\
        t \match{x}{\LC \lhole{\oc u}} & \red{\name{s}}  & \LC \lhole{t \subs{x}{u}} \ \text{if $\bv{\LC} \cap \fv{t} = \eset$}
    \end{array} \]}

The $\lam^\oc$-calculus can be fully embedded into the $\calc$-calculus through translation $\transml{\_}$. Let $\Lambda^\oc$ denote the set of terms of the $\lam^\oc$-calculus. Translation $\transml{\_} : \Lambda^\oc \mapsto \terms$ is defined as follows:
{\small \[ \begin{array}[t]{c}
        \begin{array}[t]{rcl}
            \transml{x}        & = & x
            \\
            \transml{\lam x.t} & = & \lam \const{\name{b}}{(x)}.\transml{t}
            \\
            \transml{t u}      & = & \transml{t} \transml{u}
        \end{array}
        \qquad
        \begin{array}[t]{rcl}
            \transml{\oc t}          & = & \const{\name{b}}{(\transml{t})}
            \\
            \transml{t \match{x}{u}} & = & \transml{t} \match{\const{\name{b}}{(x)}}{\transml{u}}
        \end{array}
    \end{array} \]}
where $\name{b}$ is a unary constructor used to tag banged terms. This allows us to capture the operational semantics of the $\lam^\oc$-calculus:

\begin{restatable}[Bang Simulation]{lemma}{lemsimulml}%
    For $t, u \in \Lambda^\oc$, $t \plusred{\oc} u$ implies $\transml{t} \plusred{\HC} \transml{u}$.
\end{restatable}

\section{Type System}
\label{sec:types}
The type system for the $\calc$-calculus is inspired by the non-idempotent type system in~\cite{Alves2019}. We consider the following grammars:
{\small \[ \begin{array}{rrcl}
        \textbf{(Data Types)}     & \A   & ::= & \const{\ttc}{\vect{\M_i}_n}
        \\
        \textbf{(Multiset Types)} & \M   & ::= & \mul{\sig_i}_{\iI} \text{ where $I$ is a finite set}
        \\
        \textbf{(Term Types)}     & \sig & ::= & \A \mid \atom \mid \M \ta \sigma                     \\
        \textbf{(Types)}          & \T   & ::= & \sig \mid \M
    \end{array} \]}

Notice that we distinguish between \defn{multiset types} and \defn{term types}. As will be evident once we introduce the type system, every term can be typed with the empty multiset type, so we will only consider a term to be typable if it can be assigned a term type. This is made more precise once we discuss typing derivations. The size of multiset type $\M$ (\ie its length) is denoted by $\len{\M}$. The empty multiset type is denoted by $\emul$. Multiset union is denoted by $\sqcup$. Data types are tagged products of multiset types. Type $\atom$ is a special constant type  that is used to type abstractions for which no arguments are provided.

\para{Typing contexts.} Capital Greek letters $\Gam$ and $\Del$ are used to
denote \defn{typing contexts}, which are defined as total functions from
variables to multiset types that map finitely many variables to the non-empty
multiset type. The \defn{domain of a typing context} $\Gam$ is denoted by
$\dom{\Gam}$ and defined as $\dom{\Gam} = \{x \mid \Gam(x) \neq \emul\}$. The
\defn{union of two typing contexts} $\Gam$ and $\Del$ is denoted by $\Gam +
    \Del$ and defined as {$(\Gam + \Del)(x) = \Gam(x) \sqcup \Del(x)$}. This notation can be naturally extended to an
arbitrary number of typing contexts. Whenever $\dom{\Gam} \cap \dom{\Del} =
    \eset$, the union is written as $\Gam; \Del$. In particular, $\Gam; x : \emul$
is identified as $\Gam$. The restriction of a typing context $\Gam$ to a set of
variables $X \subseteq \setofvars$ is denoted by $\Gam \restto{X}$, the
complementary is denoted by $\Gam \mminus X$, and they are defined as follows:
{\small \[ \begin{array}{l@{\hspace{.5cm}}r}
        (\Gam \restto{X})(x) = \left\{
        \begin{array}{ll}
            \Gam(x) & \text{if $x \in X$}
            \\
            \emul   & \text{otherwise}
        \end{array}
        \right.
         &
        (\Gam \mminus X)(x) = \left\{
        \begin{array}{ll}
            \emul   & \text{if $x \in X$}
            \\
            \Gam(x) & \text{otherwise}
        \end{array}
        \right.
    \end{array}
\]}
The \defn{type system} for the $\calc$-calculus is denoted by $\typsys$ and defined by the typing rules in~\cref{fig:figure2}, where $\interval{1}{n}$ is used as shorthand for $\{1, \ldots, n\}$.

\begin{figure}[ht]
    \begin{tcolorbox}[colback=white]
        {\small \[ \begin{array}{c}
                    \begin{prooftree}
                        \hypo{\phantom{BUUUUUU}}
                        \infer1[(\rulePatV)]{\seqp{x : \M}{x}{\M}}
                    \end{prooftree}
                    \vspace{-5pt}
                    \sep
                    \vspace{-5pt}
                    \begin{prooftree}
                        \hypo{(\Gam_i \Vdash p_i : \M_i)_{i \in \interval{1}{n}}}
                        \infer1[(\rulePatC)]{\seqp{\otimes_{i \in \interval{1}{n}} \Gam_i}{\const{\ttc}{(p_1, \ldots, p_n)}}{\mul{\const{\ttc}{(\M_1, \ldots, \M_n)}}}}
                    \end{prooftree}
                    \\[.7cm]
                    \hline
                    \hline
                    \\
                    \begin{prooftree}
                        \hypo{\phantom{BUUUUU}}
                        \infer1[(\ruleAx)]{\seq{x : \mul{\sig}}{x}{\sig}}
                    \end{prooftree}
                    \sep
                    \begin{prooftree}
                        \hypo{\seq{\Gam_i}{t}{\sig_i}}
                        \delims{\left(}{\right)_{\iI}}
                        \infer1[(\ruleMany)]{\seq{\otimes_{\iI} \Gam_i}{t}{\mul{\sig_i}_{\iI}}}
                    \end{prooftree}
                    \\[.5cm]
                    \begin{prooftree}
                        \hypo{\seq{\Gam}{t}{\sig}}
                        \hypo{\seqp{\Gam \restto{\var{p}}}{p}{\M}}
                        \infer2[(\ruleAbs)]{\seq{\Gam \mminus \var{p}}{\lam p.t}{\M \ta \sig}}
                    \end{prooftree}
                    \sep
                    \begin{prooftree}
                        \hypo{\phantom{BUUUUU}}
                        \infer1[(\ruleAbsStar)]{\seq{}{\lam p.t}{\atom}}
                    \end{prooftree}
                    \\[.5cm]
                    \begin{prooftree}
                        \hypo{\seq{\Gam}{t}{\M \ta \sig}}
                        \hypo{\seq{\Del}{u}{\M}}
                        \infer2[(\ruleApp)]{\seq{\Gam \otimes \Del}{t u}{\sig}}
                    \end{prooftree}
                    \\[.5cm]
                    \begin{prooftree}
                        \hypo{\seq{\Gam_i}{t_i}{\M_i}}
                        \delims{\left(}{\right)_{i \in \interval{1}{n}}}
                        \infer1[(\ruleConst)]{\seq{\otimes_{i \in \interval{1}{n}} \Gam_i}{\const{\ttc}{(t_1, \ldots, t_n)}}{\const{\ttc}{(\M_1, \ldots, \M_n)}}}
                    \end{prooftree}
                    \\[.5cm]
                    \begin{prooftree}
                        \hypo{\seq{\Gam}{t}{\sig}}
                        \hypo{\seqp{\Gam \restto{\var{p}}}{p}{\M}}
                        \hypo{\seq{\Del}{u}{\M}}
                        \infer3[(\ruleMatch)]{\seq{(\Gam \mminus \var{p}) \otimes \Del}{t \match{p}{u}}{\sig}}
                    \end{prooftree}
                    \\[.5cm]
                    \begin{prooftree}
                        \hypo{\seq{\Del}{t}{\M}}
                        \hypo{\seqp{\Gam \restto{\var{\hat{p}_k}}}{\hat{p}_k}{\M}}
                        \hypo{\seq{\Gam}{u_k}{\sig}}
                        \hypo{ \text{for some $k \in \interval{1}{n}$}}
                        \infer4[(\ruleCase)]{\seq{(\Gam \mminus \var{\hat{p}_i}) \otimes \Del}{\case{t}{\vect{\branch{\hat{p}_i}{u_i}}_n}}{\sig}}
                    \end{prooftree}
                \end{array} \]}
    \end{tcolorbox}
    \caption{Type System for the $\calc$-calculus.}
    \label{fig:figure2}
\end{figure}

\para{Typing derivations.} Capital Greek letters $\Phi$, $\Psi$, $\Sig$, and
$\Pi$ are used to denoted \defn{typing derivations}. Term type derivations
ending with the sequent $\seq{\Gam}{t}{\sig}$ (resp. $\seq{\Gam}{t}{\M}$) are
denoted by $\Phi \tr \seq{\Gam}{t}{\sig}$ (resp. $\Phi \tr \seq{\Gam}{t}{\M}$),
and pattern type derivations ending with the sequent $\seqp{\Gam}{p}{\M}$ are
denoted by $\Pi \tr \seqp{\Del}{p}{\M}$. The \defn{size of a type derivation
    $\Phi$ (resp. $\Pi$)} is denoted by $\sz{\Phi}$ (resp. $\sz{\Pi}$) and defined
as the number of rules in $\Phi$ (resp. $\Pi$) except rules (\ruleMany) and
(\ruleMatch). Indeed, rule (\ruleMany) is more like a meta-rule (it could be
removed but it makes proofs simpler); rule (\ruleMatch) types a term $t
    \match{p}{u}$, where the real information about the number of steps comes from
$t$, $p$ and $u$, but not from the explicit matching operator itself. A term $t$
(resp. pattern $p$) is \defn{typable} if there exists a derivation $\Phi \tr
    \seq{\Gam}{t}{\sig}$ (respectively, $\Pi \tr \seqp{\Gam}{p}{\M}$).

\para{The Typing Rules.} Most of the rules are straightforward. Rules
(\rulePatV) and (\rulePatC) are used to type variable patterns and data
patterns, respectively. The remaining rules are used to type terms. Rule
(\ruleAbsStar) is necessary due to evaluation being weak  \ie not occurring in the body of $\lam$-abstractions. Rule (\ruleCase) is
very similar to (\ruleMatch). However, the former appears to have a
non-deterministic flavor, but it is not. As we show (see~\cref{thm:one}), all
typable programs terminate in either a generalized $\lam$-abstraction or a data.
More specifically, programs typed with arrow types terminate in generalized
$\lam$-abstractions, and programs typed with data pattern types terminate in
data with the same top-level tag. Consider a typable case expression $t =
    \case{u}{\vect{\branch{\hat{p}_i}{s_i}}_n}$. Then, $t$ must be typed using rule
(\ruleCase), and $u$ must be typable as well. The type assigned to $u$ must
match the type assigned to one of the data patterns in the tuple of branches
$\vect{\branch{\hat{p}_i}{s_i}}_n$. Recall that all $\hat{p}_i$ are distinct by
construction. Therefore, their types are all different and the type of $t$ can
only match the type of the data pattern of \emph{exactly} one of the branches.
In sum, rule (\ruleCase), \emph{is} deterministic.

Let us recall the following term from~\cref{ex1}, that we call here $t_0$:
{\small \[ \begin{array}{c}
        t_0 = (\lam x.\case{x}{(\branch{\const{\name{pair}}{(x,y)}}{y}, \branch{\const{\name{triple}}{(x,y,z)}}{x})})\ \const{\name{triple}}{(\ttc_0, \ttc_1, \ttc_2)}
    \end{array} \]}
We can build a type derivation for $t_0$, by first considering the following type derivations for the subterms of $t_0$:
\\ \\
{\small$\Phi_1$:
\[ \begin{prooftree}
        \infer0[(\ruleAx)]{\seq{x : \mul{\const{\name{triple}}{(\mul{\ttc_0},\emul,\emul)}}}{x}{\const{\name{triple}}{(\mul{\ttc_0},\emul,\emul)}}}
        \infer1[(\ruleMany)]{\seq{x : \mul{\const{\name{triple}}{(\mul{\ttc_0},\emul,\emul)}}}{x}{\mul{\const{\name{triple}}{(\mul{\ttc_0},\emul,\emul)}}}}
    \end{prooftree} \]
$\Phi_2$:
\[ \begin{prooftree}
        \infer0[(\rulePatV)]{\seqp{x : \mul{\ttc_0}}{x}{\mul{\ttc_0}}}
        \infer0[(\rulePatV)]{\seqp{}{y}{\emul}}
        \infer0[(\rulePatV)]{\seqp{}{z}{\emul}}
        \infer3[(\rulePatC)]{\seqp{x : \mul{\ttc_0}}{\const{\name{triple}}{(x,y,z)}}{\mul{\const{\name{triple}}{(\mul{\ttc_0},\emul,\emul)}}}}
    \end{prooftree} \]
$\Psi_1$:
\[ \begin{prooftree}
        \hypo{\Phi_1}
        \hypo{\Phi_2}
        \infer0[(\ruleAx)]{\seq{x : \mul{\ttc_0}}{x}{\ttc_0}}
        \infer3[(\ruleCase)]{\seq{x : \mul{\const{\name{triple}}{(\mul{\ttc_0},\emul,\emul)}}}{\case{x}{(\branch{\const{\name{pair}}{(x,y)}}{y},\branch{\const{\name{triple}}{(x,y,z)}}{x})}}{\ttc_0}}
        \infer1[(\ruleAbs)]{\seq{}{\lam x.\case{x}{(\branch{\const{\name{pair}}{(x,y)}}{y},\branch{\const{\name{triple}}{(x,y,z)}}{x})}}{\mul{\const{\name{triple}}{(\mul{\ttc_0},\emul,\emul)}} \ra \ttc_0}}
    \end{prooftree} \]
$\Psi_2$:
\[ \begin{prooftree}
        \infer0[(\ruleConst)]{\seq{}{\ttc_0}{\ttc_0}}
        \infer1[(\ruleMany)]{\seq{}{\ttc_0}{\mul{\ttc_0}}}
        \infer0[(\ruleMany)]{\seq{}{\ttc_1}{\emul}}
        \infer0[(\ruleMany)]{\seq{}{\ttc_2}{\emul}}
        \infer3[(\ruleConst)]{\seq{}{\const{\name{triple}}{(\ttc_0,\ttc_1,\ttc_2)}}{\const{\name{triple}}{(\mul{\ttc_0},\emul,\emul)}}}
        \infer1[(\ruleMany)]{\seq{}{\const{\name{triple}}{(\ttc_0,\ttc_1,\ttc_2)}}{\mul{\const{\name{triple}}{(\mul{\ttc_0},\emul,\emul)}}}}
    \end{prooftree} \]}

The following derivation is a type derivation for $t_0$:
$\Sig:$
\[ \begin{prooftree}
        \hypo{\Psi_1}
        \hypo{\Psi_2}
        \infer2[(\ruleApp)]{\seq{}{(\lam
                x.\case{x}{(\branch{\const{\name{pair}}{(x,y)}}{y},\branch{\const{\name{triple}}{(x,y,z)}}{x})})\
                \const{\name{triple}}{(\ttc_0,\ttc_1,\ttc_2)}}{\ttc_0}} \end{prooftree}
\] Note that $\sz{\Sig} = 11$, which is bigger than that number of
evaluation steps taking $t_0$ to normal form, which is 6.

We now proceed to show our main result, which is the characterization of
termination with respect to clash-free normal forms.

\subsection{Characterizing Termination}

\para{Preliminary Properties.} Relevance holds whenever typing contexts have complete information about free variables.

\begin{restatable}[Relevance]{lemma}{lemrelevance}
    \label{lem:relevance}
    Let $t \in \calc$ and $\Phi \tr \seq{\Gam}{t}{\sig}$. Then, $\dom{\Gam} \subseteq \fv{t}$. Moreover, if $t$ is closed, then $\Gam = \eset$.
\end{restatable}

The following property tells us that terms are typable with a multiset type $\M$
if and only if, they can be typed with any sub(multi)set of $\M$ without losing
any information.

\begin{restatable}[Split and Merge for Multiset Types]{lemma}{lemsplitmerge}
    \label{lem:split-and-merge-for-multi-types}
    Let $t \in \calc$ and $\M = \sqcup_{\iI} \M_i$. Then, there exists $\Phi \tr \seq{\Gam}{t}{\M}$ iff there exist $(\Phi_i \tr \seq{\Gam_i}{t}{\M_i})_{\iI}$, s.t. $\Gam = \otimes_{\iI} \Gam_i$. Moreover, $\sz{\Phi} = +_{\iI} \sz{\Phi_i}$.
\end{restatable}

\para{Soundness.} We start by proving the correctness of type system $\typsys$.
In particular, we prove that typability implies clash-freedom
(\cref{lem:typability-implies-cfness}). We start by proving that clashes
(errors) cannot be typed.

\begin{restatable}[Clashes are not Typable]{lemma}{lemclashesnottypable}%
    \label{lem:clashes-are-not-typable}
    If $t \in \Cl$, there is no $\Phi \tr \seq{\Gam}{t}{\sig}$.
\end{restatable}

Now, we can use a \emph{quantitative} version of the subject reduction property
in order to show that all typable terms are clash-free.

\begin{restatable}[Weighted Subject Reduction]{lemma}{lemwightedsubjred}
    \label{lem:weighted-subject-reduction}
    Let $t, t' \in \calc$ and there exists $\Phi_t \tr \seq{\Gam}{t}{\sig}$. If $t \redd t'$, then there exists $\Phi_{t'} \tr \seq{\Gam}{t'}{\sig}$, s.t. $\sz{\Phi_t} > \sz{\Phi_{t'}}$.
\end{restatable}

\begin{proof}
    The proof is by induction on $\redd$. The most interesting case is $t \redd[(\ruleE)] t'$, which needs a weighted version of the usual substitution lemma.
\end{proof}

The quantitative flavor of~\cref{lem:weighted-subject-reduction} gives us a good
measure that ensures termination. Together
with~\cref{lem:clashes-are-not-typable}, the preservation of typings allows us
to prove that typable terms cannot terminate in a clash.

\begin{restatable}[Typability Implies Clash-Freedom]{lemma}{lemtypabilitycfness}
    \label{lem:typability-implies-cfness}
    Let $t \in \calc$. If $\seq{\Gam}{t}{\sig}$, then $t$ is clash-free.
\end{restatable}

\begin{proof}
    We reason by contraposition. Assume $t$ is \emph{not} clash-free. Then, there exists $t' \in \Cl$, such that $t \plusredd t'$. Therefore, either $t=t'$, in which case $\Phi \tr \seq{\Gam}{t'}{\sig}$ by hypothesis, or $t \redd t'$ and by~\cref{lem:weighted-subject-reduction}, we can conclude that there exists $\Phi \tr \seq{\Gam}{t'}{\sig}$. However, this cannot be the case according to~\cref{lem:clashes-are-not-typable}. Therefore, there is no such $t'$ and thus $t$ is clash-free.
\end{proof}

\para{Completeness.} To show the completeness of type system $\typsys$ we need
to show that all (closed) clash-free normal forms are typable.

\begin{restatable}[Typability of Closed Clash-Free Normal Forms]{lemma}{lemtypability}
    \label{lem:typability-of-closed-cf-d-normal-forms}
    Let $t \in \calc$ be a closed term. If $t \in \CF$, then there exists $\Phi\ \tr \seq{}{t}{\sig}$.
\end{restatable}

Then, we can use a \emph{quantitative} version of the subject expansion property
in order to show that all typable terms are clash-free.

\begin{restatable}[Weighted Subject Expansion]{lemma}{lemweightedsubjexp}
    \label{lem:weighted-subject-expansion}
    Let $t, t' \in \calc$ and there exist $\Phi_{t'} \tr \seq{\Gam}{t'}{\sig}$. If $t \redd t'$, thena there exists $\Phi_t \tr \seq{\Gam}{t}{\sig}$, s.t. $\sz{\Phi_t} > \sz{\Phi_{t'}}$.
\end{restatable}

\begin{proof}
    The proof is by induction on $\redd$. The most interesting case is $t \redd[(\ruleE)] t'$, which needs a weighted version of the usual anti-substitution lemma.
\end{proof}

\subsection{Characterization of Weak Head-Termination for Closed Terms}

We are finally able to state and prove the main results of this work.

\begin{restatable}[]{theorem}{thm}
    \label{thm:one}
    Let $t \in \calc$ be a closed term. Then, $t$ is typable iff $t$ is weak head terminating. Moreover, if $\Phi \tr \seq{\Gam}{t}{\sig}$, then $t$ weak head evaluation terminates in at most $\sz{\Phi}$ steps.
\end{restatable}

\begin{proof}
    The left-to-right implication follows from the weighted subject reduction property (\cref{lem:weighted-subject-reduction}). The right-to-left implication follows by the typability of all closed terms in clash-free normal form (\cref{lem:typability-of-closed-cf-d-normal-forms}) and weighted subject expansion (\cref{lem:weighted-subject-expansion}). In particular, the ``moreover'' part of the statement follows from the quantitative nature of~\cref{lem:weighted-subject-reduction} and~\cref{lem:weighted-subject-expansion}.
\end{proof}

An interesting corollary of this theorem is that typability also implies termination of the weak head reduction $\red{\HC}$, since it is not difficult to prove that $t \not\redd$ implies $t \not\red{\HC}$. Moreover,  if $t$ terminates for the weak head reduction $\red{\HC}$,
then $t$ terminates for the weak head evaluation $\redd$, so that we can conclude:
\begin{corollary}
    Let $t \in \calc$ be a closed term. Then, $t$ is typable iff $t$
    terminates for the weak head relation $\red{\HC}$.
\end{corollary}

\section{Conclusion and Future Work}
\label{sec:conclusions}

This work provides a quantitative insight into the semantics of programming
languages with pattern matching. In particular, we define a quantitative type
system $\typsys$ that characterizes weak head termination for the
$\calc$-calculus and provides upper-bounds for the number of steps needed to
fully evaluate a term to normal form. Crucially, we also prove that system
$\typsys$ ensures that ``well-typed programs do not go wrong'' or, equivalently,
by using the terminology adopted in this work, that typable programs are
clash-free.

Future work includes adopting call-by-need (CBNeed) in order to fully capture
the operational semantics of lazy functional programming languages such as
Haskell. Indeed, in~\cite{AccattoliB17} it is shown that pattern matching steps
are negligible for CBN only when using a CBNeed strategy, so it would be
fruitful to develop a CBNeed version of our calculus. We also believe that type
system $\mathcal{E}$ from~\cite{Alves2019} can be naturally extended to the
$\calc$-calculus. This would allow us to obtain exact measures (instead of
upper-bounds) for evaluation. The calculus in~\cite{Barenbaum18} includes a
fixpoint, we would like to consider such an extension as well. Something that
looks challenging is to consider an open version of $\calc$-calculus. This would
allow us to extend the results in~\cite{BucciarelliKR21} to generalized
$\lam$-abstractions and case expressions. Last, but not least, it would be
interesting, and challenging, to explore \emph{nondeterministic} versions of
case expressions. These could be used to simulate logic programming languages
(which usually have nondeterministic evaluation strategies) within a functional
setting.

\newpage

\bibliographystyle{splncs04}
\bibliography{refs}

\newpage

\appendix

\section{Proofs}
\label{sec:appendix}

\subsection{The Pattern-Matching Calculus}

\subsubsection{\textit{Confluence.}}

\begin{restatable}[Substitution Lemma]{lemma}{lemsubstitutionlemma}
    \label{lem:substitution-lemma}
    Let $s, u, r \in \calc$ and $y \notin \fv{u}$. Then, $s \subs{x}{u} \subs{y}{r \subs{x}{u}} = s \subs{y}{r} \subs{x}{u}$.
\end{restatable}


\maybehide{\begin{proof}
    By induction on $s$.
\end{proof}
}

\begin{restatable}[Preservation of Substitutions]{lemma}{lempreservationofsubstitutions}
    \label{lem:preservation-of-substitutions}
    Let $t, u, r \in \calc$. If $t \red{\HC(s)} r$ then $t \subs{x}{u} \red{\HC(s)} r \subs{x}{u}$.
\end{restatable}


\maybehide{\begin{proof}
    The proof follows by induction over $t \red{\HC} t'$:
    \begin{itemize}
        \item Case $t = \MC \lhole{\lam p.s} r \red{\HC(\ruleBeta)} \MC \lhole{s \match{p}{r}} = t'$, such that $\bv{\MC} \cap \fv{r} = \eset$. By $\alpha$-conversion we can suppose in particular that  $\var{p} \cap \fv{u} = \eset$. We proceed by induction over $\MC$:
              \begin{itemize}
                  \item Case $\MC = \ehole$. Then, $t = (\lam p.s) r \red{\HC(\ruleBeta)} s \match{p}{r} = t'$. Clearly, $((\lam p.s) r) \subs{x}{u} = (\lam p.s \subs{x}{u}) (r \subs{x}{u})$. Moreover, $(\lam p.s \subs{x}{u}) (r \subs{x}{u}) \red{\HC(\ruleBeta)} s \subs{x}{u} \match{p}{r \subs{x}{u}}$. Therefore, we can conclude since $s \subs{x}{u} \match{p}{r \subs{x}{u}} = s \match{p}{r} \subs{x}{u}$ (because $\var{p} \cap \fv{u} = \eset$).
                  \item Case $\MC = \MC_1 \match{q}{v}$. Then, $t = \MC_1 \lhole{\lam p.s} \match{q}{v} r \red{\HC(\ruleBeta)} \MC_1 \lhole{s \match{p}{r}} \match{q}{v} = t'$, such that $\bv{\MC_1 \match{q}{v}} \cap \fv{r} = \eset$. Notice that $\MC_1 \lhole{\lam p.s} r \red{\HC(\ruleBeta)} \MC_1 \lhole{s \match{p}{r}}$. So, we can apply the \ih, obtaining $(\MC_1 \lhole{\lam p.s} r) \subs{x}{u} \red{\HC(\ruleBeta)} \MC_1 \lhole{s \match{p}{r}} \subs{x}{u}$. Therefore, $t_1 = \MC_1 \lhole{\lam p.s}\subs{x}{u} \match{q}{v\subs{x}{u}} r \subs{x}{u} \red{\HC(\ruleBeta)} \MC_1 \subs{x}{u} \lhole{s \subs{x}{u} \match{p}{r \subs{x}{u}}} \match{q}{v \subs{x}{u}} = t'_1$. And we can conclude since $t_1 = (\MC_1 \lhole{\lam p.s} \match{q}{v} r) \subs{x}{u}$ and $t'_1 = (\MC_1 \lhole{s \match{p}{r}} \match{q}{v}) \subs{x}{u}$ (because $\var{q} \cap \fv{u} = \eset$ and $\var{p} \cap \fv{u} = \eset$).
              \end{itemize}
        \item Case $t = \case{\MC \lhole{\const{\ttc}{\vect{s_i}_n}}}{(\ldots, \branch{\const{\ttc}{\vect{p_i}_n}}{r}, \ldots)} \red{\HC(\ruleC)} \MC \lhole{r \match{p_1}{s_1} \cdots \match{p_n}{s_n}} = t'$. By $\alpha$-conversion we can assume $\var{\const{\ttc}{\vect{p_i}_n}} \cap \fv{u} = \eset$. We proceed by induction over $\MC$:
              \begin{itemize}
                  \item Case $\MC = \ehole$. Then, $t = \case{\const{\ttc}{\vect{s_i}_n}}{(\ldots, \branch{\const{\ttc}{\vect{p_i}_n}}{r}, \ldots)} \red{\HC(\ruleC)} r \match{p_1}{s_1} \cdots \match{p_n}{s_n} = t'$. Clearly, $(\case{\const{\ttc}{\vect{s_i}_n}}{(\ldots, \branch{\const{\ttc}{\vect{p_i}_n}}{r}, \ldots)}) \subs{x}{u} = \case{\const{\ttc}{\vect{s_i \subs{x}{u}}_n}}{(\ldots, \branch{\const{\ttc}{\vect{p_i}_n}}{r \subs{x}{u}}, \ldots)}$. Moreover, \\ $\case{\const{\ttc}{\vect{s_i \subs{x}{u}}_n}}{(\ldots, \branch{\const{\ttc}{\vect{p_i}_n}}{r \subs{x}{u}}, \ldots)} \red{\HC(\ruleC)} r \subs{x}{u} \match{p_1}{r_1 \subs{x}{u}} \cdots \match{p_n}{r_n \subs{x}{u}}$. Therefore, we can conclude since $r \subs{x}{u} \match{p_1}{r_1 \subs{x}{u}} \cdots \match{p_n}{r_n \subs{x}{u}} = r \match{p_1}{r_1} \cdots \match{p_n}{r_n} \subs{x}{u}$ (because $\var{\const{\ttc}{\vect{p_i}_n}} \cap \fv{u} = \eset$).
                  \item Case $\MC = \MC_1 \match{q}{v}$. Then, $t = \case{\MC_1 \lhole{\const{\ttc}{\vect{s_i}_n}} \match{q}{v}}{(\ldots, \branch{\const{\ttc}{\vect{p_i}_n}}{r}, \ldots)} \red{\HC(\ruleC)} \MC_1 \lhole{r \match{p_1}{s_1} \cdots \match{p_n}{s_n}} \match{q}{v} = t'$. Notice that $\case{\MC_1 \lhole{\const{\ttc}{\vect{s_i}_n}}}{(\ldots, \branch{\const{\ttc}{\vect{p_i}_n}}{r}, \ldots)} \red{\HC(\ruleC)} \MC_1 \lhole{r \match{p_1}{s_1} \cdots \match{p_n}{s_n}}$. By the \ih, $(\case{\MC_1 \lhole{\const{\ttc}{\vect{s_i}_n}}}{(\ldots, \branch{\const{\ttc}{\vect{p_i}_n}}{r}, \ldots)}) \subs{x}{u} \red{\HC(\ruleC)} \MC_1 \lhole{r \match{p_1}{s_1} \cdots \match{p_n}{s_n}} \subs{x}{u}$. Therefore, $t_1 = (\case{\MC_1 \lhole{\const{\ttc}{\vect{s_i}_n}}}{(\ldots, \branch{\const{\ttc}{\vect{p_i}_n}}{r}, \ldots)})\subs{x}{u} \match{q}{v \subs{x}{u}}$ $\red{\HC(\ruleC)} \MC_1 \lhole{r \match{p_1}{s_1} \cdots \match{p_n}{s_n}} \subs{x}{u} \match{q}{v \subs{x}{u}} = t'_1$. And we can conclude since $t_1 = (\case{\MC_1 \lhole{\const{\ttc}{\vect{s_i}_n}} \match{q}{v}}{(\ldots, \branch{\const{\ttc}{\vect{p_i}_n}}{r}, \ldots)}) \subs{x}{u}$ and $t'_1 =  (\MC_1 \lhole{r \match{p_1}{s_1} \cdots \match{p_n}{s_n}} \match{q}{v}) \subs{x}{u}$ (because $\var{q} \cap \fv{u} = \eset$ and $\var{\const{\ttc}{\vect{p_i}_n}} \cap \fv{u}  = \eset$).
              \end{itemize}
        \item Case $t = s \match{\const{\ttc}{\vect{p_i}_n}}{\MC \lhole{\const{\ttc}{\vect{r_i}_n}}} \red{\HC(\ruleM)} \MC \lhole{s \match{p_1}{r_1} \cdots \match{p_n}{r_n}} = t'$, such that $\bv{\MC} \cap \fv{s} = \eset$. Moreover, by $\alpha$-conversion we can suppose $\var{\const{\ttc}{\vect{p_i}_n}} \cap \fv{u} = \eset$.  We proceed by induction over $\MC$:
              \begin{itemize}
                  \item Case $\MC = \ehole$. Then, $t = s \match{\const{\ttc}{\vect{p_i}_n}}{\const{\ttc}{\vect{r_i}_n}} \red{\HC(\ruleM)} s \match{p_1}{r_1} \cdots \match{p_n}{r_n} = t'$. Clearly, $(s \match{\const{\ttc}{\vect{p_i}_n}}{\const{\ttc}{\vect{r_i}_n}}) \subs{x}{u} = s \subs{x}{u} \match{\const{\ttc}{\vect{p_i}_n}}{\const{\ttc}{\vect{r_i \subs{x}{u}}_n}}$. Moreover, $s \subs{x}{u} \match{\const{\ttc}{\vect{p_i}_n}}{\const{\ttc}{\vect{r_i \subs{x}{u}}_n}}$ $\red{\HC(\ruleM)} s \subs{x}{u} \match{p_1}{r_1 \subs{x}{u}} \cdots \match{p_n}{r_n \subs{x}{u}}$. Therefore, we can conclude since $s \subs{x}{u} \match{p_1}{r_1 \subs{x}{u}} \cdots \match{p_n}{r_n \subs{x}{u}} = s \match{p_1}{r_1} \cdots \match{p_n}{r_n} \subs{x}{u}$ (because $\var{\const{\ttc}{\vect{p_i}_n}} \cap \fv{u} = \eset$).
                  \item Case $\MC = \MC_1 \match{q}{v}$. Then, $t = s \match{\const{\ttc}{\vect{p_i}_n}}{\MC_1 \lhole{\const{\ttc}{\vect{r_i}_n}} \match{q}{v}} \red{\HC(\ruleM)} \MC_1 \lhole{s \match{p_1}{r_1} \cdots \match{p_n}{r_n}} \match{q}{v} = t'$, such that $\bv{\MC_1 \match{q}{v}} \cap \fv{u} = \eset$. Notice that $s \match{\const{\ttc}{\vect{p_i}_n}}{\MC_1 \lhole{\const{\ttc}{\vect{r_i}_n}}} \red{\HC(\ruleM)} \MC_1 \lhole{s \match{p_1}{r_1} \cdots \match{p_n}{r_n}}$. So, we can apply the \ih, obtaining $s \match{\const{\ttc}{\vect{p_i}_n}}{\MC_1 \lhole{\const{\ttc}{\vect{r_i}_n}}} \subs{x}{u} \red{\HC(\ruleM)} \MC_1 \lhole{s \match{p_1}{r_1} \cdots \match{p_n}{r_n}} \subs{x}{u}$. Therefore, $t_1 = s \match{\const{\ttc}{\vect{p_i}_n}}{\MC_1 \lhole{\const{\ttc}{\vect{r_i}_n}}} \subs{x}{u} \match{q}{v \subs{x}{u}} \red{\HC(\ruleM)} \MC_1 \lhole{s \match{p_1}{r_1} \cdots \match{p_n}{r_n}} \subs{x}{u} \match{q}{v \subs{x}{u}} = t'_1$. And we can conclude with $t_1 = s \match{\const{\ttc}{\vect{p_i}_n}}{\MC_1 \lhole{\const{\ttc}{\vect{r_i}_n}} \match{q}{v}} \subs{x}{u}$ and $t'_1 = \MC_1 \lhole{s \match{p_1}{r_1} \cdots \match{p_n}{r_n}} \match{q}{v} \subs{x}{u}$ (because $\var{q} \cap \fv{u} = \eset$ and $\var{\const{\ttc}{\vect{p_i}_n}} \cap \fv{u} = \eset$.
              \end{itemize}
        \item Case $t = s \match{y}{r} \red{\HC(\ruleE)} s \subs{y}{r} = t'$. We assume by $\alpha$-conversion that $y \notin \fv{u}$. Notice that $s \match{y}{r} \subs{x}{u} = s \subs{x}{u} \match{y}{r \subs{x}{u}}$. Moreover, $s \subs{x}{u} \match{y}{r \subs{x}{u}} \red{\HC(\ruleE)} s \subs{x}{u} \subs{y}{r \subs{x}{u}}$. And we can conclude since $s \subs{x}{u} \subs{y}{r \subs{x}{u}} = s \subs{y}{r} \subs{x}{u}$ by the substitution~\cref{lem:substitution-lemma}.
        \item Case $t = s \match{p}{r} \red{\HC(s)} s' \match{p}{r} = t'$, where $s \red{\HC(s)} s'$. We assume by $\alpha$-conversion that $\var{p} \cap \fv{u} = \eset$. By the \ih, we know that $s \subs{x}{u} \red{\HC(s)} s' \subs{x}{u}$. Therefore, $s \subs{x}{u} \match{p}{r \subs{x}{u}} \red{\HC(s)} s' \subs{x}{u} \match{p}{r \subs{x}{u}}$. And we can conclude since $s \subs{x}{u} \match{p}{r \subs{x}{u}} = s \match{p}{r} \subs{x}{u}$ and $s' \subs{x}{u} \match{p}{r \subs{x}{u}} = s' \match{p}{r} \subs{x}{u}$ (because $\var{p} \cap \fv{u} = \eset$).
        \item Case $t = sr \red{\HC(s)} s'r = t'$, where $s \red{\HC(s)} s'$. By the \ih, we know that $s \subs{x}{u} \red{\HC(s)} s' \subs{x}{u}$. Therefore, $s \subs{x}{u} (r \subs{x}{u}) \red{\HC(s)} s' \subs{x}{u} (r \subs{x}{u})$. And we can conclude since $s \subs{x}{u} (r \subs{x}{u}) = (sr) \subs{x}{u}$ and $s' \subs{x}{u} (r \subs{x}{u}) = (s'r) \subs{x}{u}$.
        \item Case $t = s \match{\hat{p}}{r} \red{\HC(s)} s \match{\hat{p}}{r'} = t'$, where $r \red{\HC(s)} r'$. By the \ih, we know that $r \subs{x}{u} \red{\HC(s)} r' \subs{x}{u}$. We assume by $\alpha$-conversion that $\var{\hat{p}} \cap \fv{u} = \eset$. Therefore, $s \subs{x}{u} \match{\hat{p}}{r \subs{x}{u}} \red{\HC} s \subs{x}{u} \match{\hat{p}}{r' \subs{x}{u}}$. And we can conclude since $s \subs{x}{u} \match{\hat{p}}{r \subs{x}{u}} = s \match{\hat{p}}{r} \subs{x}{u}$ and $s \subs{x}{u} \match{\hat{p}}{r' \subs{x}{u}} = s \match{\hat{p}}{r'} \subs{x}{u}$ (because $\var{\hat{p}} \cap \fv{u} = \eset$).
        \item Case $t = \case{s}{\vect{\branch{\const{\ttc_i}{\ptuple_i}}{r_i}}_n} \red{\HC(s)} \case{s'}{\vect{\branch{\const{\ttc_i}{\ptuple_i}}{r_i}}_n} = t'$, where $s \red{\HC(s)} s'$. By $\alpha$-convertion we can assume in particular that $\bigcup\limits_{i \in \interval{1}{n}} \var{\const{\ttc_i}{\ptuple_i}} \cap \fv{u} = \eset$. By the \ih, we know that $s \subs{x}{u} \red{\HC(s)} s' \subs{x}{u}$. Therefore, $\case{s \subs{x}{u}}{\vect{\branch{\const{\ttc_i}{\ptuple_i}}{r_i}}_n \subs{x}{u}} \red{\HC(s)} \case{s' \subs{x}{u}}{\vect{\branch{\const{\ttc_i}{\ptuple_i}}{r_i}}_n \subs{x}{u}}$. And we can conclude since $\case{s \subs{x}{u}}{\vect{\branch{\const{\ttc_i}{\ptuple_i}}{r_i}}_n \subs{x}{u}} = (\case{s}{\vect{\branch{\const{\ttc_i}{\ptuple_i}}{r_i}}_n}) \subs{x}{u}$ and $\case{s' \subs{x}{u}}{\vect{\branch{\const{\ttc_i}{\ptuple_i}}{r_i}}_n \subs{x}{u}} = (\case{s'}{\vect{\branch{\const{\ttc_i}{\ptuple_i}}{r_i}}_n}) \subs{x}{u}$ (because $\bigcup\limits_{i \in \interval{1}{n}} \var{\const{\ttc_i}{\ptuple_i}} \cap \fv{u} = \eset$).
    \end{itemize}
\end{proof}
}

\lemdiamond*

\maybehide{\begin{proof}
    We prove that, if  $t \red{\HC(s_1)} t_1$ and $t \red{\HC(s_2)} t_2$, with $t_1 \neq t_2$, there exists $t_3$ such that $t_1 \red{\HC(s_2)} t_3$ and $t_2 \red{\HC(s_1)} t_3$.
    The proof follows by induction over $t$:
    \begin{itemize}
        \item Case $t = x$. We can conclude immediately, since $x \not\red{\HC}$.
        \item Case $t = us$. We have to consider four possible diverging cases:
              \begin{itemize}
                  \item Let $u = \MC_1 \lhole{\MC_2 \lhole{\lam q.r} \match{x}{v}}$, $s_1 = \ruleBeta$ and $t_1 = \MC_1 \lhole{\MC_2 \lhole{r \match{q}{s}} \match{x}{v}}$, such that $\bv{\MC_2 \match{x}{v} \MC_1} \cap \fv{s} = \eset$ (so in particular $x \notin \fv{s}$), and $s_2 = \ruleE$ and $t_2 = \MC_1 \lhole{\MC_2 \lhole{\lam q.r} \subs{x}{v}} s$. We can take $t_3 = \MC_1 \lhole{\MC_2 \lhole{r \match{q}{s}} \subs{x}{v}}$ and close the diagram:
                        \[ \begin{tikzcd}
                                \MC_1 \lhole{\MC_2 \lhole{\lam q.r} \match{x}{v}} s
                                \arrow{r}{\HC(\ruleBeta)}
                                \arrow[swap]{d}{\HC(\ruleE)} &
                                \arrow{d}{\HC(\ruleE)}
                                \MC_1 \lhole{\MC_2 \lhole{r \match{q}{s}} \match{x}{v}}
                                \\
                                \begin{array}{c}
                                    \MC_1 \lhole{\MC_2 \lhole{\lam q.r} \subs{x}{v}} s
                                    \\[-.2cm]
                                    =
                                    \\[-.2cm]
                                    \MC_1 \lhole{\MC_2 \subs{x}{v} \lhole{\lam q.r \subs{x}{v}}} s
                                \end{array}
                                \arrow{r}{}[swap]{\HC(\ruleBeta)} &
                                \begin{array}{c}
                                    \MC_1 \lhole{\MC_2 \lhole{r \match{q}{s}} \subs{x}{v}}
                                    \\[-.2cm]
                                    =
                                    \\[-.2cm]
                                    \MC_1 \lhole{\MC_2 \subs{x}{v} \lhole{r \subs{x}{v} \match{q}{s \subs{x}{v}}}}
                                \end{array}
                            \end{tikzcd} \]
                        We can conclude since $s \subs{x}{v} =s $.
                  \item Let $u = \MC_1 \lhole{\MC_2 \lhole{\lam q.r} \match{\const{\ttc}{\vect{p_i}_n}}{\MC_3 \lhole{\const{\ttc}{\vect{v_i}_n}}}}$, $s_1 = \ruleBeta$ and $t_1 = \MC_1 \lhole{\MC_2 \lhole{r \match{q}{s}} \match{\const{\ttc}{\vect{p_i}_n}}{\MC_3 \lhole{\const{\ttc}{\vect{v_i}_n}}}}$, such that $\bv{\MC_2 \match{\const{\ttc}{\vect{p_i}_n}}{\MC_3 \lhole{\const{\ttc}{\vect{v_i}_n}}} \MC_1} \cap \fv{s} = \eset$, and $s_2 = \ruleM$ and $t_2 = \MC_1 \lhole{\MC_3 \lhole{\MC_2 \lhole{\lam q.r} \match{p_1}{v_1} \cdots \match{p_n}{v_n}}}$, such that $\bv{\MC_3} \cap \fv{\MC_2 \lhole{\lam q.r}} = \eset$. We can take $t_3 = \MC_1 \lhole{\MC_3 \lhole{\MC_2 \lhole{r \match{q}{s}} \match{p_1}{v_1} \cdots \match{p_n}{v_n}}}$ and close the diagram as follows:
                        \[ \begin{tikzcd}
                                \MC_1 \lhole{\MC_2 \lhole{\lam q.r} \match{\const{\ttc}{\vect{p_i}_n}}{\MC_3 \lhole{\const{\ttc}{\vect{v_i}_n}}}} s
                                \arrow{r}{\HC(\ruleBeta)}
                                \arrow[swap]{d}{\HC(\ruleM)} &
                                \arrow{d}{\HC(\ruleM)}
                                \MC_1 \lhole{\MC_2 \lhole{r \match{q}{s}} \match{\const{\ttc}{\vect{p_i}_n}}{\MC_3 \lhole{\const{\ttc}{\vect{v_i}_n}}}}
                                \\
                                \MC_1 \lhole{\MC_3 \lhole{\MC_2 \lhole{\lam q.r} \match{p_1}{v_1} \cdots \match{p_n}{v_n}}} s
                                \arrow{r}{}[swap]{\HC(\ruleBeta)} &
                                \MC_1 \lhole{\MC_3 \lhole{\MC_2 \lhole{r \match{q}{s}} \match{p_1}{v_1} \cdots \match{p_n}{v_n}}}
                            \end{tikzcd} \]
              \end{itemize}
        \item Case $t = u \match{x}{s}$. Then, $t_1 = u \subs{x}{s}$, $s_1 = \ruleE$, and $t_2 = u' \match{x}{s}$, such that $u \red{\HC(s_2)} u'$. We can pick $t_3 = u' \subs{x}{s}$ and close the diagram using~\cref{lem:preservation-of-substitutions}:
              \[ \begin{tikzcd}
                      u \match{x}{s}
                      \arrow{r}{\HC(\ruleE)}
                      \arrow[swap]{d}{\HC(s_2)} &
                      \arrow{d}{\HC(s_2)}
                      u \subs{x}{s}
                      \\
                      u' \match{x}{s}
                      \arrow{r}{}[swap]{\HC(\ruleE)} &
                      u' \subs{x}{s}
                  \end{tikzcd} \]
        \item Case $t = u \match{\hat{p}}{s}$. We have to consider four possible diverging cases:
              \begin{itemize}
                  \item Let $\hat{p} = \const{\ttc}{\vect{p_i}_n}$ and $s = \MC_1 \lhole{\MC_2 \lhole{\const{\ttc}{\vect{s_i}_n}} \match{\const{\ttc'}{\vect{q_i}_m}}{\MC_3 \lhole{\const{\ttc'}{\vect{r_i}_m}}}}$. Then, \\ $t_1 = u \match{\const{\ttc}{\vect{p_i}_n}}{\MC_1 \lhole{\MC_3 \lhole{\MC_2 \lhole{\const{\ttc}{\vect{s_i}_n}} \match{q_1}{r_1} \cdots \match{q_m}{r_m}}}}$ and $s_1 = \ruleM$, such that $\bv{\MC_3} \cap \fv{\MC_2 \lhole{\const{\ttc}{\vect{s_i}_n}}} = \eset$, and $t_2 = \MC_1 \lhole{\MC_2 \lhole{u \match{p_1}{s_1} \cdots \match{p_n}{s_n}} \match{\const{\ttc'}{\vect{q_i}_m}}{\MC_3 \lhole{\const{\ttc'}{\vect{r_i}_m}}}}$, and $s_2 = \ruleM$, such that $\bv{\MC_2 \match{\const{\ttc'}{\vect{q_i}_m}}{\MC_3 \lhole{\const{\ttc'}{\vect{r_i}_m}}}} \cap \fv{u} = \eset$. We can pick $t_3 = \MC_1 \lhole{\MC_3 \lhole{\MC_2 \lhole{u \match{p_1}{s_1} \cdots \match{p_n}{s_n}} \match{q_1}{r_1} \cdots \match{q_m}{r_m}}}$ and close the diagram:
                        \[ \hspace{-1cm} \begin{tikzcd}
                                u \match{\const{\ttc}{\vect{p_i}_n}}{\MC_1 \lhole{\MC_2 \lhole{\const{\ttc}{\vect{s_i}_n}} \match{\const{\ttc'}{\vect{q_i}_m}}{\MC_3 \lhole{\const{\ttc'}{\vect{r_i}_m}}}}}
                                \arrow{r}{\HC(\ruleM)}
                                \arrow[swap]{d}{\HC(\ruleM)} &
                                \arrow{d}{\HC(\ruleM)}
                                u \match{\const{\ttc}{\vect{p_i}_n}}{\MC_1 \lhole{\MC_3 \lhole{\MC_2 \lhole{\const{\ttc}{\vect{s_i}_n}} \match{q_1}{r_1} \cdots \match{q_m}{r_m}}}}
                                \\
                                \MC_1 \lhole{\MC_2 \lhole{u \match{p_1}{s_1} \cdots \match{p_n}{s_n}} \match{\const{\ttc'}{\vect{q_i}_m}}{\MC_3 \lhole{\const{\ttc'}{\vect{r_i}_m}}}}
                                \arrow{r}{}[swap]{\HC(\ruleM)} &
                                \MC_1 \lhole{\MC_3 \lhole{\MC_2 \lhole{u \match{p_1}{s_1} \cdots \match{p_n}{s_n}} \match{q_1}{r_1} \cdots \match{q_m}{r_m}}}
                            \end{tikzcd} \]
                  \item Let $t_1 = u_1 \match{\hat{p}}{s}$, such that $u \red{\HC(s_1)} u_1$, $t_2 = u_2 \match{\hat{p}}{s}$, such that $u \red{\HC(s_2)} u_2$. By the \ih, we know there exists $u_3$, such that $u_1 \red{\HC(s_2)} u_3$ and $u_2 \red{\HC(s_1)} u_3$. Therefore, we can take $t_3 = u_3 \match{\hat{p}}{s}$ and close the diagram:
                        \[ \begin{tikzcd}
                                u \match{\hat{p}}{s}
                                \arrow{r}{\HC(s_1)}
                                \arrow[swap]{d}{\HC(s_2)} &
                                \arrow{d}{\HC(s_2)}
                                u_1 \match{\hat{p}}{s}
                                \\
                                u_2 \match{\hat{p}}{s}
                                \arrow{r}{}[swap]{\HC(s_1)} &
                                u_3 \match{\hat{p}}{s}
                            \end{tikzcd} \]
                  \item Let $t_1 = u \match{\hat{p}}{r_1}$, such that $s \red{\HC(s_1)} r_1$, and $t_2 = u \match{\hat{p}}{r_2}$, such that $s \red{\HC(s_2)} r_2$. By the \ih, we know there exists $r_3$, such that $r_1 \red{\HC(s_2)} r_3$ and $r_2 \red{\HC(s_1)} r_3$. Therefore, we can take $t_3 = u \match{\hat{p}}{r_3}$ and close the diagram:
                        \[ \begin{tikzcd}
                                u \match{\hat{p}}{s}
                                \arrow{r}{\HC(s_1)}
                                \arrow[swap]{d}{\HC(s_2)} &
                                \arrow{d}{\HC(s_2)}
                                u \match{\hat{p}}{r_1}
                                \\
                                u \match{\hat{p}}{r_2}
                                \arrow{r}{}[swap]{\HC(s_1)} &
                                u \match{\hat{p}}{r_3}
                            \end{tikzcd} \]
                  \item Let $t_1 = u' \match{\hat{p}}{s}$, such that $u \red{\HC(s_1)} u'$, and $t_2 = u \match{\hat{p}}{s'}$, such that $s \red{\HC(s_2)} s'$. We can pick $t_3 = u' \match{\hat{p}}{s'}$ and close the diagram:
                        \[ \begin{tikzcd}
                                u \match{\hat{p}}{s}
                                \arrow{r}{\HC(s_1)}
                                \arrow[swap]{d}{\HC(s_2)} &
                                \arrow{d}{\HC(s_2)}
                                u' \match{\hat{p}}{s}
                                \\
                                u \match{\hat{p}}{s}
                                \arrow{r}{}[swap]{\HC(s_1)} &
                                u' \match{\hat{p}}{s'}
                            \end{tikzcd} \]
              \end{itemize}
        \item Case $t = \case{u}{\btuple}$. We have to consider two diverging cases:
              \begin{itemize}
                  \item Let $u = \MC_1 \lhole{\MC_2 \lhole{\const{\ttc}{\vect{s_i}_m}} \match{\const{\ttc'}{\vect{p_i}_n}}{\MC_3 \lhole{\const{\ttc'}{\vect{r_i}_n}}}}$ and $\btuple = (\ldots, \branch{\const{\ttc}{\vect{q_i}_m}}{v}, \ldots)$. Then, $t_1 = \case{\MC_1 \lhole{\MC_3 \lhole{\MC_2 \lhole{\const{\ttc}{\vect{s_i}_m}} \match{p_1}{r_1} \cdots \match{p_n}{r_n}}}}{\btuple}$ and $s_1 = \ruleM$, such that $\bv{\MC_3} \cap \fv{\MC_2 \lhole{\const{\ttc}{\vect{s_i}_m}}} = \eset$, and $t_2 = \MC_1 \lhole{\MC_2 \lhole{v \match{q_1}{s_1} \cdots \match{q_m}{s_m}} \match{\const{\ttc'}{\vect{p_i}_n}}{\MC_3 \lhole{\const{\ttc'}{\vect{r_i}_n}}}}$ and $s_2 = \ruleC$, such that $\bv{\MC_2 \match{\const{\ttc'}{\vect{p_i}_n}}{\MC_3 \lhole{\const{\ttc'}{\vect{r_i}_n}}} \MC_1} \cap \fv{v} = \eset$. We can pick $t_3 = \MC_1 \lhole{\MC_3 \lhole{\MC_2 \lhole{v \match{q_1}{v_1} \cdots \match{q_m}{v_m}} \match{p_1}{r_1} \cdots \match{p_n}{r_n}}}$ and close the diagram:
                        \[ \hspace{-1cm} \begin{tikzcd}
                                \case{\MC_1 \lhole{\MC_2 \lhole{\const{\ttc}{\vect{s_i}_m}} \match{\const{\ttc'}{\vect{p_i}_n}}{\MC_3 \lhole{\const{\ttc'}{\vect{r_i}_n}}}}}{\btuple}
                                \arrow{r}{\HC(\ruleM)}
                                \arrow[swap]{d}{\HC(\ruleC)} &
                                \arrow{d}{\HC(\ruleC)}
                                \case{\MC_1 \lhole{\MC_3 \lhole{\MC_2 \lhole{\const{\ttc}{\vect{s_i}_m}} \match{p_1}{r_1} \cdots \match{p_n}{r_n}}}}{\btuple}
                                \\
                                \MC_1 \lhole{\MC_2 \lhole{v \match{q_1}{s_1} \cdots \match{q_m}{s_m}} \match{\const{\ttc'}{\vect{p_i}_n}}{\MC_3 \lhole{\const{\ttc'}{\vect{r_i}_n}}}}
                                \arrow{r}{}[swap]{\HC(\ruleM)} &
                                \MC_1 \lhole{\MC_3 \lhole{\MC_2 \lhole{v \match{q_1}{v_1} \cdots \match{q_m}{v_m}} \match{p_1}{r_1} \cdots \match{p_n}{r_n}}}
                            \end{tikzcd} \]
                  \item Let $t_1 = \case{u_1}{\btuple}$, such that $u \red{\HC(s_1)} u_1$, $t_2 = \case{u_2}{\btuple}$, such that $u \red{\HC(s_2)} u_2$. By the \ih, we know there exists $u_3$, such that $u_1 \red{\HC(s_2)} u_3$ and $u_2 \red{\HC(s_1)} u_3$. Therefore, we can take $t_3 = \case{u_3}{\btuple}$ and close the diagram:
                        \[ \begin{tikzcd}
                                \case{u}{\btuple}
                                \arrow{r}{\HC(s_1)}
                                \arrow[swap]{d}{\HC(s_2)} &
                                \arrow{d}{\HC(s_2)}
                                \case{u_1}{\btuple}
                                \\
                                \case{u_2}{\btuple}
                                \arrow{r}{}[swap]{\HC(s_1)} &
                                \case{u_3}{\btuple}
                            \end{tikzcd} \]
              \end{itemize}
    \end{itemize}
\end{proof}
}

\lemconfluence*


\subsubsection{\textit{Determinism and (Clash-Free) Normal Forms.}}

\propdeterminism*

\maybehide{\begin{proof}
    Reduction relation $\redd$ is deterministic if, for any term $t$, either $t \not\redd$ or there exists exactly one $t'$, such that $t \redd t'$. We show, by induction over the structure of $t$, that this is always the case:
    \begin{itemize}
        \item Cases $t = x$, $t = \lam p.u$, and $t = \const{\ttc}{\ttuple}$. No rule applies to $t$, so we can conclude.
        \item Case $t = us$. Let us assume that $t \redd t'$. Rules (\ruleBeta) and (\ruleAppL) may apply to $t$. However, either $u$ is of the form $\MC \lhole{\lam p.r}$ and only rule (\ruleBeta) applies, or $\neg\isabs{u}$ and only rule (\ruleAppL) applies. Let us assume that only rule (\ruleBeta) applies. Then, we can trivially conclude. Now, let us assume that only rule (\ruleAppL) applies. Then, $us \redd u's$, such that $u \redd u'$. We can conclude by the \ih.
        \item Case $t = u \match{p}{s}$. Let us assume that $t \redd t'$. Rules (\ruleM), (\ruleE), (\ruleEsL), and (\ruleEsR) may apply to $t$. However, either $p$ is a variable and only rule (\ruleE) applies, or $p$ is of the form $\const{\ttc}{\vect{q_i}_n}$ and only rules (\ruleM), (\ruleEsL) and (\ruleEsR) may apply. If rule (\ruleE) applies, then we can trivially conclude. Otherwise, either $\isconst[\ttc]{s}$ or $\neg\isconst[\ttc]{s}$. If $\isconst[\ttc]{s}$ holds, then only rule (\ruleM) applies, and we can trivially conclude. If $\neg\isconst[\ttc]{s}$ holds, then either $s \not\redd$ or $s \redd s'$. Indeed, if $s \not\redd$ then only rule (\ruleEsL) applies. In which case, $u \redd u'$, and we can conclude by the \ih over $u$. If $s \redd s'$, then only rule (\ruleEsR) applies, and we can conclude by the \ih over $s$.
        \item Case $t = \case{u}{\btuple}$. Let us assume $t \redd t'$. Rules (\ruleC) and (\ruleCaseIn) may apply to $t$. However, either $u \not\redd$ and only rule (\ruleC) applies, or $u \redd u'$ and only rule (\ruleCaseIn) applies. If rule (\ruleC) applies, then we can conclude immediately. If rule (\ruleCaseIn) applies, then $u \redd u'$, and we can conclude by the \ih  over $u$.
    \end{itemize}
\end{proof}
}


\begin{restatable}[]{lemma}{lemsyntactic}
    \label{lem:syntactic-char-of-d-normal-forms}
    $t \in \no[\ttc]$ iff $t \not\redd$, and either $\neg\isconst{t}$ or $\isconst[\ttc]{t}$.
\end{restatable}

\maybehide{\begin{proof}
    In order to show the original statement, we are going to refine it into the three following statements:
    \begin{enumerate}
        \item $t \in \ne$ iff $t \not\redd$, $\neg\isabs{t}$, and $\neg\isconst{t}$.
        \item $t \in \na[\ttc]$ iff $t \not\redd$, $\neg\isabs{t}$, and either $\neg\isconst[]{t}$ or $\isconst[\ttc]{t}$.
        \item $t \in \no[\ttc]$ iff $t \not\redd$, and either $\neg\isconst{t}$ or $\isconst[\ttc]{t}$.
    \end{enumerate}
    The proof of the left-to-right implication follows by induction over $t \in \no[\ttc]$:
    \begin{enumerate}
        \item Let $t \in \ne$:
              \begin{itemize}
                  \item Case $t = x$. Then, $x \not\redd$, $\neg\isabs{x}$, and $\neg\isconst{x}$ trivially hold.
                  \item Case $t = us$, where $u \in \na$. Since $\na = \bigcup\limits_{\ttc \in \setofconsts} \na[\ttc]$, we can assume, without loss of generality, that $u \in \na[\ttc]$, for some $\ttc \in \setofconsts$. By the \ih~(2) over $u$, we know $u \not\redd$, $\neg\isabs{u}$, and either $\neg\isconst{u}$ or $\isconst[\ttc]{u}$. Since $\neg\isabs{u}$ and $u \not\redd$, then $us \not\redd$.  Moreover, $\neg\isabs{us}$ and $\neg\isconst{us}$ trivially hold.
                  \item Case $t = u \match{\const{\ttc'}{\ptuple}}{s}$, where $u \in \ne$ and $s \in \no[\neg \ttc']$. By the \ih~(1)  over $u$, we know $u \not\redd$, $\neg\isabs{u}$, and $\neg\isconst{u}$. Since $s \in \no[\neg\ttc']$, we can assume, without loss of generality, that $s \in \no[\ttc_0]$ for some $\ttc_0 \neq \ttc'$. By the \ih~(3) over $s$, we know $s \not\redd$, and either $\neg\isconst{s}$ or $\isconst[\ttc_0]{s}$. Clearly, rule (\ruleE) does not apply to $t$. Additionally, since either $\neg\isconst{s}$ or $\isconst[\ttc_0]{s}$ for $\ttc_0 \neq \ttc'$, rule (\ruleM) also does not apply to $t$. Finally, since $s \not\redd$ and $u \not\redd$, then $u \match{\const{\ttc'}{\ptuple}}{s} \not\redd$. Moreover, $\neg\isabs{u \match{\const{\ttc'}{\ptuple}}{s}}$ and $\neg\isconst{u \match{\const{\ttc'}{\ptuple}}{s}}$ hold, since $\neg\isabs{u}$ and $\neg\isconst{u}$ hold.
                  \item Case $t = \case{u}{\vect{\branch{\const{\ttc_i}{\ptuple_i}}{u_i}}_n}$, where $u \in \no[\neg \{\ttc_1, \ldots, \ttc_n\}]$. Since $u \in \no[\neg \{\ttc_1, \ldots, \ttc_n\}]$, we can assume, without loss of generality, that $u \in \no[\ttc_0]$ for some $\ttc_0 \not\in \{\ttc_1, \ldots, \ttc_n\}$. By the \ih~(3), we know $u \not\redd$, and either $\neg\isdata{u}$ or $\isdata[\ttc_0]{u}$. Since $\neg\isdata{u}$ or $\isdata[\ttc_0]{u}$ for $\ttc_0 \not\in \{\ttc_1, \ldots, \ttc_n\}$, then rule (\ruleC) does not apply to $t$. Additionally, since $u \not\redd$, then rule (\ruleCaseIn) also does not apply to $t$. Therefore, $\case{u}{\vect{\branch{\const{\ttc_i}{\ptuple_i}}{u_i}}_n} \not\redd$. Moreover, $\neg\isabs{\case{u}{\vect{\branch{\const{\ttc_i}{\ptuple_i}}{u_i}}_n}}$ and $\neg\isdata{\case{u}{\vect{\branch{\const{\ttc_i}{\ptuple_i}}{u_i}}_n}}$ trivially hold.
              \end{itemize}
        \item Let $t \in \na[\ttc]$:
              \begin{itemize}
                  \item Case $t \in \ne$. Then, $t \not\redd$, $\neg\isabs{t}$, and $\neg\isdata{t}$ by statement (1). So, we can conclude.
                  \item Case $t = \const{\ttc}{\ttuple}$. Then, $\const{\ttc}{\ttuple} \not\redd$, $\neg\isabs{\const{\ttc}{\ttuple}}$, and $\isdata[\ttc]{\const{\ttc}{\ttuple}}$.
                  \item Case $t = u \match{\const{\ttc'}{\ptuple}}{s}$, where $u \in \na[\ttc]$ and $s \in \no[\neg \ttc']$. This case is very similar to the corresponding case when assuming $t \in \ne$.
              \end{itemize}
        \item Let $t \in \no[\ttc]$:
              \begin{itemize}
                  \item Case $t \in \na[\ttc]$. Then, $t \not\redd$, $\neg\isabs{t}$, and either $\neg\isdata{t}$ or $\isdata[\ttc]{t}$ by statement (2). So, we can conclude.
                  \item Case $t = \lam p.u$. Then, $\lam p.u \not\redd$ and $\neg\isdata{\lam p.u}$.
                  \item Case $t = u \match{\const{\ttc'}{\ptuple}}{s}$, where $u \in \no[\ttc]$ and $s \in \no[\neg \ttc']$. This case is very similar to the corresponding case when assuming $t \in \na[\ttc]$ or $t \in \ne$.
              \end{itemize}
    \end{enumerate}
    The right-to-left implication follows by induction over the structure of $t$:
    \begin{itemize}
        \item Case $t = x$. Then, $x \not\redd$, $\neg\isabs{x}$, and $\neg\isconst{x}$, and thus the hypothesis of points (1), (2), and (3)  apply. However, since $\ne \subseteq \na[\ttc] \subseteq \no[\ttc]$ for any $\ttc$, it is enough to show point (1). And we can conclude since $x \in \ne$, by definition.
        \item Case $t = us$. Then, suppose $us \not\redd$. We always have $\neg\isabs{us}$, and $\neg\isconst{us}$. Then, the hypothesis of points (1), (2), and (3) apply. But, $\ne \subseteq \na[\ttc] \subseteq \no[\ttc]$ for any $\ttc$, it is enough to show point (1). Since $us \not\redd$, then in particular $\neg\isabs{u}$ holds (otherwise rule (\ruleBeta) would apply), and $u \not\redd$ (otherwise rule (\ruleAppL) would apply). Note that it is always the case that either $\neg\isdata{u}$, or there exists $\ttc' \in \setofconsts$,  such that $\isdata[\ttc']{u}$. By the \ih~(2) over $u$, we know that $u \in \na[\ttc']$ for some $\ttc' \in \setofconsts$. Therefore, $u \in \na[\ttc'] \subseteq \na$ and $us \in \ne$.
        \item Case $t = \lam p.u$. Then, $\lam p.u \not\redd$ and $\neg\isdata{\lam p.u}$. Therefore, the hypothesis of point (3) applies. And we can conclude since $\lam p.u \in \no[\ttc]$, by definition.
        \item Case $t = u \match{p}{s}$. Then, suppose $u \match{p}{s} \not\redd$. We always have $\neg\isdata{u \match{p}{s}}$ or $\isdata[\ttc]{u \match{p}{s}}$ for some $\ttc \in \setofconsts$, which implies either $\neg\isdata{u}$ or $\isdata[\ttc]{u}$. In any case, $p$ must not be a variable (otherwise rule (\ruleE) would apply), so let us assume that $p = \const{\ttc'}{\ptuple}$, for some $\ttc' \in \setofconsts$. It is also the case that $\neg\isconst[\ttc']{s}$ must hold (otherwise rule (\ruleM) would apply), $u \not\redd$ (otherwise rule (\ruleEsL) would apply), and $s \not\redd$ (otherwise rule (\ruleEsR) would apply). Moreover, since $\neg\isdata[\ttc']{s}$, then either $\neg\isdata{s}$, or $\isdata[\ttc_0]{s}$, for some $\ttc_0 \in \setofconsts$ such that $\ttc_0 \neq \ttc'$. By the \ih~(3) over $s$, we know $s \in \no[\ttc_0]$ \ie that $s \in \no[\neg \ttc']$.
              \begin{itemize}
                  \item If we make no further assumptions regarding the form of $u$, then only the hypothesis of point (3) applies. By the induction \ih~(1) over $u$, we know $u \in \no[\ttc]$. Therefore, $u \match{p}{s} \in \no[\ttc]$ by definition, and this concludes point (3).
                  \item If we also assume that $\neg\isabs{u \match{p}{s}}$ holds, then $\neg\isabs{u}$ also holds and the hypothesis of point (2) also applies. By the \ih~(2) over $u$, we know $u \in \na[\ttc]$, for some $\ttc \in \setofconsts$. Therefore, $u \match{p}{s} \in \na[\ttc]$ by definition, and this concludes point (2).
                  \item If we assume both $\neg\isabs{u \match{p}{s}}$ and $\neg\isdata{u \match{p}{s}}$, then $\neg\isdata{u}$ also holds and the hypothesis of point (1) also applies. By the \ih (1) over $u$, we know that $u \in \ne$. Therefore, $u \match{p}{s} \in \ne$ by definition, and this concludes point (1).
              \end{itemize}
        \item Case $t = \const{\ttc}{\ttuple}$. Then, $\const{\ttc}{\ttuple} \not\redd$, $\neg\isabs{\const{\ttc}{\ttuple}}$ and $\isconst[\ttc]{\const{\ttc}{\ttuple}}$. Therefore, the hypothesis of points (2) and (3) apply. We have $\const{\ttc}{\ttuple} \in \na[\ttc]$ by definition, thus point (2) holds. Therefore, point (3) also holds since $\na[\ttc] \subseteq \no[\ttc]$.
        \item Case $t = \case{u}{\vect{\branch{\const{\ttc_i}{\ptuple_i}}{s_i}}_n}$. Then, suppose that $\case{u}{\vect{\branch{\const{\ttc_i}{\ptuple_i}}{s_i}}_n} \not\redd$. We also have $\neg\isabs{\case{u}{\vect{\branch{\const{\ttc_i}{\ptuple_i}}{s_i}}_n}}$, and $\neg\isdata{\case{u}{\vect{\branch{\const{\ttc_i}{\ptuple_i}}{s_i}}_n}}$. Therefore, the hypothesis of points (1), (2), and (3) apply. As before, it is enough to show point (1). Since $\case{u}{\vect{\branch{\const{\ttc_i}{\ptuple_i}}{s_i}}_n} \not\redd$, then $\neg\isdata[\ttc_0]{u}$, for $\ttc_0 \in \{\ttc_1, \ldots, \ttc_n\}$ must hold (otherwise rule (\ruleC) would apply), and $u \not\redd$ (otherwise rule (\ruleCaseIn) would apply). Thus, either $\neg\isdata{u}$ or $\isdata[\ttc']{u}$ for $\ttc' \not\in \{\ttc_1, \ldots, \ttc_n\}$. By the \ih~(3) over $u$, we know $u \in \no[\ttc']$, \ie $u \in \no[\neg \{\ttc_1, \ldots, \ttc_n\}]$. Therefore, $\case{u}{\vect{\branch{\const{\ttc_i}{\ptuple_i}}{s_i}}_n} \in \ne$, by definition.
    \end{itemize}
\end{proof}
}

\lemsyntnfs*

\maybehide{\begin{proof}
        Simple corollary of~\cref{lem:syntactic-char-of-d-normal-forms}.
    \end{proof}}

\lemsyntacticclashfree*

\maybehide{\begin{proof}
    To show the original statement, we are going to refine it into the following two statements:
    \begin{enumerate}
        \item $t \in \ncf$ iff $t \not\redd$, $t \not\in \Cl$, $\neg\isabs{t}$, and $\neg\isdata{t}$.
        \item $t \in \cf$ iff $t \not\redd$ and $t \not\in \Cl$.
    \end{enumerate}
    We start by showing the left-to-right implication, by induction over $t \in \cf$:
    \begin{enumerate}
        \item Let $t \in \ncf$:
              \begin{itemize}
                  \item Case $t = x$. Then, $x \not\redd$, $x \not\in \Cl$, $\neg\isabs{x}$, and $\neg\isdata{x}$ trivially hold.
                  \item Case $t = us$, such that $u \in \ncf$. By the \ih~(1) over $u$, we know $u \not\redd$, $u \not\in \Cl$, $\neg\isabs{u}$, and $\neg\isdata{u}$. We can easily conclude that $us \not\redd$ by a simple inspection of the rules. Let us assume $us \in \Cl$. Then, $us$ must be of one of the following forms:
                        \begin{itemize}
                            \item $u = \MC \lhole{\const{\ttc}{\ttuple}}$, for some $c \in \setofconsts$. Contradiction with the fact that $\neg\isdata{u}$.
                            \item $u \in \Cl$. Contradiction with the fact that $u \not\in \Cl$.
                        \end{itemize}
                        Thus, we can conclude that $us \not\in \Cl$. Moreover, $\neg\isabs{us}$ and $\neg\isdata{us}$ trivially hold.
                  \item Case $t = u \match{p}{s}$, such that $u, s \in \ncf$ and $p = \const{\ttc}{\ptuple}$. By the \ih~(1) over $u$ (resp. $s$), we know $u \not\redd$ (resp. $s \not\redd$), $u \not\in \Cl$ (resp. $s \not\in \Cl$), $\neg\isabs{u}$ (resp. $\neg\isabs{s}$), and $\neg\isdata{u}$ (resp. $\neg\isdata{s}$). We can easily conclude that $u \match{p}{s} \not\redd$ by a simple inspection of the rules. Let us assume $u \match{p}{s} \in \Cl$. Then, $u \match{p}{s}$ must be of one of the following forms:
                        \begin{itemize}
                            \item $s = \MC \lhole{\lam q.r}$. Contradiction with the fact that $\neg\isabs{s}$.
                            \item $s = \MC \lhole{\const{\ttc'}{\ttuple}}$, such that $\ttc \neq \ttc'$. Contradiction with the fact that $\neg\isdata{s}$.
                            \item $u \in \Cl$. Contradiction with the fact that $u \not\in \Cl$.
                            \item $s \in \Cl$. Contradiction with the fact that $s \not\in \Cl$.
                        \end{itemize}
                        Thus, we can conclude that $u \match{p}{s} \not\in \Cl$. Moreover, $\neg\isabs{u \match{p}{s}}$ and $\neg\isdata{u \match{p}{s}}$ hold, since $\neg\isabs{u}$ and $\neg\isdata{u}$ hold.
                  \item Case $t = \case{u}{\vect{\branch{\const{\ttc_i}{\ptuple_i}}{u_i}}_n}$, such that $u \in \ncf$. By the \ih~(1), we know $u \not\redd$, $u \not\in \Cl$, $\neg\isabs{u}$, and $\neg\isdata{u}$. We can easily conclude that $\case{u}{\vect{\branch{\const{\ttc_i}{\ptuple_i}}{u_i}}_n} \not\redd$ by a simple inspection of the rules. Let us assume $\case{u}{\vect{\branch{\const{\ttc_i}{\ptuple_i}}{u_i}}_n} \in \Cl$. Then, $\case{u}{\vect{\branch{\const{\ttc_i}{\ptuple_i}}{u_i}}_n}$ must be of one of the following forms:
                        \begin{itemize}
                            \item $u = \MC \lhole{\lam p.t}$. Contradiction with the fact that $\neg\isabs{u}$.
                            \item $u = \const{\ttc}{\ttuple}$, such that $\ttc \not\in \{\ttc_1, \ldots, \ttc_n\}$.
                                  Contradiction with the fact $\neg\isdata{u}$.
                            \item $u \in \Cl$. Contradiction with the fact that $u \not\in \Cl$.
                        \end{itemize}
                        Thus, we can conclude that $\case{u}{\vect{\branch{\const{\ttc_i}{\ptuple_i}}{u_i}}_n} \not\in \Cl$. Moreover, it is the case that $\neg\isabs{\case{u}{\vect{\branch{\const{\ttc_i}{\ptuple_i}}{u_i}}_n}}$ and $\neg\isdata{\case{u}{\vect{\branch{\const{\ttc_i}{\ptuple_i}}{u_i}}_n}}$ hold.
              \end{itemize}
        \item Case $t \in \cf$:
              \begin{itemize}
                  \item Case $t \in \ncf$. This case follows from statement (1).
                  \item Case $t = \lam p.u$. Then, $\lam p.u \not\redd$ and $\lam p.u \not\in \Cl$, by definition.
                  \item Case $t = \const{\ttc}{\ttuple}$. Then, $\const{\ttc}{\ttuple} \not\redd$ and $\const{\ttc}{\ttuple} \not\in \Cl$, by definition.
                  \item Case $t = u \match{\const{\ttc}{\ptuple}}{s}$, such that $u \in \cf$ and $s \in \ncf$. This case is very similar to the corresponding case when $t \in \ncf$.
              \end{itemize}
    \end{enumerate}
    Now, we show the right-to-left implication, by induction over the structure of $t$:
    \begin{itemize}
        \item Case $t = x$. Then $x \in \ncf \subseteq \cf$ by definition so that both points (1) and (2) hold.
        \item Case $t = \lam p.u$. Then, $\lam p.u \not\redd$, $\lam p.u \not\in \Cl$, and $\isabs{\lam p.u}$ and $\neg\isdata{\lam p.u}$. Therefore, only the hypothesis of point (2) applies. And we can conclude since $\lam p.u \in \cf$, by definition.
        \item Case $t = us$. Then, suppose $us \not\redd$ and $us \not\in \Cl$. We always have $\neg\isabs{us}$ and $\neg\isdata{us}$. Then, the hypothesis of point (1) and (2) apply. As before, it is enough to show point (1) which implies (2). Since $us \not\redd$, then $\neg\isabs{u}$ must hold (otherwise rule (\ruleBeta) would apply), and $u \not\redd$ (otherwise rule (\ruleAppL) would apply). Additionally, since $us \not\in \Cl$, then $\neg\isdata{u}$ and $u \not\in \Cl$. By the \ih~(1), we know $u \in \ncf$. Therefore, $us \in \ncf$, by definition.
        \item Case $t = u \match{p}{s}$. Then, suppose $u \match{p}{s} \not\redd$, $u \match{p}{s} \not\in \Cl$. Since $u \match{p}{s} \not\redd$, the following must be the case: $p$ must be of the form $p = \const{\ttc}{\ptuple}$ for some $\ttc \in \setofconsts$ (otherwise rule (\ruleE) would apply); $\neg\isdata[\ttc]{s}$ (otherwise rule (\ruleM) would apply); $u \not\redd$ (otherwise rule (\ruleEsL) would apply); and $s \not\redd$ (otherwise rule (\ruleEsR) would apply). Additionally, since $u \match{p}{s} \not\in \Cl$, then $\neg\isabs{s}$, $\neg\isdata[\ttc']{s}$ for any $\ttc' \neq \ttc$, $u \not\in \Cl$, and $s \not\in \Cl$. Thus, $\neg\isdata{s}$ must necessarily hold. By \ih~(1), we know $s \in \ncf$. If we make no further assumptions regarding the form of $u$, then only the hypothesis of point (2) applies. By \ih~(2), $u \in \cf$. Therefore, $u \match{p}{s} \in \cf$, by definition. If we also assume that $\neg\isabs{u \match{p}{s}}$ and $\neg\isdata{u \match{p}{s}}$ hold, then $\neg\isabs{u}$ and $\neg\isdata{u}$ also hold and the hypothesis of point (1) also applies. By \ih~(1), $u \in \ncf$. Therefore, $u \match{p}{s} \in \ncf$, by definition.
        \item Case $t = \const{\ttc}{\ttuple}$. Then, $\const{\ttc}{\ttuple} \not\redd$, $\const{\ttc}{\ttuple} \not\in \Cl$, $\neg\isabs{\const{\ttc}{\ttuple}}$ and $\isdata{\const{\ttc}{\ttuple}}$. Therefore, only the hypothesis of point (2) applies. And we can conclude since $\const{\ttc}{\ttuple} \in \cf$, by definition.
        \item Case $t = \case{u}{\vect{\branch{\const{\ttc}{\ptuple_i}}{s_i}}_n}$. Then, suppose $\case{u}{\vect{\branch{\const{\ttc}{\ptuple_i}}{s_i}}_n} \not\redd$ and $\case{u}{\vect{\branch{\const{\ttc}{\ptuple_i}}{s_i}}_n} \not\in \Cl$. We always have $\neg\isabs{\case{u}{\vect{\branch{\const{\ttc}{\ptuple_i}}{s_i}}_n}}$ and $\neg\isdata{\case{u}{\vect{\branch{\const{\ttc}{\ptuple_i}}{s_i}}_n}}$. Then, the hypothesis of points (1) and (2) apply. It is enough to show point (1) which implies point (2). Since $\case{u}{\vect{\branch{\const{\ttc}{\ptuple_i}}{s_i}}_n} \not\redd$, the following must be the case: either $\neg\isdata{u}$ or $\isdata[\ttc']{u}$ for $\ttc' \not\in \{\ttc_1, \ldots, \ttc_n\}$ (otherwise rule (\ruleC) would apply); and $u \not\redd$ (otherwise rule (\ruleCaseIn) would apply). Additionally, since $\case{u}{\vect{\branch{\const{\ttc}{\ptuple_i}}{s_i}}_n} \not\in \Cl$, then $\neg\isabs{u}$, $\neg\isdata[\ttc']{u}$ for $\ttc' \notin \{\ttc_1, \ldots, \ttc_n\}$, and $u \not\in \Cl$. Thus, $\neg\isdata{u}$ must necessarily hold. By the \ih~(1) over $u$, we know that $u \in \ncf$. Therefore, $\case{u}{\vect{\branch{\const{\ttc}{\ptuple_i}}{s_i}}_n} \in \ncf$, by definition.
    \end{itemize}
\end{proof}
}

\lemsyntacticclosed*

\maybehide{\begin{proof}\
    \begin{itemize} 
        \item[$\Ra)$] Let $t \not\redd$ and $t \not\in \Cl$. The proof follows by induction over $t$:
            \begin{itemize}
                \item Case $t = x$ does not apply because $x$ is not a closed term.
                \item Case $t = \lam p.u$ or $t = \const{\ttc}{\ttuple}$ for some $\ttc \in \setofconsts$, we can conclude immediately.
                \item Case $t = u s$. Since $us \not\redd$, we know that $\neg\isabs{u}$ and $u \not\redd$ by a simple inspection of the reduction rules. Since $us \not\in \Cl$, we know $u \not\in \Cl$. By the \ih over $u$, we know $u = \lam p.u$ or $u = \const{\ttc}{\ttuple}$ for some $\ttc \in \setofconsts$. However, we know that $\neg\isabs{u}$ must hold. Therefore, $u = \const{\ttc}{\ttuple}$. But, then $\const{\ttc}{\ttuple} \in \Cl$, which contradicts the hypothesis. Thus, this case does not apply.
                \item Case $t = u \match{p}{s}$. Since $u \match{p}{s} \not\redd$, we know that $p$ cannot be a variable because of rule (\ruleE). Let us assume, without any loss of generality, that $p = \const{\ttc}{\ptuple}$ for some $\ttc \in \setofconsts$. Then, we also know that $\neg\isdata[\ttc]{s}$, $s \not\redd$ and $u \not\redd$ by a simple inspection of the remaining reduction rules. Since $u \match{p}{s} \not\in \Cl$, we know $\neg\isabs{s}$, $\neg\isdata[\ttc']{s}$ with $\ttc' \neq \ttc$, $u \not\in \Cl$, and $s \not\in \Cl$. Therefore, we can immediately conclude that $\neg\isdata{s}$. Note that, since $u \match{p}{s}$ is closed, then so is $s$. By the \ih over $s$, we know $s = \lam p.r$ or $s = \const{\ttc_0}{\ttuple}$ for some $\ttc_0$. However, we know that $\neg\isabs{s}$ and $\neg\isdata{s}$ must hold. Thus, this case does not apply.
                \item Case $t = \case{u}{\vect{\branch{\const{\ttc_i}{\ptuple_i}}{s_i}}_n}$. Since $\case{u}{\vect{\branch{\const{\ttc_i}{\ptuple_i}}{s_i}}_n} \not\redd$, we know that $\neg\isdata[\ttc_i]{u}$ for $i \in \interval{1}{n}$ and $u \not\redd$ by a simple inspection of the reduction rules. Since $\case{u}{\vect{\branch{\const{\ttc_i}{\ptuple_i}}{s_i}}_n} \not\in \Cl$, we know $\neg\isabs{u}$, $\neg\isdata[\ttc_0]{u}$ for some $\ttc_0 \not\in \{\ttc_1, \ldots, \ttc_n\}$, and $u \not\in \Cl$. Therefore, we can immediately conclude that $\neg\isdata{u}$. Note that, since $\case{u}{\vect{\branch{\const{\ttc_i}{\ptuple_i}}{s_i}}_n}$ is closed, then so is $u$. By the \ih over $u$, we know $u = \lam x.s$ or $u = \const{\ttc}{\ttuple}$. However, we also know that $\neg\isabs{u}$ and $\neg\isdata{u}$ must hold. Thus, this case does not apply.
            \end{itemize}
        \item[$\La)$] Let $t = \lam p.u$ or $t = \const{\ttc}{\ttuple}$. Then $t \not\redd$ and $t \not\in \Cl$ hold trivially by definition.
    \end{itemize}
\end{proof}
}

\subsection{The Encoding Capabilities of the \texorpdfstring{$\calc$-Calculus}{Pattern Matching Calculus}}

\subsubsection{\textit{Plotkin's CBN.}}

\lemsimulcbn*

\begin{proof}
    Let $t \plusred{\name{n}}^n u$, where $n$ is the number of evaluation steps between $t$ and $u$. The proof follows by induction over $n$:
    \begin{itemize}
        \item Let $n = 0$. Then, $u = t$ and thus $t \plusred{\HC} u$.
        \item Let $n > 0$. Then, there exists $t'$, such that $t \red{\name{n}} t' \plusred{\name{n}} u$. By the \ih $t' \plusred{\HC} u$, therefore, it is now enough to show that $t \plusred{\HC} t'$. We reason by induction over CBN contexts:
              \begin{itemize}
                  \item Case $\name{N} = \square$. Then, $t = (\lam x.r) s \red{\beta_{\name{n}}} r \subs{x}{s} = t'$. Then, $(\lam x.r) s \red{\ruleBeta} r \match{x}{s} \red{\ruleE} r \subs{x}{s}$.
                  \item Case $\name{N} = \name{N}' r$. Then, $\name{N} \lhole{s} = \name{N}' \lhole{s} r$ and $\name{N}' \lhole{s} \red{\name{n}} \name{N}' \lhole{s'}$. By the \ih we know $\name{N}' \lhole{s}  \plusred{\HC} \name{N}' \lhole{s'}$. Therefore, $t = \name{N} \lhole{s} = \name{N}' \lhole{s} r \plusred{\HC} \name{N}' \lhole{s'} r = \name{N} \lhole{s'} = t'$.
              \end{itemize}
    \end{itemize}
\end{proof}

\subsubsection{\textit{Plotkin's CBV.}}

\begin{restatable}[Substitution Preserves Simulation for CBV]{lemma}{lemsubssimul}
    \label{lem:subs-simul}
    Let $t, v \in \Lambda_{\name{v}}$. Then, $\trans{t \subs{x}{v}} = \trans{t} \subs{x}{\transtwo{v}}$.
\end{restatable}


\begin{proof}
  By induction over $t$:
  \begin{itemize}
    \item Case $t = x$. Then, $x \subs{x}{v} = v$. Therefore, $\trans{t \subs{x}{v}} = \trans{v} = \const{\name{v}}{(\transtwo{v})}$. Thus, $\trans{t} \subs{x}{\transtwo{v}} = \const{\name{v}}{(x)} \subs{x}{\transtwo{v}} = \const{\name{v}}{(\transtwo{v})} = \trans{v}$.
    \item Case $t = y \neq x$. Then, $y \subs{x}{v} = y$. Therefore, $\trans{t \subs{x}{v}} = \trans{y}$. Also, $\trans{t} \subs{x}{\transtwo{v}} = \trans{y} \subs{x}{\transtwo{v}} = \trans{y}$.
    \item Case $t = \lambda y.t'$. Then $(\lambda y.t') \subs{v}{v}= \lambda y.t'\subs{x}{v}$, so that
          \[ \begin{array}{rll}

              \trans{(\lambda y.t') \subs{x}{v}} & =                & \name{v}(\transtwo{\lambda y.t'\subs{x}{v}})            \\
                                                 & =                & \name{v}(\lambda y.\trans{t'\subs{x}{v}})               \\
                                                 & \overset{\ih}{=} & \name{v}(\lambda y.\trans{t'}\subs{x}{\transtwo{v}})
              \\
                                                 & =                & \name{v}(\lambda y.\trans{t'})\subs{x}{\transtwo{v}}    \\
                                                 & =                & \name{v}(\transtwo{\lambda y.t'})\subs{x}{\transtwo{v}} \\
                                                 & =                & \trans{\lambda y.t'}\subs{x}{\transtwo{v}}.
            \end{array} \]
    \item Case $t = ur$. Then, $(ur) \subs{x}{v} = (u \subs{x}{v}) (r \subs{x}{v})$, so that
          \[ \begin{array}{rll}

              \trans{(ur) \subs{x}{v}} & =                & (yz) \match{\const{\name{v}}{(z)}}{\trans{r \subs{x}{v}}} \match{\const{\name{v}}{(y)}}{\trans{u \subs{x}{v}}}                       \\
                                       & \overset{\ih}{=} & (yz) \match{\const{\name{v}}{(z)}}{\trans{r} \subs{x}{\transtwo{v}}} \match{\const{\name{v}}{(y)}}{\trans{u} \subs{x}{\transtwo{v}}}
              \\
                                       & =                & ((yz) \match{\const{\name{v}}{(z)}}{\trans{r}} \match{\const{\name{v}}{(y)}}{\trans{u}}) \subs{x}{\transtwo{v}}
              \\
                                       & =                & \trans{ur} \subs{x}{\transtwo{v}}.
            \end{array} \]
  \end{itemize}
\end{proof}

\lemsimul*
\begin{proof}
    Let $t \plusred{\name{v}}^n u$, where $n$ is the number of evaluation steps between $t$ and $u$. The proof follows by induction over $n$:
    \begin{itemize}
        \item Let $n = 0$. Then, $u = t$ and thus $\trans{t} \plusred{\HC} \trans{u}$.
        \item Let $n > 0$. Then, there exists $t'$, such that $t \red{\name{v}} t' \plusred{\name{v}} u$. By the \ih  $\trans{t'} \plusred{\HC} \trans{u}$. Therefore, it is now enough to show that $\trans{t} \plusred{\HC} \trans{t'}$. We reason by induction over CBV contexts:
              \begin{itemize}
                  \item Case $\name{V} = \square$. Then, $t = (\lam x.r) v \red{\beta_{\name{v}}} r \subs{x}{v} = t'$. Therefore,
                        \[ \begin{array}{rcl}
                                \trans{(\lam x.r) v}
                                 & =                                  & (yz) \match{\const{\name{v}}{(z)}}{\trans{v}} \match{\const{\name{v}}{(y)}}{\trans{\lam x.r}}
                                \\
                                 & =                                  & (yz) \match{\const{\name{v}}{(z)}}{\const{\name{v}}{(\transtwo{v})}} \match{\const{\name{v}}{(y)}}{\const{\name{v}}{(\lam x.\trans{r})}}
                                \\
                                 & \red{\HC(\ruleM)}                  & (yz) \match{\const{\name{v}}{(z)}}{\const{\name{v}}{(\transtwo{v})}} \match{y}{\lam x.\trans{r}}
                                \\
                                 & \red{\HC(\ruleE)}                  & ((\lam x.\trans{r}) z) \match{\const{\name{v}}{(z)}}{\const{\name{v}}{(\transtwo{v})}}
                                \\
                                 & \red{\HC(\ruleM)}                  & ((\lam x.\trans{r}) z) \match{z}{\transtwo{v}}
                                \\
                                 & \red{\HC(\ruleE)}                  & (\lam x.\trans{r}) \transtwo{v}
                                \\
                                 & \red{\HC(\ruleBeta)}               & \trans{r} \match{x}{\transtwo{v}}
                                \\
                                 & \red{\HC(\ruleE)}                  & \trans{r} \subs{x}{\transtwo{v}}
                                \\
                                 & \overset{\cref{lem:subs-simul}}{=} & \trans{r \subs{x}{v}}.
                            \end{array} \]
                  \item Case $\name{V} = \name{V}' r$. Then, $t = \name{V} \lhole{s} = \name{V}' \lhole{s} r$ and $\name{V}' \lhole{s} \red{\name{v}} \name{V}' \lhole{s'}$. Moreover,
                        \[ \begin{array}{rcl}
                                \trans{t} & = & \trans{\name{V} \lhole{s}}
                                \\
                                          & = & \trans{\name{V}' \lhole{s}\ r}
                                \\
                                          & = & (xy) \match{\const{\name{v}}{(y)}}{\trans{r}} \match{\const{\name{v}}{(x)}}{\trans{\name{V}' \lhole{s}}}.
                            \end{array} \]
                        By the \ih, we know $\trans{\name{V}' \lhole{s}} \plusred{\HC} \trans{\name{V}' \lhole{s'}}$. Therefore,
                        \[ \begin{array}{rcl}
                                \trans{t} & =             & (xy) \match{\const{\name{v}}{(y)}}{\trans{r}} \match{\const{\name{v}}{(x)}}{\trans{\name{V}' \lhole{s}}}
                                \\
                                          & \plusred{\HC} & (xy) \match{\const{\name{v}}{(y)}}{\trans{r}} \match{\const{\name{v}}{(x)}}{\trans{\name{V}' \lhole{s'}}}
                                \\
                                          & =             & \trans{\name{V}' \lhole{s'}\ r}
                                \\
                                          & =             & \trans{\name{V} \lhole{s'}}
                                \\
                                          & =             & \trans{t'}.
                            \end{array} \]
                  \item Case $\name{V} = v \name{V}'$. Then, $t = \name{V} \lhole{r} = v\ \name{V}' \lhole{r}$ and $\name{V}' \lhole{r} \red{\name{v}} \name{V}' \lhole{r'}$. Moreover,
                        \[ \begin{array}{rcl}
                                \trans{t} & = & \trans{\name{V} \lhole{r}}
                                \\
                                          & = & \trans{v\ \name{V}' \lhole{r}}
                                \\
                                          & = & (xy) \match{\const{\name{v}}{(y)}}{\trans{\name{V}' \lhole{r}}} \match{\const{\name{v}}{(x)}}{\trans{v}}.
                            \end{array} \]
                        By the \ih, we know $\trans{\name{V}' \lhole{r}} \plusred{\HC} \trans{\name{V}' \lhole{r'}}$. Therefore,
                        \[ \begin{array}{rcl}
                                \trans{t} & =             & (xy) \match{\const{\name{v}}{(y)}}{\trans{\name{V}' \lhole{r}}} \match{\const{\name{v}}{(x)}}{\trans{v}}
                                \\
                                          & \plusred{\HC} & (xy) \match{\const{\name{v}}{(y)}}{\trans{\name{V}' \lhole{r'}}} \match{\const{\name{v}}{(x)}}{\trans{v}}
                                \\
                                          & =             & \trans{v\ \name{V}' \lhole{r'}}
                                \\
                                          & =             & \trans{\name{V} \lhole{r'}}
                                \\
                                          & =             & \trans{t'}.
                            \end{array} \]
                        Note that $\red{\HC}$ makes it possible to do a $\red{\HC(\ruleE)}$ for the matching between $\const{\name{v}}{(x)}$ and $\const{\name{v}}{(\transtwo{v})}$. However, since $\red{\HC}$ is nondeterministic, it is also possible to reduce $r$ first.
              \end{itemize}
    \end{itemize}
\end{proof}

\subsubsection{\textit{The Bang-Calculus.}}

\begin{restatable}[Substitution Preserves Simulation for Bang]{lemma}{lemsubssimulml}
    \label{lem:subs-simul-ml}
    Let $t, u \in \Lambda^\oc$. Then, $\transml{t \subs{x}{u}} = \transml{t} \subs{x}{\transml{u}}$
\end{restatable}


\begin{proof}
  The proof follows by induction over $t$:
  \begin{itemize}
    \item Case $t = x$. Then, $x \subs{x}{u} = u$. Therefore, $\transml{u} = x \subs{x}{\transml{u}} = \transml{x} \subs{x}{\transml{u}}$.
    \item Case $t = y \neq x$. Then $y \subs{x}{u} = y$. Therefore, $\transml{y} = \transml{y} \subs{x}{\transml{u}}$.
    \item Case $t = \lam y.r$. Then, $(\lam y.r) \subs{x}{u} = \lam y.(r \subs{x}{u})$. Therefore,
          \[ \begin{array}{rcl}
              \transml{\lam y. (r \subs{x}{u})} & =                & \lam \const{\name{b}}{(y)}.\transml{r \subs{x}{u}}
              \\
                                                & \overset{\ih}{=} & \lam \const{\name{b}{(y)}}.(\transml{r} \subs{x}{\transml{u}})
              \\
                                                & =                & (\lam \const{\name{b}}{(y)}.\transml{r}) \subs{x}{\transml{u}}
              \\
                                                & =                & \transml{\lam y.r} \subs{x}{\transml{u}}.
            \end{array} \]
    \item Case $t = rs$. Then, $(rs) \subs{x}{u} = (r \subs{x}{u}) (s \subs{x}{u})$. Therefore,
          \[ \begin{array}{rcl}
              \transml{(rs) \subs{x}{u}} & =                & \transml{r \subs{x}{u}} \transml{s \subs{x}{u}}
              \\
                                         & \overset{\ih}{=} & (\transml{r} \subs{x}{\transml{u}}) (\transml{s} \subs{x}{\transml{u}})
              \\
                                         & =                & (\transml{r}  \transml{s}) \subs{x}{\transml{u}}
              \\
                                         & =                & \transml{rs} \subs{x}{\transml{u}}.
            \end{array} \]
    \item Case $t = \oc r$. Then, $(\oc r) \subs{x}{u} = \oc (r \subs{x}{u})$. Therefore,
          \[ \begin{array}{rcl}
              \transml{(\oc r) \subs{x}{u}} & =                & \const{\name{b}}{(\transml{r \subs{x}{u}})}
              \\
                                            & \overset{\ih}{=} & \const{\name{b}}{(\transml{r} \subs{x}{\transml{u}})}
              \\
                                            & =                & (\const{\name{b}}{\transml{r}}) \subs{x}{\transml{u}}
              \\
                                            & =                & \transml{!r} \subs{x}{\transml{u}}.
            \end{array} \]
    \item Case $t = r \match{y}{s}$. Then $(r \match{y}{s}) \subs{x}{u} = (r \subs{x}{u}) \match{y}{s \subs{x}{u}}$. Therefore,
          \[ \begin{array}{rcl}
              \transml{(r \match{y}{s}) \subs{x}{u}}
               & =                & \transml{r \subs{x}{u}} \match{\const{\name{b}}{(y)}}{\transml{s \subs{x}{u}}}
              \\
               & \overset{\ih}{=} & \transml{r} \subs{x}{\transml{u}} \match{\const{\name{b}}{(y)}}{\transml{s} \subs{x}{\transml{u}}}
              \\
               & =                & (\transml{r} \match{\const{\name{b}}{(y)}}{\transml{s}}) \subs{x}{\transml{u}}
              \\
               & =                & \transml{r \match{y}{s}} \subs{x}{\transml{u}}.
            \end{array} \]
  \end{itemize}
\end{proof}

\lemsimulml*

\begin{proof}
    Let $t \plusred{\oc}^n u$, where $n$ is the number of evaluation steps between $t$ and $u$. The proof follows by induction over $n$:
    \begin{itemize}
        \item Let $n = 0$. Then, $u = t$ and $\transml{t} \plusred{\HC} \transml{u}$.
        \item Let $n > 0$. Then, there exists $t'$, such that $t \red{\oc} t' \plusred{\oc} u$. By the \ih, $\transml{t'} \plusred{\HC} \transml{u}$. Therefore, it is now enough to show that $\transml{t} \plusred{\HC} \transml{t'}$. We reason by induction over ML contexts:
              \begin{itemize}
                  \item Case $\WC = \square$. We have to consider two cases:
                        \begin{itemize}
                            \item Let $t = \LC \lhole{(\lam x.r)} s \red{\name{dB}} \LC \lhole{r \match{x}{s}} = t'$. We proceed by induction over $\LC$:
                                  \begin{itemize}
                                      \item Case $\LC = \square$. Then,
                                            \[ \begin{array}{rcl}
                                                    \transml{t} = \transml{(\lam x.r) s} & =                    & \transml{\lam x.r} \transml{s}
                                                    \\
                                                                                         & =                    & (\lam \const{\name{b}}{(x)}.\transml{r}) \transml{s}
                                                    \\
                                                                                         & \red{\HC(\ruleBeta)} & \transml{r} \match{\const{\name{b}}{(x)}}{\transml{s}}
                                                    \\
                                                                                         & =                    & \transml{r \match{x}{s}} = \transml{t'}.
                                                \end{array} \]
                                      \item Case $\LC = \LC' \match{y}{p}$. Then,
                                            \[ \begin{array}{rcl}
                                                    \transml{t} = \transml{\LC' \lhole{\lam x.r} \match{y}{p} s} & = & \transml{\LC' \lhole{\lam x.r} \match{y}{p}} \transml{s}
                                                    \\
                                                                                                                 & = & \transml{\LC' \lhole{\lam x.r}} \match{\const{\name{b}}{(y)}}{\transml{p}} \transml{s}.
                                                \end{array} \]
                                            Note that $\transml{\LC' \lhole{\lam x.r} s} = \transml{\LC' \lhole{\lam x.r}} \transml{s}$. By the \ih, we know
                                            \[ \begin{array}{rcl}
                                                    \transml{\LC' \lhole{\lam x.r}} \transml{s} & \plusred{\HC} & \transml{\LC ' \lhole{r \match{x}{s}}}.
                                                \end{array} \]
                                            Therefore,
                                            \[ \begin{array}{rcl}
                                                    \transml{t} = \transml{\LC' \lhole{\lam x.r}} \match{\const{\name{b}}{(y)}}{\transml{p}} \transml{s} & \plusred{\HC} & \transml{\LC' \lhole{r \match{x}{s}}} \match{\const{\name{b}}{(y)}}{\transml{p}}
                                                    \\
                                                                                                                                                         & =             & \transml{\LC' \lhole{r \match{x}{s}} \match{y}{p}}                               \\
                                                                                                                                                         & =             & \transml{\LC \lhole{r \match{x}{s}}} = \transml{t'}.
                                                \end{array} \]
                                  \end{itemize}
                            \item Let $t = r \match{x}{\LC \lhole{\oc s}} \red{\name{s}} \LC \lhole{r \subs{x}{s}} = t'$. We proceed by induction over $\LC$:
                                  \begin{itemize}
                                      \item Case $\LC = \square$. Then,
                                            \[ \begin{array}{rcl}
                                                    \transml{t} = \transml{r \match{x}{\oc s}} & =                                     & \transml{r} \match{\const{\name{b}}{(x)}}{\transml{\oc s}}
                                                    \\
                                                                                               & =                                     & \transml{r} \match{\const{\name{b}}{(x)}}{\const{\name{b}}{(\transml{s})}}
                                                    \\
                                                                                               & \red{\HC(\ruleM)}                     & \transml{r} \match{x}{\transml{s}}
                                                    \\
                                                                                               & \red{\HC(\ruleE)}                     & \transml{r} \subs{x}{\transml{s}}
                                                    \\
                                                                                               & \overset{\cref{lem:subs-simul-ml}}{=} & \transml{r \subs{x}{s}} = \transml{t'}.
                                                    \\
                                                \end{array} \]
                                      \item Case $\LC = \LC' \match{y}{p}$. Then,
                                            \[ \begin{array}{rcl}
                                                    \transml{t} = \transml{r \match{x}{\LC' \lhole{\oc s} \match{y}{p}}} & = & \transml{r} \match{\const{\name{b}}{(x)}}{\transml{\LC' \lhole{\oc s} \match{y}{p}}}
                                                    \\
                                                                                                                         & = & \transml{r} \match{\const{\name{b}}{(x)}}{\transml{\LC' \lhole{\oc s}} \match{\const{\name{b}}{(y)}}{\transml{p}}}.
                                                \end{array} \]
                                            Note that $\transml{r \match{x}{\LC' \lhole{\oc s}}} = \transml{r} \match{\const{\name{b}}{(x)}}{\transml{\LC' \lhole{\oc s}}}$. By the \ih, we know
                                            \[ \begin{array}{rcl}
                                                    \transml{r} \match{\const{\name{b}}{(x)}}{\transml{\LC' \lhole{\oc s}}} \plusred{\HC} \transml{\LC' \lhole{r \subs{x}{s}}}.
                                                \end{array} \]
                                            Therefore,
                                            \[ \begin{array}{rcl}
                                                    \transml{t} & =             & \transml{r} \match{\const{\name{b}}{(x)}}{\transml{\LC' \lhole{\oc s}} \match{\const{\name{b}}{(y)}}{\transml{p}}}
                                                    \\
                                                                & \plusred{\HC} & \transml{\LC' \lhole{r \subs{x}{s}}} \match{\const{\name{b}}{(y)}}{\transml{p}}
                                                    \\
                                                                & =             & \transml{\LC' \lhole{r \subs{x}{s}} \match{y}{p}} = \transml{t'}.
                                                \end{array} \]
                                  \end{itemize}
                        \end{itemize}
                  \item Case $\WC = \WC' r$. Then, $t = \WC \lhole{s} = \WC' \lhole{s} r$ and $\WC' \lhole{s} \plusred{\oc} \WC' \lhole{s'}$. Moreover,
                        \[ \begin{array}{rcl}
                                \transml{t} & = & \transml{\WC \lhole{s}}
                                \\
                                            & = & \transml{\WC' \lhole{s} r}
                                \\
                                            & = & \transml{\WC' \lhole{s}} \transml{r}.
                            \end{array} \]
                        By the \ih, we know $\transml{\WC' \lhole{s}} \plusred{\HC} \transml{\WC' \lhole{s'}}$. Therefore,
                        \[ \begin{array}{rcl}
                                \transml{t} & =             & \transml{\WC' \lhole{s}} \transml{r}
                                \\
                                            & \plusred{\HC} & \transml{\WC' \lhole{s'}} \transml{r}
                                \\
                                            & =             & \transml{\WC' \lhole{s'} r}
                                \\
                                            & =             & \transml{t'}.
                            \end{array} \]
                  \item Case $\WC = \WC' \match{y}{r}$. Then, $t = \WC \lhole{s} = \WC' \lhole{s} \match{y}{r}$ and $\WC' \lhole{s} \plusred{\oc} \WC' \lhole{s'}$. Moreover,
                        \[ \begin{array}{rcl}
                                \transml{t} & = & \transml{\WC \lhole{s}}
                                \\
                                            & = & \transml{\WC' \lhole{s} \match{y}{r}}
                                \\
                                            & = & \transml{\WC' \lhole{s}} \match{\const{\name{b}}{(y)}}{\transml{r}}.
                            \end{array} \]
                        By the \ih, we know $\transml{\WC' \lhole{s}} \plusred{\HC} \transml{\WC' \lhole{s'}}$. Therefore,
                        \[ \begin{array}{rcl}
                                \transml{t} & =             & \transml{\WC' \lhole{s}} \match{\const{\name{b}}{(y)}}{\transml{r}}
                                \\
                                            & \plusred{\HC} & \transml{\WC' \lhole{s'}} \match{\const{\name{b}}{(y)}}{\transml{r}}
                                \\
                                            & =             & \transml{\WC' \lhole{s'} \match{y}{r}} = \transml{t'}.
                            \end{array} \]
                  \item Case $\WC = r \match{x}{\WC'}$. Then, $t = \WC \lhole{s} = r \match{y}{\WC' \lhole{s}}$ and $\WC' \lhole{s} \plusred{\oc} \WC' \lhole{s'}$. Moreover,
                        \[ \begin{array}{rcl}
                                \transml{t} & = & \transml{\WC \lhole{s}}
                                \\
                                            & = & \transml{r \match{y}{\WC' \lhole{s}}}
                                \\
                                            & = & \transml{r} \match{\const{\name{b}}{(y)}}{\transml{\WC' \lhole{s}}}.
                            \end{array} \]
                        By the \ih, we know $\transml{\WC' \lhole{s}} \plusred{\HC} \transml{\WC' \lhole{s'}}$. Therefore,
                        \[ \begin{array}{rcl}
                                \transml{t} & =             & \transml{r} \match{\const{\name{b}}{(y)}}{\transml{\WC' \lhole{s}}}
                                \\
                                            & \plusred{\HC} & \transml{r} \match{\const{\name{b}}{(y)}}{\transml{\WC' \lhole{s'}} }
                                \\
                                            & =             & \transml{r \match{y}{\WC' \lhole{s'}}} = \transml{t'}.
                            \end{array} \]

              \end{itemize}
    \end{itemize}
\end{proof}

\subsection{Type System}

\subsubsection{\textit{Preliminary Properties.}}

\lemrelevance*

\maybehide{\begin{proof}
    We generalize the original statement by allowing $\Phi_{t}$ to conclude with either a term type $\sig$ or a multiset type $\M$.

    The proof follows by induction over $\Phi$. We reason according to the last rule:
    \begin{itemize}
        \item Rule (\ruleAx). Then, $t = x$ and $\Gam = (x : \mul{\sig})$. Therefore, we can conclude immediately.
        \item Rule (\ruleMany). Suppose $\Phi_i \tr \seq{\Gam_i}{t}{\sig_i}$ are the premises of the rule. Then, $\Gam = \otimes_{\iI} \Gam_i$. By the \ih over each $\Phi_i$, we know $\dom{\Gam_i} \subseteq \fv{t}$. Therefore, $\dom{\Gam} = \bigcup\limits_{\iI} \dom{\Gam_i} \subseteq \bigcup\limits_{\iI} \fv{t} = \fv{t}$.
        \item Rule (\ruleAbs). Then, $t = \lam p.u$. Suppose $\Phi_t \tr \seq{\Gam_u}{u}{\sig}$ and $\Pi \tr \seqp{\Gam \restto{p}}{p}{\M}$ are the premises of the rule for $u$ and $p$, respectively. Then, $\Gam = \Gam_u \mminus \var{p}$. By the \ih over $\Phi_t$, we know $\dom{\Gam_u} \subseteq \fv{u}$. Therefore, $\dom{\Gam_u} \setminus \var{p} \subseteq \fv{u} \setminus \var{p}$. And we can conclude since $\dom{\Gam_u} \setminus \var{p} = \dom{\Gam_u \mminus \var{p}} = \dom{\Gam}$ and $\fv{u} \setminus \var{p} = \fv{\lam p.u}$.
        \item Rule (\ruleAbsStar). Then, $t = \lam p.u$ and $\Gam = \eset$. So we can conclude immediately.
        \item Rule (\ruleApp). Then, $t = us$. Suppose $\Phi_u \tr \seq{\Gam_u}{u}{\M \ta \sig}$ and $\Phi_s \tr \seq{\Gam_s}{s}{\M}$ are the premises of the rule for $u$ and $s$, respectively. Then, $\Gam = \Gam_u \otimes \Gam_s$. By the \ih over $\Phi_u$ (resp. $\Phi_s$), we know $\dom{\Gam_u} \subseteq \fv{u}$ (resp. $\dom{\Gam_s} \subseteq \fv{s}$). Therefore, $\dom{\Gam} = \dom{\Gam_u} \cup \dom{\Gam_s} \subseteq \fv{u} \cup \fv{s} = \fv{us}$.
        \item Rule (\ruleConst). Then, $t = \const{\ttc}{\vect{u_i}_n}$. Suppose $\Phi_i \tr \seq{\Gam_i}{u_i}{\M_i}$ are the premises of the rule for each $u_i$. Then, $\Gam = \otimes_{i \in \interval{1}{n}} \Gam_i$. By the \ih over each $\Phi_i$, we know $\dom{\Gam_i} \subseteq \fv{u_i}$. Therefore, $\dom{\Gam} = \bigcup\limits_{i \in \interval{1}{n}} \dom{\Gam_i} \subseteq \bigcup\limits_{i \in \interval{1}{n}} \fv{u_i} = \fv{\const{\ttc}{\vect{u_i}_n}}$.
        \item Rule (\ruleMatch). Then, $t = u \match{p}{s}$. Suppose $\Phi_u \tr \seq{\Gam_u}{u}{\sig}$, $\Pi \tr \seqp{\Gam \restto{p}}{p}{\M}$, and $\Phi_s \tr \seq{\Gam_s}{s}{\M}$ are the premises of the rule for $u$, $p$, and $s$, respectively. Then, $\Gam = (\Gam_u \mminus \var{p}) \otimes \Gam_s$. By the \ih over $\Phi_u$ (resp. $\Phi_s$), we know $\dom{\Gam_u} \subseteq \fv{u}$ (resp. $\dom{\Gam_s} \subseteq \fv{s}$). Therefore, $\dom{\Gam} = \dom{\Gam_u \mminus \var{p}} \cup \dom{\Gam_s} = (\dom{\Gam_u} \setminus \var{p}) \cup \dom{\Gam_s} \subseteq (\fv{u} \setminus \var{p}) \cup \fv{s} = \fv{u \match{p}{s}}$.
        \item Rule (\ruleCase). Then, $t = \case{u}{\vect{\branch{\hat{p_i}}{s_i}}_n}$. Let $k \in \interval{1}{n}$. Now, suppose $\Phi_u \tr \seq{\Gam_u}{u}{\M}$, $\Pi \tr \seqp{\Gam_u \restto{\hat{p}_k}}{\hat{p}_k}{\M}$ and $\Phi_{s_k} \tr \seq{\Gam_{s_k}}{s_k}{\sig}$ are the premises of the rule for $u$, $\hat{p_k}$, and $s_k$, respectively. Then, $\dom{\Gam} = \dom{\Gam_{s_k} \mminus \var{\hat{p}_k}} \cup \dom{\Gam_u}$. By the \ih over $\Phi_u$ (resp. $\Phi_{s_k}$), we know $\dom{\Gam_u} \subseteq \fv{u}$ (resp. $\dom{\Gam_{s_k}} \subseteq \fv{s_k}$). Therefore, $\dom{\Gam} = \dom{\Gam_{s_k} \mminus \var{\hat{p}_k}} \cup \dom{\Gam_u} = (\dom{\Gam_{s_k}} \setminus \var{\hat{p}_k}) \cup \dom{\Gam_u} = \fv{u} \cup (\fv{s_k} \setminus \var{\hat{p}_k}) \subseteq \fv{u} \bigcup\limits_{i \in \interval{1}{n}} (\fv{s_i} \setminus \var{\hat{p}_i}) = \fv{\case{u}{\vect{\branch{\hat{p}_i}{s_i}}_n}}$.
    \end{itemize}
    If $t$ is closed, then $\fv{t} = \eset$ by definition. Therefore, $\Gam = \eset$ by the previous induction.
\end{proof}}

\lemsplitmerge*

\maybehide{\begin{proof}
    Let $\M = \sqcup_{\iI} \M_i$.
    \begin{itemize}
        \item[$\Ra$)] If there exists $\Phi \tr \seq{\Gam}{t}{\M}$, then $\Phi$ it must be of the following form:
            \[ \begin{prooftree}
                    \hypo{\Phi_k \tr \seq{\Gam_k}{t}{\sig_k}}
                    \delims{\left(}{\right)_{\kK}}
                    \infer1[(\ruleMany)]{\hspace{.6cm}\seq{\otimes_{\kK} \Gam_k}{t}{\mul{\sig_k}_{\kK}}\hspace{.6cm}}
                \end{prooftree} \]
            where $\Gam = \otimes_{\kK} \Gam_k$ and $\M = \mul{\sig_k}_{\kK}$.
            Since also $\M = \sqcup_{\iI} \M_i$, then there exist disjoint sets
            $(K_i)_{\iI} \subseteq K$ such that $\cup_{\iI} K_i = K$
            and $\M_i = \mul{\sig_k}_{\kK_i}$. One concludes this case by
            constructing $\Phi_i$ and $\Gam_i =  \otimes_{\kK_i} \Gam_k$ for any $\iI$ as follows:
            \[ \begin{prooftree}
                    \hypo{\Phi_k \tr \seq{\Gam_k}{t}{\sig_k}}
                    \delims{\left(}{\right)_{\kK_i}}
                    \infer1[(\ruleMany)]{\hspace{.6cm}\seq{\otimes_{\kK_i} \Gam_k}{t}{\mul{\sig_k}_{\kK_i}}\hspace{.6cm}}
                \end{prooftree} \]
            Indeed, one has:
            \[ \Gam =  \otimes_{\kK} \Gam_k = \otimes_{\iI}  \otimes_{\kK_i} \Gam_k  =
                \otimes_{\iI} \Gam_i    \]
            \[ \sz{\Phi} =  +_{\kK} \sz{\Phi_k} =
                +_{\iI} +_{\kK_i} \sz{\Phi}_k  =
                +_{\iI} \sz{\Phi_i}    \]
        \item[$\La$)] If there exist $(\Phi_i \tr \seq{\Gam_i}{t}{\M_i})_{\iI}$, then each $\Phi_i$ must be of the form:
            \[ \begin{prooftree}
                    \hypo{\Phi_k \tr \seq{\Gam_k}{t}{\sig_k}}
                    \delims{\left(}{\right)_{\kK_i}}
                    \infer1[(\ruleMany)]{\hspace{.6cm}\seq{\otimes_{\kK_i} \Gam_k}{t}{\mul{\sig_k}_{\kK_i}}\hspace{.6cm}}
                \end{prooftree} \]
            where $\Gam_i = \otimes_{\kK_i} \Gam_k$,
            $\M_i = \mul{\sig_k}_{\kK_i}$, and
            $\sz{\Phi_i} = +_{\kK_i} \sz{\Phi_k}$.
            Then, we can build $\Phi$ as follows:
            \[ \begin{prooftree}
                    \hypo{\Phi_k \tr \seq{\Gam_k}{t}{\sig_k}}
                    \delims{\left(}{\right)_{\kK_i, \iI}}
                    \infer1[(\ruleMany)]{\hspace{1cm}\seq{\otimes_{\kK_i, \iI} \Gam_k}{t}{\mul{\sig_k}_{\kK_i, \iI}}\hspace{1cm}}
                \end{prooftree} \]
            where $\Gam = \otimes_{\kK_i, \iI} \Gam_k = +_{\iI} \Gam_i$, and $\M = \mul{\sig_k}_{\kK_i, \iI} = \mul{\sig_i}_{\iI}$. And we can conclude with $+_{\iI} \sz{\Phi_i} = +_{\kK_i, \iI} \sz{\Phi_k} = \sz{\Phi}$.
    \end{itemize}
\end{proof}

}

\subsubsection{\textit{Soundness.}}

\lemclashesnottypable*

\maybehide{\begin{proof}
  The proof follows by induction over $t \in \Cl$. For each case, we reason by contradiction, assuming there exist $\Phi$:
  \begin{enumerate}
    \item Let $t \in \BCl$:
          \begin{itemize}
            \item Case $t = \MC \lhole{\const{\ttc}{\ttuple}} s$. Then, $\Phi$ must end with rule (\ruleApp) and $\MC \lhole{\const{\ttc}{\ttuple}}$ must be assigned an arrow type. However, we can easily conclude by a simple inspection of the rules that $\const{\ttc}{\ttuple}$ can only be assigned a data type. Therefore, we can also conclude by that $\MC \lhole{\const{\ttc}{\ttuple}}$ can only be assigned data type. Thus, we can conclude that there is no $\Phi$, by contradiction.
            \item Case $t = u \match{\const{\ttc}{\ptuple}}{\MC \lhole{\lam q.s}}$. Then, $\Phi$ must end with rule (\ruleMatch) and $\MC \lhole{\lam q.s}$ must be assigned the same type as $\const{\ttc}{\ptuple}$. A simple inspection of the rules tells us that $\const{\ttc}{\ptuple}$ can only be assigned a data type. However, it is also clear from the rules that $\lam q.s$ can only be assigned an arrow type. Therefore, $\MC \lhole{\lam q.s}$ can only be assigned an arrow type. Thus, we can conclude that there is no $\Phi$, by contradiction.
            \item Case $t = u \match{\const{\ttc}{\ptuple}}{\MC \lhole{\const{\ttc'}{\ttuple}}}$, where $\ttc \neq \ttc'$. Then, $\Phi$ must end with rule (\ruleMatch) and $\MC \lhole{\const{\ttc'}{\ttuple}}$ must be assigned the same type as $\const{\ttc}{\ptuple}$. A simple inspection of the rules tell us that $\const{\ttc}{\ptuple}$ can only be assigned a data type with top symbol $\ttc$. Consequently, $\MC \lhole{\const{\ttc'}{\ttuple}}$ will also have to be assigned a data type with top symbol $\ttc$. However, it is also clear from the rules that $\const{\ttc'}{\ttuple}$ can only be assigned a data type with top symbol $\ttc'$. Therefore, $\MC \lhole{\const{\ttc'}{\ttuple}}$ can only be assigned a data type starting constructor $\ttc'$. Thus, we can conclude that there is not $\Phi$, by contradiction.
            \item Case $t = \case{\MC \lhole{\lam p.u}}{\vect{\branch{\const{\ttc_i}{\ptuple_i}}{s_i}}_n}$. Then, $\Phi$ must end with rule (\ruleCase) and $\MC \lhole{\lam p.u}$ must be assigned the same type one of the patterns $\const{\ttc_i}{\ptuple_i}$. A simple inspection of the rules tells us that all data patterns must be typed with data types. However, it is also clear from the rules that $\lam p.u$ can only be assigned an arrow type. Therefore, $\MC \lhole{\lam p.u}$ can only be assigned an arrow type. Thus, we can conclude that there is not $\Phi$, by contradiction.
            \item Case $t = \case{\MC \lhole{\const{\ttc}{\ttuple}}}{\vect{\branch{\const{\ttc_i}{\ptuple_i}}{u_i}}_n}$, where $\ttc \not\in \{\ttc_1, \ldots, \ttc_n\}$. Then, $\Phi$ must end with rule (\ruleCase) and $\MC \lhole{\const{\ttc}{\ttuple}}$ must be assigned the same type as one of the patterns $\const{\ttc_i}{\ptuple_i}$. A simple inspection of the rules tells us that data patterns must be typed with data types starting with matching constructors. However, it is also clear from the rules that $\const{\ttc}{\ttuple}$ must be assigned a data type with top symbol $\ttc$. Therefore, $\MC \lhole{\const{\ttc}{\ttuple}}$ can only be assigned a data type with top symbol $\ttc$. Thus, we can conclude there is no $\Phi$, by contradiction.
          \end{itemize}
    \item Let $t \in \Cl$:
          \begin{itemize}
            \item Case $t \in \BCl$. We can conclude by the previous point.
            \item Let $t \not\in \BCl$. These cases follow easily from the \ih
          \end{itemize}
  \end{enumerate}
\end{proof}
}

\begin{restatable}[Weighted Substitution]{lemma}{lemweightedsubs}
    \label{lem:weighted-substitution}
    Let $t, u \in \calc$ and assume $\Phi_{t} \tr \seq{\Gam; x : \M}{t}{\sig}$ and $\Phi_{u}\ \tr \seq{\Del}{u}{\M}$. Then, there exists $\Phi_{t \subs{x}{u}} \tr \seq{\Gam \otimes \Del}{t \subs{x}{u}}{\sig}$, such that $\sz{\Phi_{t \subs{x}{u}}} = \sz{\Phi_{t}} + \sz{\Phi_{u}} - \sizeof{\M}$.
\end{restatable}


\maybehide{\begin{proof}
    We generalize the original statement by allowing $\Phi_{t}$ to conclude with either a term type $\sig$ or a multiset type $\M$.

    If $\Phi_{t} \tr \seq{\Gam; x : \M}{t}{\T}$ and $\Phi_{u}\tr \seq{\Del}{u}{\M}$, then there exists $\Phi_{t \subs{x}{u}} \tr \seq{\Gam \otimes \Del}{t \subs{x}{u}}{\T}$, such that $\sz{\Phi_{t \subs{x}{u}}} = \sz{\Phi_{t}} + \sz{\Phi_{u}} - \sizeof{\M}$.

    The proof follows by induction over $\Phi_t$. We reason according to the last rule:
    \begin{itemize}
        \item Rule (\ruleAx). Then, $t = y$ and we must consider two different cases:
              \begin{itemize}
                  \item  If $y = x$, then $t \subs{x}{u} = u$ and $\Phi_t$ must be of the following form:
                        \[ \begin{prooftree}
                                \hypo{\phantom{BUUUUU}}
                                \infer1[(\ruleAx)]{\seq{x : \mul{\sig}}{x}{\sig}}
                            \end{prooftree} \]
                        where $\T = \sig$, $\Gam = \eset$, and $\M = \mul{\sig}$. Therefore, $\Phi_u$ must be of the following form:
                        \[ \begin{prooftree}
                                \hypo{\Phi'_u\tr \seq{\Del}{u}{\sig}}
                                \infer1[(\ruleMany)]{\seq{\Del}{u}{\mul{\sig}}}
                            \end{prooftree} \]
                        So, we can pick $\Phi_{t \subs{x}{u}} = \Phi'_u$ and conclude since $\Gam \otimes \Del = \Del$, and $\sz{\Phi_{t \subs{x}{u}}} = \sz{\Phi'_u} = \sz{\Phi_u} = 1 + \sz{\Phi_u} - 1 = \sz{\Phi_t} + \sz{\Phi_u} - \len{\mul{\sig}}$.
                  \item If $y \neq x$, then $y \subs{x}{u} = y$ and $\Phi_t$ must be of the following form:
                        \[ \begin{prooftree}
                                \hypo{\phantom{BUUUUU}}
                                \infer1[(\ruleAx)]{\seq{y : \mul{\sig}; x : \emul}{y}{\sig}}
                            \end{prooftree} \]
                        where $\T = \sig$, $\Gam = (y : \mul{\sig})$, and $\M = \emul$. Therefore, $\Phi_u$ must be of the following form:
                        \[ \begin{prooftree}
                                \infer0[(\ruleMany)]{\seq{}{u}{\emul}}
                            \end{prooftree} \]
                        where $\Del = \eset$. So, we can pick $\Phi_{t \subs{x}{u}} = \Phi_t$ and conclude since $\Gam \otimes \Del = \Gam$ and $\sz{\Phi_{t \subs{x}{u}}} = \sz{\Phi_t} = \sz{\Phi_t} + 0 - 0 = \sz{\Phi_t} + \sz{\Phi_u} - \len{\M}$.
              \end{itemize}
        \item Rule (\ruleMany). Then, $\Phi_t$ must be of the following form:
              \[ \begin{prooftree}
                      \hypo{\Phi^i_t \tr \seq{\Gam_i; x : \M_i}{t}{\sig_i}}
                      \delims{\left(}{\right)_{\iI}}
                      \infer1[(\ruleMany)]{\seq{\otimes_{\iI} (\Gam_i; x : \M_i)}{t}{\mul{\sig_i}_{\iI}}}
                  \end{prooftree} \]
              where $\T = \mul{\sig_i}_{\iI}$, $(\Gam; x : \M) = \otimes_{\iI} (\Gam_i; x : \M_i)$, and thus $\Gam = \otimes_{\iI} \Gam_i$ and $\M = \sqcup_{\iI} \M_i$. By~\cref{lem:split-and-merge-for-multi-types} over $\Phi_u$, we know there exist $(\Phi^i_u\tr \seq{\Del_i}{u}{\M_i})_{\iI}$, such that $\Del = \otimes_{\iI} \Del_i$ and $\sz{\Phi_u} = +_{\iI} \sz{\Phi^i_u}$. By the induction hypothesis over each $\Phi^i_t$, we know there exists $\Phi^i_{t \subs{x}{u}} \tr \seq{\Gam_i \otimes \Del_i}{t \subs{x}{u}}{\sig_i}$, for each $\iI$, such that $\sz{\Phi^i_{t \subs{x}{u}}} = \sz{\Phi^i_t} + \sz{\Phi^i_u} - \len{\M_i}$. Therefore, we can build $\Phi_{t \subs{x}{u}}$ as follows:
              \[ \begin{prooftree}
                      \hypo{\Phi^i_{t \subs{x}{u}}}
                      \delims{\left(}{\right)_{\iI}}
                      \infer1[(\ruleMany)]{\seq{\otimes_{\iI} \Gam_i \otimes_{\iI} \Del_i}{t \subs{x}{u}}{\mul{\sig_i}_{\iI}}}
                  \end{prooftree} \]
              We can conclude since $\Gam \otimes \Del = \otimes_{\iI} \Gam_i \otimes_{\iI} \Del_i$ and $\sz{\Phi_{t \subs{x}{u}}} = \sz{\Phi_{t \subs{x}{u}}} + \len{\M} - \len{\M} = +_{\iI} \sz{\Phi^i_{t \subs{x}{u}}} +_{\iI} \len{\M_i} - \len{\M} = +_{\iI} (\sz{\Phi^i_{t \subs{x}{u}}} + \len{\M_i}) - \len{\M} = +_{\iI} (\Phi^i_t + \Phi^i_u) - \len{\M} = (+_{\iI} \Phi^i_t) + (+_{\iI} \Phi^i_u) - \len{\M} = \sz{\Phi_t} + \sz{\Phi_u} - \len{\M}$.
        \item Rule (\ruleAbs). Then, $\Phi_t$ must be of the following form:
              \[ \begin{prooftree}
                      \hypo{\Phi_s \tr \seq{\Gam_s; x : \M}{s}{\sig}}
                      \hypo{\Pi \tr \seqp{(\Gam_s; x : \M) \restto{p}}{p}{\M_0}}
                      \infer2[(\ruleAbs)]{\seq{(\Gam_s; x : \M) \mminus \var{p}}{\lam p.s}{\M_0 \ta \sig}}
                  \end{prooftree} \]
              where $t = \lam p.s$, $\T = \M_0 \ta \sig$, and $\Gam = (\Gam_s; x : \M) \mminus \var{p}$. Moreover, by $\alpha$-conversion we can assume $\var{p} \cap \{ x \} = \eset$ so that $(\Gam_s; x : \M) \mminus \var{p} = (\Gam_s \mminus \var{p}); x : \M = (\Gam; x : \M) $, and thus $\Gam = \Gam_s \mminus \var{p}$ and $(\Gam_s; x : \M) \restto{p} = \Gam_s \restto{p}$. By the induction hypothesis over $\Phi_s$, we know there exists $\Phi_{s \subs{x}{u}} \tr \seq{\Gam_s \otimes \Del}{s \subs{x}{u}}{\sig}$, such that $\sz{\Phi_{s \subs{x}{u}}} = \sz{\Phi_s} + \sz{\Phi_u} - \len{\M}$. Therefore, we can build $\Phi_{t \subs{x}{u}}$ as follows:
              \[ \begin{prooftree}
                      \hypo{\Phi_{s \subs{x}{u}}}
                      \hypo{\Pi}
                      \infer2[(\ruleAbs)]{\seq{(\Gam_s \otimes \Del) \mminus \var{p}}{\lam p.s \subs{x}{u}}{\M_0 \ta \sig}}
                  \end{prooftree} \]
              By $\alpha$-conversion, we may assume that $\var{p} \cap \fv{u} = \eset$. Thus, $\lam p.s \subs{x}{u} = (\lam p.s) \subs{x}{u}$ and $\dom{\Del} \cap \var{p} = \eset$ (by~\cref{lem:relevance}). Therefore, $(\Gam_s \otimes \Del) \mminus \var{p} = (\Gam_s \mminus \var{p}) \otimes \Del$, and we can conclude since $\Gam \otimes \Del = (\Gam_s \mminus \var{p}) \otimes \Del$ and $\sz{\Phi_{t \subs{x}{u}}} = \sz{\Phi_{s \subs{x}{u}}} + \sz{\Pi} + 1 = \sz{\Phi_s} + \sz{\Phi_u} - \len{\M} + \sz{\Pi} + 1 = \sz{\Phi_t} + \sz{\Phi_u} - \len{\M}$.
        \item Rule (\ruleAbsStar). Then, $\Phi_t$ must be of the following form:
              \[ \begin{prooftree}
                      \hypo{\phantom{BUUUUU}}
                      \infer1[(\ruleAbsStar)]{\seq{}{\lam p.s}{\atom}}
                  \end{prooftree} \]
              where $\T = \atom$, $t = \lam p.s$, and $(\Gam; x : \M) = \eset$, and thus $\Gam = \eset$ and $\M = \emul$. Therefore, $\Phi_u$ must be of the following form:
              \[ \begin{prooftree}
                      \hypo{\phantom{BUUUUUU}}
                      \infer1[(\ruleMany)]{\seq{}{u}{\emul}}
                  \end{prooftree} \]
              where $\Del = \eset$. So, we can build $\Phi_{t \subs{x}{u}}$ as follows:
              \[ \begin{prooftree}
                      \hypo{\phantom{BUUUUU}}
                      \infer1[(\ruleAbsStar)]{\seq{}{\lam p.(s \subs{x}{u})}{\atom}}
                  \end{prooftree} \]
              By $\alpha$-conversion, we may assume that $\var{p} \cap \fv{u} = \eset$. Thus, $\lam p.s \subs{x}{u} = (\lam p.s) \subs{x}{u}$. We can conclude since $\Gam \otimes \Del = \eset$ and $\sz{\Phi_{t \subs{x}{u}}} = 1 = \sz{\Phi_t} + \sz{\Phi_u} - \len{\M}$.
        \item Rule (\ruleApp). Then, $\Phi_t$ must be of the following form:
              \[ \begin{prooftree}
                      \hypo{\Phi_s \tr \seq{\Gam_s; x : \M_1}{s}{\M_0 \ta \sig}}
                      \hypo{\Phi_r \tr \seq{\Gam_r; x : \M_2}{r}{\M_0}}
                      \infer2[(\ruleApp)]{\seq{(\Gam_s; x : \M_1) \otimes (\Gam_r; x : \M_2)}{sr}{\sig}}
                  \end{prooftree} \]
              where $t = sr$, $\T = \sig$, $(\Gam; x : \M) = (\Gam_s; x : \M_1) \otimes (\Gam_r; x : \M_2) = (\Gam_s \otimes \Gam_r); x : \M_1 \sqcup \M_2$, and thus $\Gam = \Gam_s \otimes \Gam_r$, and $\M = \M_1 \sqcup \M_2$. By~\cref{lem:split-and-merge-for-multi-types} over $\Phi_u$, we know there exist $(\Phi^i_u\tr \seq{\Del_i}{u}{\M_i})_{i \in \interval{1}{2}}$, such that $\Del = \Del_1 \otimes \Del_2$, $\M = \M_1 \sqcup \M_2$ and $\sz{\Phi_u} = \sz{\Phi^1_u} + \sz{\Phi^2_u}$. By the induction hypothesis over $\Phi_s$ (resp. $\Phi_r$), we know there exists $\Phi_{s \subs{x}{u}} \tr \seq{\Gam_s \otimes \Del_1}{s \subs{x}{u}}{\M_0 \ta \sig}$ (resp. $\Phi_{r \subs{x}{u}} \tr \seq{\Gam_r \otimes \Del_2}{r \subs{x}{u}}{\M_0}$), such that $\sz{\Phi_{s \subs{x}{u}}} = \sz{\Phi_s} + \sz{\Phi^1_u} - \len{\M_1}$ (resp. $\sz{\Phi_{r \subs{x}{u}}} = \sz{\Phi_r} + \sz{\Phi^2_u} - \len{\M_2}$). Therefore, we can build $\Phi_{t \subs{x}{u}}$ as follows:
              \[ \begin{prooftree}
                      \hypo{\Phi_{s \subs{x}{u}}}
                      \hypo{\Phi_{r \subs{x}{u}}}
                      \infer2[(\ruleApp)]{\seq{\Gam_s \otimes \Del_1 \otimes \Gam_r \otimes \Del_2}{s \subs{x}{u} r \subs{x}{u}}{\sig}}
                  \end{prooftree} \]
              We can conclude since $s \subs{x}{u}r \subs{x}{u} = (sr) \subs{x}{u}$, $\Gam \otimes \Del = \Gam_s \otimes \Gam_r \otimes \Del_1 \otimes \Del_2$, and $\sz{\Phi_{t \subs{x}{u}}} = \sz{\Phi_{s \subs{x}{u}}} + \sz{\Phi_{r \subs{x}{u}}} + 1 = (\sz{\Phi_s} + \sz{\Phi_r} + 1) + (\sz{\Phi^1_u} + \sz{\Phi^2_u}) - \len{\M_1} - \len{\M_2} = \sz{\Phi_t} + \sz{\Phi_u} - \len{\M}$.
        \item Rule (\ruleConst). Then, $\Phi_t$ must be of the following form:
              \[ \begin{prooftree}
                      \hypo{\Phi_{s_i} \tr \seq{\Gam_i; x : \M_i}{s_i}{\M^0_i}}
                      \delims{\left(}{\right)_{i \in \interval{1}{n}}}
                      \infer1[(\ruleConst)]{\hspace{.5cm}\seq{\otimes_{i \in \interval{1}{n}} (\Gam_i; x :  \M_i)}{\const{\ttc}{\vect{s_i}_n}}{\const{\ttc}{\vect{\M^0_i}_{n}}}\hspace{.5cm}}
                  \end{prooftree} \]
              where $t = \const{\ttc}{\vect{s_i}_n}$, $\T = \const{\ttc}{\vect{\M^0_i}_n}$, $(\Gam; x : \M) = \otimes_{i \in \interval{1}{n}} (\Gam_i; x :  \M_i) = \otimes_{i \in \interval{1}{n}} \Gam_i; x : \sqcup_{i \in \interval{1}{n}} \M_i$, and thus $\Gam = \otimes_{i \in \interval{1}{n}} \Gam_i$ and $\M = \sqcup_{i \in \interval{1}{n}} \M_i$. By~\cref{lem:split-and-merge-for-multi-types} over $\Phi_u$, we know there exist $(\Phi^i_u\tr \seq{\Del_i}{u}{\M_i})_{\interval{1}{n}}$, such that $\Del = \otimes_{\iI} \Del_i$ and $\sz{\Phi_u} = +_{i \in \interval{1}{n}} \sz{\Phi^i_u}$. By the induction hypothesis over each $\Phi_{s_i}$, we know there exists $\Phi_{s_i \subs{x}{u}} \tr \seq{\Gam_i \otimes \Del_i}{s_i \subs{x}{u}}{\M^0_i}$, such that $\sz{\Phi_{s_i \subs{x}{u}}} = \sz{\Phi_{s_i}} + \sz{\Phi^i_u} - \len{\M_i}$. Therefore, we can build $\Phi_{t \subs{x}{u}}$ as follows:
              \[ \begin{prooftree}
                      \hypo{\Phi_{s_i \subs{x}{u}}}
                      \delims{\left(}{\right)_{i \in \interval{1}{n}}}
                      \infer1[(\ruleConst)]{\seq{\otimes_{i \in \interval{1}{n}} (\Gam_i \otimes \Del_i)}{\const{\ttc}{\vect{s_i \subs{x}{u}}_n}}{\const{\ttc}{\vect{\M^0_i}_n}}}
                  \end{prooftree} \]
              We can conclude since $\const{\ttc}{\vect{s_i \subs{x}{u}}_n} = \const{\ttc}{\vect{s_i}_n} \subs{x}{u}$, $\Gam \otimes \Del = \otimes_{i \in \interval{1}{n}} ( \Gam_i \otimes \Del_i)$, and $\sz{\Phi_{t \subs{x}{u}}} = 1 +_{i \in \interval{1}{n}} \sz{\Phi_{s_i \subs{x}{u}}} = 1 +_{i \in \interval{1}{n}} (\sz{\Phi_{s_i}} + \sz{\Phi^i_u} - \len{\M_i}) = 1 +_{i \in \interval{1}{n}} \sz{\Phi_{s_i}} +_{i \in \interval{1}{n}} \sz{\Phi^i_u} - \len{\M} = \sz{\Phi_t} + \sz{\Phi_u} - \len{\M}$.
        \item Rule (\ruleMatch). Then, $\Phi_t$ must be of the following form:
              \[ \begin{prooftree}
                      \hypo{\Phi_s \tr \seq{\Gam_s; x : \M_1}{s}{\sig}}
                      \hypo{\Pi \tr \seqp{(\Gam_s; x : \M_1) \restto{p}}{p}{\M_0}}
                      \hypo{\Phi_r \tr \seq{\Gam_r; x : \M_2}{r}{\M_0}}
                      \infer3[(\ruleMatch)]{\seq{(\Gam_s; x : \M_1 )\mminus \var{p} \otimes \Gam_r; x : \M_2}{s \match{p}{r}}{\sig}}
                  \end{prooftree} \]
              where $t = s \match{p}{r}$, $\T = \sig$. Moreover, by $\alpha$-conversion, we may assume that $\var{p} \cap \fv{u} = \eset$ so that $(\Gam; x : \M) = (\Gam_s; x : \M_1) \mminus \var{p} \otimes \Gam_r; x : \M_2 = ((\Gam_s \mminus \var{p}) \otimes \Gam_r ); x : \M_1 \sqcup \M_2$, and thus $\Gam = (\Gam_s \mminus \var{p}) \otimes \Gam_r$, $\M = \M_1 \sqcup \M_2$ and $(\Gam_s; x : \M_1) \restto{p} = \Gam_s \restto{p}$. By~\cref{lem:split-and-merge-for-multi-types} over $\Phi_u$, we know there exist $(\Phi^i_u\tr \seq{\Del_i}{u}{\M_i})_{i \in \interval{1}{2}}$, such that $\Del = \otimes_{i \in \interval{1}{2}} \Del_i$ and $\sz{\Phi_u} = \sz{\Phi^1_u} + \sz{\Phi^2_u}$. By the induction hypothesis over $\Phi_s$ (resp. $\Phi_r$), we know there exists $\Phi_{s \subs{x}{u}} \tr \seq{\Gam_s \otimes \Del_1}{s \subs{x}{u}}{\sig}$ (resp. $\Phi_{r \subs{x}{u}} \tr \seq{\Gam_r \otimes \Del_2}{r \subs{x}{u}}{\M_0}$), such that $\sz{\Phi_{s \subs{x}{u}}} = \sz{\Phi_s} + \sz{\Phi^1_u} - \len{\M_1}$ (resp. $\sz{\Phi_{r \subs{x}{u}}} = \sz{\Phi_r} + \sz{\Phi^2_u} - \len{\M_2}$). Therefore, we can build $\Phi_{t \subs{x}{u}}$ as follows:
              \[ \begin{prooftree}
                      \hypo{\Phi_{s \subs{x}{u}}}
                      \hypo{\Pi}
                      \hypo{\Phi_{r \subs{x}{u}}}
                      \infer3[(\ruleMatch)]{\seq{((\Gam_s \otimes \Del_1) \mminus \var{p}) \otimes \Gam_r \otimes \Del_2}{s \subs{x}{u} \match{p}{r\subs{x}{u}}}{\sig}}
                  \end{prooftree} \]
              Thus, $s \subs{x}{u} \match{p}{r \subs{x}{u}} = (s \match{p}{r}) \subs{x}{u}$ and $\dom{\Del_1} \cap \var{p} = \eset$ (by~\cref{lem:relevance}). Therefore, $(\Gam_s \otimes \Del_1) \mminus \var{p} = (\Gam_s \mminus \var{p}) \otimes \Del_1$, and we can conclude since $\Gam \otimes \Del = (\Gam_s \mminus \var{p}) \otimes \Gam_r \otimes \Del_1 \otimes \Del_2$ and $\sz{\Phi_{t \subs{x}{u}}} = \sz{\Phi_{s \subs{x}{u}}} + \sz{\Pi} + \sz{\Phi_{r \subs{x}{u}}} = \sz{\Phi_s} + \sz{\Phi^1_u} - \len{\M_1} + \sz{\Pi} + \sz{\Phi_r}  + \sz{\Phi^2_u} - \len{\M_2} = \sz{\Phi_t} + \sz{\Phi_u} - \len{\M}$.
        \item Rule (\ruleCase). Then, $\Phi_t$ must be of the following form:
              \[ \begin{prooftree}
                      \hypo{\Phi_r \tr \seq{\Gam_r; x : \M_2}{r}{\M_0}}
                      \hypo{\Pi \tr \seqp{(\Gam_{s_k}; x : \M_1) \restto{\hat{p}_k}}{\hat{p}_k}{\M_0}}
                      \hypo{\Phi_{s_k} \tr \seq{\Gam_{s_k}; x : \M_1}{s_k}{\sig}}
                      \infer3[(\ruleCase)]{\seq{((\Gam_{s_k}; x : \M_1) \mminus \var{\hat{p}_k}) \otimes \Gam_r; x : \M_2}{\case{r}{\vect{\branch{\hat{p}_i}{s_i}}_n}}{\sig}}
                  \end{prooftree} \]
              where $t = \case{r}{\vect{\branch{\hat{p}_i}{s_i}}_n}$, $\T = \sig$, $(\Gam; x : \M) = ((\Gam_{s_k}; x : \M_1) \mminus \var{\hat{p}_k}) \otimes \Gam_r; x : \M_2 = ((\Gam_{s_k} \mminus \var{\hat{p}_k}) \otimes \Gam_r); x : \M_1 \sqcup \M_2$, and thus $\Gam = (\Gam_{s_k} \mminus \var{\hat{p}_k}) \otimes \Gam_r$, $\M = \M_1 \sqcup \M_2$ and $(\Gam_{s_k}; x : \M_1) \restto{\hat{p}_k} = \Gam_{s_k} \restto{\hat{p}_k}$. By~\cref{lem:split-and-merge-for-multi-types} over $\Phi_u$, we know there exist $(\Phi^i_u \tr \seq{\Del_i}{u}{\M_i})_{i \in \interval{1}{2}}$, such that $\Del = \Del_1 \otimes \Del_2$ and $\sz{\Phi_u} = \sz{\Phi^1_u} + \sz{\Phi^2_u}$. By the induction hypothesis over $\Phi_{s_k}$ (resp. $\Phi_r$), we know there exists $\Phi_{s_k \subs{x}{u}} \tr \seq{\Gam_{s_k} \otimes \Del_1}{s_k \subs{x}{u}}{\sig}$ (resp. $\Phi_{r \subs{x}{u}} \tr \seq{\Gam_r \otimes \Del_2}{r \subs{x}{u}}{\M_0}$), such that $\sz{\Phi_{s_k \subs{x}{u}}} = \sz{\Phi_{s_k}} + \sz{\Phi^1_u} - \len{\M_1}$ (resp. $\sz{\Phi_{r \subs{x}{u}}} = \sz{\Phi_r} + \sz{\Phi^2_u} - \len{\M_2}$). Therefore, we can build $\Phi_{t \subs{x}{u}}$ as follows:
              \[ \begin{prooftree}
                      \hypo{\Phi_{r \subs{x}{u}}}
                      \hypo{\Pi}
                      \hypo{\Phi_{s_k \subs{x}{u}}}
                      \infer3[(\ruleCase)]{\seq{((\Gam_{s_k} \otimes \Del_1) \mminus \var{\hat{p}_k}) \otimes \Gam_r \otimes \Del_2}{\case{r \subs{x}{u}}{\vect{\branch{\hat{p}_i}{s_i \subs{x}{u}}}_n}}{\sig}}
                  \end{prooftree} \]
              By $\alpha$-conversion, we may assume $\var{\hat{p}_i} \cap \fv{u} = \eset$, for all $i \in \interval{1}{n}$. Thus, $\case{r \subs{x}{u}}{\vect{\branch{\hat{p}_i}{s_i \subs{x}{u}}}_n} = (\case{r}{\vect{\branch{\hat{p}_i}{s_i}}_n}) \subs{x}{u}$ and $\dom{\Del_1} \cap \var{\hat{p}_k} = \eset$ (by~\cref{lem:relevance}). Therefore, $(\Gam_{s_k} \otimes \Del_1) \mminus \var{\hat{p}_k} = (\Gam_s \mminus \var{\hat{p}_k}) \otimes \Del_1$, and we can conclude since $\Gam \otimes \Del = (\Gam_s \mminus \var{\hat{p}_k}) \otimes \Gam_r \otimes \Del_1 \otimes \Del_2$ and $\sz{\Phi_{t \subs{x}{u}}} = \sz{\Phi_{s_k \subs{x}{u}}} + \sz{\Pi} + \sz{\Phi_{r \subs{x}{u}}} + 1 = \sz{\Phi_{s_k}} + \sz{\Phi^1_u} - \len{\M_1} + \sz{\Pi} + \sz{\Phi_r} + \sz{\Phi^2_u} - \len{\M_2} + 1 = \sz{\Phi_t} + \sz{\Phi_u} - \len{\M}$.
    \end{itemize}
\end{proof}
}

\lemwightedsubjred*

\maybehide{\begin{proof}
    The proof follows by induction over $t \redd t'$:
    \begin{itemize}
        \item Case $t = \MC_0 \lhole{\lam p.u} s \redd[(\ruleBeta)] \MC_0 \lhole{u \match{p}{s}} = t'$, such that $\bv{\MC_0} \cap \fv{s} = \eset$. We proceed by induction over the list matching context $\MC_0$:
              \begin{itemize}
                  \item Let $\MC_0 = \ehole$. Then, $\Phi_t$ must be of the following form:
                        \[ \begin{prooftree}
                                \hypo{\Phi_u \tr \seq{\Gam_u}{u}{\sig}}
                                \hypo{\Pi \tr \seqp{\Gam_u \restto{p}}{p}{\M}}
                                \infer2[(\ruleAbs)]{\seq{\Gam_u \mminus \var{p}}{\lam p.u}{\M \ta \sig}}
                                \hypo{\Phi_s \tr \seq{\Gam_s}{s}{\M}}
                                \infer2[(\ruleApp)]{\seq{(\Gam_u \mminus \var{p}) \otimes \Gam_s}{(\lam p.u) s}{\sig}}
                            \end{prooftree} \]
                        where $\Gam = (\Gam_u \mminus \var{p}) \otimes \Gam_s$. Therefore, we can build $\Phi_{t'}$ as follows:
                        \[ \begin{prooftree}
                                \hypo{\Phi_u}
                                \hypo{\Pi}
                                \hypo{\Phi_s}
                                \infer3[(\ruleMatch)]{\seq{(\Gam_u \mminus \var{p}) \otimes \Gam_s}{u \match{p}{s}}{\sig}}
                            \end{prooftree} \]
                        We can conclude with $\sz{\Phi_t} = \sz{\Phi_u} + \sz{\Pi} + 1 + \sz{\Phi_s} + 1 > \sz{\Phi_u} + \sz{\Pi} + \sz{\Phi_s} = \sz{\Phi_{t'}}$.
                  \item Let $\MC_0 = \MC_1 \match{q}{r}$. Then, $\Phi_t$ must be of the following form:
                        \[ \hspace{-2cm} \begin{prooftree}
                                \hypo{\Phi_{\MC_1 \lhole{\lam p.u}} \tr \seq{\Gam_{\MC_1 \lhole{\lam p.u}}}{\MC_1 \lhole{\lam p.u}}{\M_p \ta \sig}}
                                \infer[no rule]1{\Pi \tr \seqp{\Gam_{\MC_1 \lhole{\lam p.u}} \restto{q}}{q}{\M_q}}
                                \infer[no rule]1{\Phi_r \tr \seq{\Gam_r}{r}{\M_q}}
                                \infer1[(\ruleMatch)]{\seq{(\Gam_{\MC_1 \lhole{\lam p.u}} \mminus \var{q}) \otimes \Gam_r}{\MC_1 \lhole{\lam p.u} \match{q}{r}}{\M_p \ta \sig}}
                                \hypo{\Phi_s \tr \seq{\Gam_s}{s}{\M_p}}
                                \infer2[(\ruleApp)]{\seq{(\Gam_{\MC_1 \lhole{\lam p.u}} \mminus \var{q}) \otimes \Gam_r \otimes \Gam_s}{((\MC_1 \lhole{\lam p.u}) \match{q}{r}) s}{\sig}}
                            \end{prooftree} \]
                        where $\Gam = (\Gam_{\MC_1 \lhole{\lam p.u}} \mminus \var{q}) \otimes \Gam_r \otimes \Gam_s$. Since $\bv{\MC_1} \subseteq \bv{\MC_0}$, we can do the step $\MC_1 \lhole{\lam p.u} s \redd \MC_1 \lhole{u \match{p}{s}}$. Moreover, we can build derivation $\Psi_{\MC_1 \lhole{\lam p.u} s}$ for $\MC_1 \lhole{\lam p.u} s$ as follows:
                        \[ \begin{prooftree}
                                \hypo{\Phi_{\MC_1 \lhole{\lam p.u}}}
                                \hypo{\Phi_s}
                                \infer2[(\ruleApp)]{\seq{\Gam_{\MC_1 \lhole{\lam p.u}} \otimes \Gam_s}{\MC_1 \lhole{\lam p.u} s}{\sig}}
                            \end{prooftree} \]
                        Therefore, by the \ih over $\MC_1 \lhole{\lam p.u} s \redd \MC_1 \lhole{u \match{p}{s}}$, we know there exists $\Psi_{\MC_1 \lhole{u \match{p}{s}}} \tr \seq{\Gam_{\MC_1 \lhole{\lam p.u}} \otimes \Gam_s}{\MC_1 \lhole{u \match{p}{s}}}{\sig}$, such that $\sz{\Psi_{\MC_1 \lhole{\lam p.u} s}} > \sz{\Psi_{\MC_1 \lhole{u \match{p}{s}}}}$. Since $\bv{\MC_0} \cap \fv{s} = \eset$, we know that $\fv{s} \cap \var{q} = \eset$ and thus $\var{q} \cap \dom{\Gam_s} = \eset$ (\cref{lem:relevance}). Therefore, $\Gam_{\MC_1 \lhole{\lam p.u}} \otimes \Gam_s \restto{q} = \Gam_{\MC_1 \lhole{\lam p.u}} \restto{q}$, $(\Gam_{\MC_1 \lhole{\lam p.u}} \otimes \Gam_s) \mminus \var{q} = (\Gam_{\MC_1 \lhole{\lam p.u}}) \mminus \var{q} \otimes \Gam_s$, and we can build $\Phi_{t'}$ as follows:
                        \[ \begin{prooftree}
                                \hypo{\Psi_{\MC_1 \lhole{u \match{p}{s}}}}
                                \hypo{\Pi}
                                \hypo{\Phi_r}
                                \infer3[(\ruleMatch)]{\seq{((\Gam_{\MC_1 \lhole{\lam p.u}} \otimes \Gam_s) \mminus \var{q}) \otimes \Gam_r}{(\MC_1 \lhole{u \match{p}{s}}) \match{q}{r}}{\sig}}
                            \end{prooftree} \]
                        We can conclude since $\Gam = (\Gam_{\MC_1 \lhole{\lam p.u}} \mminus \var{q}) \otimes \Gam_r \otimes \Gam_s = ((\Gam_{\MC_1 \lhole{\lam p.u}} \otimes \Gam_s) \mminus \var{q}) \otimes \Gam_r$ and $\sz{\Phi_t} = \sz{\Phi_{\MC_1 \lhole{\lam p.u}}} + \sz{\Pi} + \sz{\Phi_r} + \sz{\Phi_s} + 1 = \sz{\Psi_{\MC_1 \lhole{\lam p.u} s}} + \sz{\Pi} + \sz{\Phi_r} > \sz{\Psi_{\MC_1 \lhole{u \match{p}{s}}}} + \sz{\Pi} + \sz{\Phi_r} = \sz{\Phi_{t'}}$.
              \end{itemize}
        \item Case $t = u \match{\const{\ttc}{\vect{p_i}_n}}{\MC_0 \lhole{\const{\ttc}{\vect{s_i}_n}}} \redd[(\ruleM)] \MC_0 \lhole{u \match{p_1}{s_1} \cdots \match{p_n}{s_n}} = t'$, such that $\bv{\MC_0} \cap \fv{u} = \eset$. We proceed by induction over the list matching context $\M_0$:
              \begin{itemize}
                  \item Let $\MC_0 = \ehole$. Then, $\Phi_t$ must be of the following form:
                        \[ \begin{prooftree}
                                \hypo{\Phi_u \tr \seq{\Gam_u}{u}{\sig}}
                                \hypo{(\Pi_{p_i} \tr \Gam_u \restto{p_i} \Vdash p_i : \M_i)_{i \in \interval{1}{n}}}
                                \infer1[(\rulePatC)]{\seqp{\Gam_u \restto{\const{\ttc}{\vect{p_i}_n}}}{\const{\ttc}{\vect{p_i}_n}}{\mul{\const{\ttc}{\vect{\M_i}_n}}}}
                                \hypo{\Phi_{s_i} \tr \seq{\Gam_{s_i}}{s_i}{\M_i}}
                                \delims{\left(}{\right)_{i \in \interval{1}{n}}}
                                \infer1[(\ruleConst)]{\hspace{.4cm}\seq{\otimes_{i \in \interval{1}{n}} \Gam_{s_i}}{\const{\ttc}{\vect{s_i}_n}}{\const{\ttc}{\vect{\M_i}_n}}\hspace{.4cm}}
                                \infer1[(\ruleMany)]{\seq{\otimes_{i \in \interval{1}{n}} \Gam_{s_i}}{\const{\ttc}{\vect{s_i}_n}}{\mul{\const{\ttc}{\vect{\M_i}_n}}}}
                                \infer3[(\ruleMatch)]{\seq{(\Gam_u \mminus \var{\const{\ttc}{\vect{p_i}_n}}) \otimes_{i \in \interval{1}{n}} \Gam_{s_i}}{u \match{\const{\ttc}{\vect{p_i}_n}}{\const{\ttc}{\vect{s_i}_n}}}{\sig}}
                            \end{prooftree} \]
                        where $\Gam = (\Gam_u \mminus \var{\const{\ttc}{\vect{p_i}_n}}) \otimes_{i \in \interval{1}{n}} \Gam_{s_i}$, and $\Gam_u \restto{\const{\ttc}{\vect{p_i}_n}} = \otimes_{i \in \interval{1}{n}} (\Gam_u \restto{p_i})$. By $\alpha$-conversion, we can assume that $\fv{\const{\ttc}{\vect{s_i}_n}} \cap \var{\const{\ttc}{\vect{p_i}_n}} = \eset$ and thus $\dom{\otimes_{\interval{1}{n}} \Gam_{s_i}} \cap \var{\const{\ttc}{\vect{p_i}_n}} = \eset$ (\cref{lem:relevance}). Therefore, $(\otimes_{\interval{1}{n}} \Gam_{s_i}) \mminus \var{\const{\ttc}{\vect{p_i}_n}} = \otimes_{\interval{1}{n}} \Gam_{s_i}$, $(\otimes_{\interval{1}{n}} \Gam_{s_i}) \restto{\const{\ttc}{\vect{p_i}_n}} = \eset$, and we can build $\Phi_{t'}$ as follows:
                        \[ \begin{prooftree}
                                \hypo{\Phi_u}
                                \hypo{\Pi_{p_1}}
                                \hypo{\Phi_{s_1}}
                                \infer3[(\ruleMatch)]{\seq{\Gam_u \mminus \var{p_1} \otimes \Gam_{s_1}}{u \match{p_1}{s_1}}{\sig}}
                                \infer[no rule]1[]{\vdots}
                                \hypo{\Pi_{p_n}}
                                \hypo{\Phi_{s_n}}
                                \infer3[(\ruleMatch)]{\seq{(\Gam_u \mminus \var{p_1} \cdots \mminus \var{p_n}) \otimes_{i \in \interval{1}{n}} \Gam_{s_i}}{u \match{p_1}{s_1} \cdots \match{p_n}{s_n}}{\sig}}
                            \end{prooftree} \]
                        We can conclude since $\var{\const{\ttc}{\vect{p_i}_n}} = \var{p_1} \cup \cdots \cup \vect{p_n}$, $\sz{\Phi_t} = \sz{\Phi_u} + 1 +_{i \in \interval{1}{n}} \sz{\Phi_{p_i}} + 1 +_{i \in \interval{1}{n}} \sz{\Phi_{s_i}} = \sz{\Phi_{t'}} + 2 > \sz{\Phi_{t'}}$.
                  \item Let $\MC_0 = \MC_1 \match{q}{r}$. Then, $\Phi_t$ must be of the following form:
                        \[ \hspace{-2cm} \begin{prooftree}
                                \hypo{\Phi_u \tr \seq{\Gam_u}{u}{\sig}}
                                \infer[no rule]1{\Pi \tr \seqp{\Gam_u \restto{\const{\ttc}{\vect{p_i}_n}}}{\const{\ttc}{\vect{p_i}_n}}{\mul{\const{\ttc}{\vect{\M_i}_n}}}}
                                \hypo{\Phi_{\MC_1 \lhole{\const{\ttc}{\vect{s_i}_n}}} \tr \seq{\Gam_{\MC_1 \lhole{\const{\ttc}{\vect{s_i}_n}}}}{\MC_1 \lhole{\const{\ttc}{\vect{s_i}_n}}}{\mul{\const{\ttc}{\vect{\M_i}_n}}}}
                                \infer[no rule]1{\Pi_q \tr \seqp{\Gam_{\MC_1 \lhole{\const{\ttc}{\vect{s_i}_n}}} \restto{q}}{q}{\M_q}}
                                \infer[no rule]1{\Phi_r \tr \seq{\Gam_r}{r}{\M_q}}
                                \infer1[(\ruleMatch)]{\seq{(\Gam_{\MC_1 \lhole{\const{\ttc}{\vect{s_i}_n}}} \mminus \var{q}) \otimes \Gam_r}{\MC_1 \lhole{\const{\ttc}{\vect{s_i}_n}} \match{q}{r}}{\mul{\const{\ttc}{\vect{\M_i}_n}}}}
                                \infer2[(\ruleMatch)]{\seq{(\Gam_u \mminus \var{\const{\ttc}{\vect{p_i}_n}}) \otimes (\Gam_{\MC_1 \lhole{\const{\ttc}{\vect{s_i}_n}}} \mminus \var{q}) \otimes \Gam_r}{u \match{\const{\ttc}{\vect{p_i}_n}}{\MC_1 \lhole{\const{\ttc}{\vect{s_i}_n}} \match{q}{r}}}{\sig}}
                            \end{prooftree} \]
                        where $\Gam = (\Gam_u \mminus \var{\const{\ttc}{\vect{p_i}_n}}) \otimes (\Gam_{\MC_1 \lhole{\const{\ttc}{\vect{s_i}_n}}} \mminus \var{q}) \otimes \Gam_r$. Since $\bv{\MC_1} \subseteq \bv{\MC_0}$, we can do the step $u \match{\const{\ttc}{\vect{p_i}_n}}{\MC_1 \lhole{\const{\ttc}{\vect{s_i}_n}}} \redd[(\ruleM)] \MC_1 \lhole{u \match{p_1}{s_1} \cdots \match{p_n}{s_n}}$. Moreover, we can build derivation $\Psi_{u \match{\const{\ttc}{\vect{p_i}_n}}{\MC_1 \lhole{\const{\ttc}{\vect{s_i}_n}}}}$ for $u \match{\const{\ttc}{\vect{p_i}_n}}{\MC_1 \lhole{\const{\ttc}{\vect{s_i}_n}}}$ as follows:
                        \[ \begin{prooftree}
                                \hypo{\Phi_u}
                                \hypo{\Pi}
                                \hypo{\Phi_{\MC_1 \lhole{\const{\ttc}{\vect{s_i}_n}}}}
                                \infer3[(\ruleMatch)]{\seq{(\Gam_u \mminus \var{\const{\ttc}{\vect{p_i}_n}}) \otimes \Gam_{\MC_1 \lhole{\const{\ttc}{\vect{s_i}_n}}}}{u \match{\const{\ttc}{\vect{p_i}_n}}{\MC_1 \lhole{\const{\ttc}{\vect{s_i}_n}}}}{\sig}}
                            \end{prooftree} \]
                        Therefore, by the \ih over $u \match{\const{\ttc}{\vect{p_i}_n}}{\MC_1 \lhole{\const{\ttc}{\vect{s_i}_n}}} \redd[(\ruleM)] \MC_1 \lhole{u \match{p_1}{s_1} \cdots \match{p_n}{s_n}}$, we know there exists $\Psi_{\MC_1 \lhole{u \match{p_1}{s_1} \cdots \match{p_n}{s_n}}} \tr \seq{(\Gam_u \mminus \var{\const{\ttc}{\vect{p_i}_n}}) \otimes \Gam_{\MC_1 \lhole{\const{\ttc}{\vect{s_i}_n}}}}{\MC_1 \lhole{u \match{p_1}{s_1} \cdots \match{p_n}{s_n}}}{\sig}$, such that $\sz{\Psi_{u \match{\const{\ttc}{\vect{p_i}_n}}{\MC_1 \lhole{\const{\ttc}{\vect{s_i}_n}}}}} > \sz{\Psi_{\MC_1 \lhole{u \match{p_1}{s_1} \cdots \match{p_n}{s_n}}}}$. Since $\bv{\MC_0} \cap \fv{u} = \eset$, we know that $\fv{u} \cap \var{q} = \eset$ and thus $\var{q} \cap \dom{\Gam_u} = \eset$ (\cref{lem:relevance}). Therefore, $(\Gam_u \mminus \var{\const{\ttc}{\vect{p_i}_n}}) \otimes \Gam_{\MC_1 \lhole{\const{\ttc}{\vect{s_i}_n}}} \restto{q} = \Gam_{\MC_1 \lhole{\const{\ttc}{\vect{s_i}_n}}} \restto{q}$, $((\Gam_u \mminus \var{\const{\ttc}{\vect{p_i}_n}}) \otimes \Gam_{\MC_1 \lhole{\const{\ttc}{\vect{s_i}_n}}}) \mminus \var{q} = (\Gam_u \mminus \var{\const{\ttc}{\vect{p_i}_n}}) \otimes (\Gam_{\MC_1 \lhole{\const{\ttc}{\vect{s_i}_n}}} \mminus \var{q})$, and we can build $\Phi_{t'}$ as follows:
                        \[ \begin{prooftree}
                                \hypo{\Psi_{\MC_1 \lhole{u \match{p_1}{s_1} \cdots \match{p_n}{s_n}}}}
                                \hypo{\Pi_q}
                                \hypo{\Phi_r}
                                \infer3[(\ruleMatch)]{\seq{((\Gam_u \mminus \var{\const{\ttc}{\vect{p_i}_n}}) \otimes \Gam_{\MC_1 \lhole{\const{\ttc}{\vect{s_i}_n}}}) \mminus \var{q} \otimes \Gam_r}{\MC_1 \lhole{u \match{p_1}{s_1} \cdots \match{p_n}{s_n}} \match{q}{r}}{\sig}}
                            \end{prooftree} \]
                        We can conclude since $\Gam = (\Gam_u \mminus \var{\const{\ttc}{\vect{p_i}_n}}) \otimes (\Gam_{\MC_1 \lhole{\const{\ttc}{\vect{s_i}_n}}} \mminus \var{q}) \otimes \Gam_r = ((\Gam_u \mminus \var{\const{\ttc}{\vect{p_i}_n}}) \otimes \Gam_{\MC_1 \lhole{\const{\ttc}{\vect{s_i}_n}}}) \mminus \var{q} \otimes \Gam_r$ and $\sz{\Phi_t} = \sz{\Phi_u} + \sz{\Pi} + \sz{\Phi_{\MC_1 \lhole{\const{\ttc}{\vect{s_i}_n}}}} + \sz{\Pi_q} + \sz{\Phi_r} = \sz{\Psi_{u \match{\const{\ttc}{\vect{p_i}_n}}{\MC_1 \lhole{\const{\ttc}{\vect{s_i}_n}}}}} + \sz{\Pi_q} + \sz{\Phi_r} > \sz{\Psi_{\MC_1 \lhole{u \match{p_1}{s_1} \cdots \match{p_n}{s_n}}}} + \sz{\Pi_q} + \sz{\Phi_r} = \sz{\Phi_{t'}}$.
              \end{itemize}
        \item Case $t = u \match{x}{s} \redd[(\ruleE)] u \subs{x}{s} = t'$. Then, $\Phi_t$ must be of the following form:
              \[ \begin{prooftree}
                      \hypo{\Phi_u \tr \seq{\Gam_u; x : \M}{u}{\sig}}
                      \infer0[(\rulePatV)]{\seqp{x : \M}{x}{\M}}
                      \hypo{\Phi_s \tr \seq{\Gam_s}{s}{\M}}
                      \infer3[(\ruleMatch)]{\seq{\Gam_u \otimes \Gam_s}{u \match{x}{s}}{\sig}}
                  \end{prooftree} \]
              where $\Gam = \Gam_u \otimes \Gam_s$. By~\cref{lem:weighted-substitution} over $\Phi_u$ and $\Phi_s$, we know there exists $\Phi_{u \subs{x}{s}} \tr \seq{\Gam_u \otimes \Gam_s}{u \subs{x}{s}}{\sig}$, such that $\sz{\Phi_{u \subs{x}{s}}} = \sz{\Phi_u} + \sz{\Phi_s} - \sizeof{\M}$. Therefore, we can pick $\Phi_{t'} = \Phi_{u \subs{x}{s}}$ and can conclude since $\sz{\Phi_t} = \sz{\Phi_u} + 1 + \sz{\Phi_s} > \sz{\Phi_u} + \sz{\Phi_s} - \sizeof{\M} = \sz{\Phi_{t'}}$.
        \item Case $t = \case{\MC_0 \lhole{\const{\ttc}{\vect{s_i}_n}}}{(\ldots, \branch{\const{\ttc}{\vect{p_i}_n}}{u_i}, \ldots)} \redd[(\ruleC)] \MC_0 \lhole{u_i \match{p_1}{s_1} \cdots \match{p_n}{s_n}} = t'$, such that $\bv{\MC_0} \cap \fv{u_i} = \eset$. We proceed by induction over the list matching context $\M_0$:
              \begin{itemize}
                  \item Let $\MC_0 = \ehole$. Then, $\Phi_t$ must be of the following form:
                        \[ \hspace{-1cm} \begin{prooftree}
                                \hypo{\Phi_{s_i} \tr \seq{\Gam_{s_i}}{s_i}{\M_i}}
                                \delims{\left(}{\right)_{i \in \interval{1}{n}}}
                                \infer1[(\ruleConst)]{\seq{\otimes_{i \in \interval{1}{n}} \Gam_{s_i}}{\const{\ttc}{\vect{s_i}_n}}{\const{\ttc}{\vect{\M_i}_n}}}
                                \infer1[(\ruleMany)]{\seq{\otimes_{i \in \interval{1}{n}} \Gam_{s_i}}{\const{\ttc}{\vect{s_i}_n}}{\mul{\const{\ttc}{\vect{\M_i}_n}}}}
                                \hypo{(\Pi_{p_i} \tr \Gam_{u_i} \restto{p_i} \Vdash p_i : \M_i)_{i \in \interval{1}{n}}}
                                \infer1[(\rulePatC)]{\seqp{\Gam_{u_i} \restto{\const{\ttc}{\vect{p_i}_n}}}{\const{\ttc}{\vect{p_i}_n}}{\mul{\const{\ttc}{\vect{\M_i}_n}}}}
                                \hypo{\Phi_{u_i} \tr \seq{\Gam_{u_i}}{u_i}{\sig}}
                                \infer3[(\ruleCase)]{\seq{(\Gam_{u_i} \mminus \var{\const{\ttc}{\vect{p_i}_n}}) \otimes_{i \in \interval{1}{n}} \Gam_{s_i}}{\case{\const{\ttc}{\vect{s_i}_n}}{(\ldots, \branch{\const{\ttc}{\vect{p_i}_n}}{u_i}, \ldots)}}{\sig}}
                            \end{prooftree} \]
                        where $\Gam = (\Gam_{u_i} \mminus \var{\const{\ttc}{\vect{p_i}_n}}) \otimes_{i \in \interval{1}{n}} \Gam_{s_i}$, and $\Gam_{u_i} \restto{\const{\ttc}{\vect{p_i}_n}} = \otimes_{i \in \interval{1}{n}} (\Gam_{u_i} \restto{p_i})$. By $\alpha$-conversion, we can assume that $\fv{\const{\ttc}{\vect{s_i}_n}} \cap \var{\const{\ttc}{\vect{p_i}_n}} = \eset$ and thus $\dom{\otimes_{\interval{1}{n}}} \Gam_{s_i} \cap \var{\const{\ttc}{\vect{p_i}_n}} = \eset$ (\cref{lem:relevance}). Therefore, $(\otimes_{\interval{1}{n}} \Gam_{s_i}) \mminus \var{\const{\ttc}{\vect{p_i}_n}} = \eset$, $(\otimes_{\interval{1}{n}} \Gam_{s_i}) \restto{\const{\ttc}{\vect{p_i}_n}} = \eset$, and we can build $\Phi_{t'}$ as follows:
                        \[ \begin{prooftree}
                                \hypo{\Phi_{u_i}}
                                \hypo{\Pi_{p_1}}
                                \hypo{\Phi_{s_1}}
                                \infer3[(\ruleMatch)]{\seq{\Gam_{u_i} \mminus \var{p_1} \otimes \Gam_{s_1}}{u \match{p_1}{s_1}}{\sig}}
                                \infer[no rule]1[]{\vdots}
                                \hypo{\Pi_{p_n}}
                                \hypo{\Phi_{s_n}}
                                \infer3[(\ruleMatch)]{\seq{(\Gam_{u_i} \mminus \var{p_1} \cdots \mminus \var{p_n}) \otimes_{i \in \interval{1}{n}} \Gam_{s_i}}{u_i \match{p_1}{s_1} \cdots \match{p_n}{s_n}}{\sig}}
                            \end{prooftree} \]
                        We can conclude since $\var{\const{\ttc}{\vect{p_i}_n}} = \var{p_1} \cup \cdots \cup \var{p_n}$, $\sz{\Phi_t} = 1 +_{i \in \interval{1}{n}} \sz{\Phi_{s_i}} + 1 +_{i \in \interval{1}{n}} \sz{\Pi_{p_i}} + \sz{\Phi_{u_i}} + 1 = \sz{\Phi_{t'}} + 3 > \sz{\Phi_{t'}}$.
                  \item Let $\MC_0 = \MC_1 \match{q}{r}$. Then, $\Phi_t$ must be of the following form:
                        \[ \hspace{-2cm} \begin{prooftree}
                                \hypo{\Phi_{\MC_1 \lhole{\const{\ttc}{\vect{s_i}_n}}} \tr \seq{\Gam_{\MC_1 \lhole{\const{\ttc}{\vect{s_i}_n}}}}{\MC_1 \lhole{\const{\ttc}{\vect{s_i}_n}}}{\mul{\const{\ttc}{\vect{\M_i}_n}}}}
                                \infer[no rule]1{\Pi_q \tr \seqp{\Gam_{\MC_1 \lhole{\const{\ttc}{\vect{s_i}_n}}} \restto{q}}{q}{\M_q}}
                                \infer[no rule]1{\Phi_r \tr \seq{\Gam_r}{r}{\M_q}}
                                \infer1[(\ruleMatch)]{\seq{(\Gam_{\MC_1 \lhole{\const{\ttc}{\vect{s_i}_n}}} \mminus \var{q}) \otimes \Gam_r}{\MC_1 \lhole{\const{\ttc}{\vect{s_i}_n}} \match{q}{r}}{\mul{\const{\ttc}{\vect{\M_i}_n}}}}
                                \hypo{\Pi \tr \seqp{\Gam_{u_i} \restto{\const{\ttc}{\vect{p_i}_n}}}{\const{\ttc}{\vect{p_i}_n}}{\mul{\const{\ttc}{\vect{\M_i}_n}}}}
                                \infer[no rule]1{\Phi_{u_i} \tr \seq{\Gam_{u_i}}{u_i}{\sig}}
                                \infer2[(\ruleCase)]{\seq{(\Gam_{u_i} \mminus \var{\const{\ttc}{\vect{p_i}_n}}) \otimes (\Gam_{\MC_1 \lhole{\const{\ttc}{\vect{s_i}_n}}} \mminus \var{q}) \otimes \Gam_r}{\case{\MC_1 \lhole{\const{\ttc}{\vect{s_i}_n}} \match{q}{r}}{(\ldots, \branch{\const{\ttc}{\vect{p_i}_n}}{u_i}, \ldots)}}{\sig}}
                            \end{prooftree} \]
                        where $\Gam = (\Gam_{u_i} \mminus \var{\const{\ttc}{\vect{p_i}_n}}) \otimes (\Gam_{\MC_1 \lhole{\const{\ttc}{\vect{s_i}_n}}} \mminus \var{q}) \otimes \Gam_r$. Since $\bv{\MC_1} \subseteq \bv{\MC_0}$, we can do the step $\case{\MC_1 \lhole{\const{\ttc}{\vect{s_i}_n}}}{(\ldots, \branch{\const{\ttc}{\vect{p_i}_n}}{u_i}, \ldots)} \redd[(\ruleC)] \MC_1 \lhole{u_i \match{p_1}{s_1} \cdots \match{p_n}{s_n}}$. Moreover, we can build derivation $\Psi_{\case{\MC_1 \lhole{\const{\ttc}{\vect{s_i}_n}}}{(\ldots, \branch{\const{\ttc}{\vect{p_i}_n}}{u_i}, \ldots)}}$ for $\case{\MC_1 \lhole{\const{\ttc}{\vect{s_i}_n}}}{(\ldots, \branch{\const{\ttc}{\vect{p_i}_n}}{u_i}, \ldots)}$ as follows:
                        \[ \begin{prooftree}
                                \hypo{\Phi_{\MC_1 \lhole{\const{\ttc}{\vect{s_i}_n}}}}
                                \hypo{\Pi}
                                \hypo{\Phi_{u_i}}
                                \infer3[(\ruleCase)]{\seq{(\Gam_{u_i} \mminus \var{\const{\ttc}{\vect{p_i}_n}}) \otimes \Gam_{\MC_1 \lhole{\const{\ttc}{\vect{s_i}_n}}}}{\case{\MC_1 \lhole{\const{\ttc}{\vect{s_i}_n}}}{(\ldots, \branch{\const{\ttc}{\vect{p_i}_n}}{u_i}, \ldots)}}{\sig}}
                            \end{prooftree} \]
                        Therefore, by the \ih over $\case{\MC_1 \lhole{\const{\ttc}{\vect{s_i}_n}}}{(\ldots, \branch{\const{\ttc}{\vect{p_i}_n}}{u_i}, \ldots)} \redd[(\ruleC)] \MC_1 \lhole{u_i \match{p_1}{s_1} \cdots \match{p_n}{s_n}}$, we know there exists $\Psi_{\MC_1 \lhole{u_i \match{p_1}{s_1} \cdots \match{p_n}{s_n}}} \tr \seq{(\Gam_{u_i} \mminus \var{\const{\ttc}{\vect{p_i}_n}}) \otimes \Gam_{\MC_1 \lhole{\const{\ttc}{\vect{s_i}_n}}}}{\MC_1 \lhole{u_i \match{p_1}{s_1} \cdots \match{p_n}{s_n}}}{\sig}$, such that $\sz{\Psi_{\case{\MC_1 \lhole{\const{\ttc}{\vect{s_i}_n}}}{(\ldots, \branch{\const{\ttc}{\vect{p_i}_n}}{u_i}, \ldots)}}} > \sz{\Psi_{\MC_1 \lhole{u_i \match{p_1}{s_1} \cdots \match{p_n}{s_n}}}}$. Since $\bv{\MC_0} \cap \fv{u_i} = \eset$, we know that $\fv{u_i} \cap \var{q} = \eset$ and thus $\var{q} \cap \dom{\Gam_{u_i}} = \eset$ (\cref{lem:relevance}). Therefore, $(\Gam_{u_i} \mminus \var{\const{\ttc}{\vect{p_i}_n}}) \otimes \Gam_{\MC_1 \lhole{\const{\ttc}{\vect{s_i}_n}}} \restto{q} = \Gam_{\MC_1 \lhole{\const{\ttc}{\vect{s_i}_n}}} \restto{q}$, $((\Gam_{u_i} \mminus \var{\const{\ttc}{\vect{p_i}_n}}) \otimes \Gam_{\MC_1 \lhole{\const{\ttc}{\vect{s_i}_n}}}) \mminus \var{q} = (\Gam_{u_i} \mminus \var{\const{\ttc}{\vect{p_i}_n}}) \otimes (\Gam_{\MC_1 \lhole{\const{\ttc}{\vect{s_i}_n}}} \mminus \var{q})$, and we can build $\Phi_{t'}$ as follows:
                        \[ \begin{prooftree}
                                \hypo{\Psi_{\MC_1 \lhole{u_i \match{p_1}{s_1} \cdots \match{p_n}{s_n}}}}
                                \hypo{\Pi_q}
                                \hypo{\Phi_r}
                                \infer3[(\ruleMatch)]{\seq{((\Gam_{u_i} \mminus \var{\const{\ttc}{\vect{p_i}_n}}) \otimes \Gam_{\MC_1 \lhole{\const{\ttc}{\vect{s_i}_n}}}) \mminus \var{q} \otimes \Gam_r}{\MC_1 \lhole{u_i \match{p_1}{s_1} \cdots \match{p_n}{s_n}} \match{q}{r}}{\sig}}
                            \end{prooftree} \]
                        We can conclude since $\Gam = (\Gam_{u_i} \mminus \var{\const{\ttc}{\vect{p_i}_n}}) \otimes (\Gam_{\MC_1 \lhole{\const{\ttc}{\vect{s_i}_n}}} \mminus \var{q}) \otimes \Gam_r = ((\Gam_{u_i} \mminus \var{\const{\ttc}{\vect{p_i}_n}}) \otimes \Gam_{\MC_1 \lhole{\const{\ttc}{\vect{s_i}_n}}}) \mminus \var{q} \otimes \Gam_r$ and $\sz{\Phi_t} = \sz{\Phi_{\MC_1 \lhole{\const{\ttc}{\vect{s_i}_n}}}} + \sz{\Pi_q} + \sz{\Phi_r} + \sz{\Pi} + \sz{\Phi_{u_i}} + 1 = \sz{\Psi_{\case{\MC_1 \lhole{\const{\ttc}{\vect{s_i}_n}}}{(\ldots, \branch{\const{\ttc}{\vect{p_i}_n}}{u_i}, \ldots)}}} + \sz{\Pi_q} + \sz{\Phi_r} > \Psi_{\MC_1 \lhole{u \match{p_1}{s_1} \cdots \match{p_n}{s_n}}} + \sz{\Pi_q} + \sz{\Phi_r} = \sz{\Phi_{t'}}$.
              \end{itemize}
        \item Case $t = us \redd[(\ruleAppL)] u's = t'$, such that $\neg\isabs{u}$ and $u \redd u'$. Then, $\Phi_t$ must be of the following form:
              \[ \begin{prooftree}
                      \hypo{\Phi_u \tr \seq{\Gam_u}{u}{\M \ta \sig}}
                      \hypo{\Phi_s \tr \seq{\Gam_s}{s}{\M}}
                      \infer2[(\ruleApp)]{\seq{\Gam_u \otimes \Gam_s}{us}{\sig}}
                  \end{prooftree} \]
              where $\Gam = \Gam_u \otimes \Gam_s$. By the \ih over $u \redd u'$, we know there exists $\Phi_{u'} \tr \seq{\Gam_u}{u}{\M \ta \sig}$, such that $\Phi_u > \Phi_{u'}$. Therefore, we can build $\Phi_{t'}$ as follows:
              \[ \begin{prooftree}
                      \hypo{\Phi_{u'}}
                      \hypo{\Phi_s}
                      \infer2[(\ruleApp)]{\seq{\Gam_u \otimes \Gam_s}{u's}{\sig}}
                  \end{prooftree} \]
              We can conclude since $\sz{\Phi_t} = \sz{\Phi_u} + \sz{\Phi_s} + 1 > \sz{\Phi_{u'}} + \sz{\Phi_s} + 1 = \sz{\Phi_{t'}}$.
        \item Case $t = u \match{\const{\ttc}{\ptuple}}{s} \redd[(\ruleEsL)] u' \match{\const{\ttc}{\ptuple}}{s} = t'$, such that $\neg\isdata[\ttc]{s}$ and $u \redd u'$. Then, $\Phi_t$ must be of the following form:
              \[ \begin{prooftree}
                      \hypo{\Phi_u \tr \seq{\Gam_u}{u}{\sig}}
                      \hypo{\Pi \tr \seqp{\Gam_u \restto{\const{\ttc}{\ptuple}}}{\const{\ttc}{\ptuple}}{\M}}
                      \hypo{\Phi_s \tr \seq{\Gam_s}{s}{\M}}
                      \infer3[(\ruleMatch)]{\seq{(\Gam_u \mminus \var{\const{\ttc}{\ptuple}}) \otimes \Gam_s}{u \match{\const{\ttc}{\ptuple}}{s}}{\sig}}
                  \end{prooftree} \]
              where $\Gam = (\Gam_u \mminus \var{\const{\ttc}{\ptuple}}) \otimes \Gam_s$. By the \ih over $u \redd u'$, we know there exists $\Phi_{u'} \tr \seq{\Gam_u}{u'}{\sig}$, such that $\sz{\Phi_u} > \sz{\Phi_{u'}}$. Therefore, we can build $\Phi_{t'}$ as follows:
              \[ \begin{prooftree}
                      \hypo{\Phi_{u'}}
                      \hypo{\Pi}
                      \hypo{\Phi_s}
                      \infer3[(\ruleMatch)]{\seq{(\Gam_u \mminus \var{\const{\ttc}{\ptuple}}) \otimes \Gam_s}{u' \match{\const{\ttc}{\ptuple}}{s}}{\sig}}
                  \end{prooftree} \]
              We can conclude since $\sz{\Phi_t} = \sz{\Phi_u} + \sz{\Pi} + \sz{\Phi_s} > \sz{\Phi_{u'}} \sz{\Pi} + \sz{\Phi_s} = \sz{\Phi_{t'}}$.
        \item Case $t = u \match{\const{\ttc}{\ptuple}}{s} \redd[(\ruleEsR)] u \match{\const{\ttc}{\ptuple}}{s'} = t'$, such that $\neg\isdata[\ttc]{u}$, $u \not\redd$, and $s \redd s'$. Then, $\Phi_t$ must be of the following form:
              \[ \begin{prooftree}
                      \hypo{\Phi_u \tr \seq{\Gam_u}{u}{\sig}}
                      \hypo{\Pi \tr \seqp{\Gam_u \restto{\const{\ttc}{\ptuple}}}{\const{\ttc}{\ptuple}}{\M}}
                      \hypo{\Phi_s \tr \seq{\Gam_s}{s}{\M}}
                      \infer3[(\ruleMatch)]{\seq{(\Gam_u \mminus \var{\const{\ttc}{\ptuple}}) \otimes \Gam_s}{u \match{\const{\ttc}{\ptuple}}{s}}{\sig}}
                  \end{prooftree} \]
              where $\Gam = (\Gam_u \mminus \var{\const{\ttc}{\ptuple}}) \otimes \Gam_s$. By the \ih over $s \redd s'$, we know there exists $\Phi_{s'} \tr \seq{\Gam_u}{s'}{\M}$, such that $\sz{\Phi_s} > \sz{\Phi_{s'}}$. Therefore, we can build $\Phi_{t'}$ as follows:
              \[ \begin{prooftree}
                      \hypo{\Phi_u}
                      \hypo{\Pi}
                      \hypo{\Phi_{s'}}
                      \infer3[(\ruleMatch)]{\seq{(\Gam_u \mminus \var{\const{\ttc}{\ptuple}}) \otimes \Gam_s}{u \match{\const{\ttc}{\ptuple}}{s'}}{\sig}}
                  \end{prooftree} \]
              We can conclude since $\sz{\Phi_t} = \sz{\Phi_u} + \sz{\Pi} + \sz{\Phi_s} > \sz{\Phi_u} + \sz{\Pi} + \sz{\Phi_{s'}} = \sz{\Phi_{t'}}$.
        \item Case $t = \case{s}{\vect{\branch{\const{\ttc_i}{\ptuple_i}}{u_i}}_n} \redd[(\ruleCaseIn)] \case{s'}{\vect{\branch{\const{\ttc_i}{\ptuple_i}}{u_i}}_n} = t'$. Then, $\Phi_t$ must be of the following form:
              \[ \begin{prooftree}
                      \hypo{\Phi_s \tr \seq{\Gam_s}{s}{\M}}
                      \hypo{\Pi \tr \seqp{\Gam_{u_k} \restto{\const{\ttc_k}{\ptuple_k}}}{\const{\ttc_k}{\ptuple_k}}{\M}}
                      \hypo{\Phi_{u_k} \tr \seq{\Gam_{u_k}}{u_k}{\sig}}
                      \hypo{ \text{for some $k \in \interval{1}{n}$}}
                      \infer4[(\ruleCase)]{\seq{(\Gam_{u_k} \mminus \var{\const{\ttc_k}{\ptuple_k}}) \otimes \Gam_s}{\case{s}{\vect{\branch{\const{\ttc_i}{\ptuple_i}}{u_i}}_n}}{\sig}}
                  \end{prooftree} \]
              where $\Gam = (\Gam_{u_k} \mminus \var{\const{\ttc_k}{\ptuple_k}}) \otimes \Gam_s$. By the \ih over $s \redd s'$, we know there exists $\Phi_{s'} \tr \seq{\Gam_s}{s'}{\M}$, such that $\sz{\Phi_s} > \sz{\Phi_{s'}}$. Therefore, we can build $\Phi_{t'}$ as follows:
              \[ \begin{prooftree}
                      \hypo{\Phi_{s'}}
                      \hypo{\Pi}
                      \hypo{\Phi_{u_k}}
                      \infer3[(\ruleCase)]{\seq{(\Gam_{u_k} \mminus \var{\const{\ttc_k}{\ptuple_k}}) \otimes \Gam_s}{\case{s'}{\vect{\branch{\const{\ttc_i}{\ptuple_i}}{u_i}}_n}}{\sig}}
                  \end{prooftree} \]
              We can conclude since $\sz{\Phi_t} = \sz{\Phi_s} + \sz{\Pi} + \sz{\Phi_{u_k}} + 1 > \sz{\Phi_{s'}} + \sz{\Pi} + \sz{\Phi_{u_k}} + 1 = \sz{\Phi_{t'}}$.
    \end{itemize}
\end{proof}
}



\subsubsection{\textit{Completeness.}}

\begin{restatable}[Patterns are Always Typable]{lemma}{lempatterns}
    \label{lem:patterns-are-always-typable}
    Given any typing context $\Gam$ and pattern $p$, there exists $\Pi \tr \seqp{\Gam \restto{p}}{p}{\M}$. Moreover, if $p = \const{\ttc}{\vect{q_i}_n}$, then there exist $\Pi_i \tr \seqp{\Gam \restto{q_i}}{q_i}{\M_i}$, such that $\M = \mul{\const{\ttc}{\vect{\M_i}_n}}$.
\end{restatable}


\maybehide{\begin{proof}
    The proof follows by induction over $p$:
    \begin{itemize}
        \item Case $p = x$. Then, it is always the case that $\Gam \restto{x} (x) = \M$ from some $\M$ (possibly $\M = \emul$ if $x \not\in \Gam$). So we can build $\Pi$ using rule (\rulePatV).
        \item Case $p = \const{\ttc}{\vect{q_i}_n}$. By the \ih, we know there exist $\Pi_i \tr \seqp{\Gam \restto{q_i}}{q_i}{\M_i}$ for each $q_i$. So we can build $\Pi \tr \seqp{\otimes_{i \in \interval{1}{n}} \Gam \restto{q_i}}{\const{\ttc}{\vect{q_i}_n}}{\mul{\const{\ttc}{\vect{\M_i}_n}}}$ using rule (\rulePatC). We can conclude since $\Gam \restto{p} = \otimes_{\interval{1}{n}} (\Gam \restto{q_i})$.
    \end{itemize}
\end{proof}}

\lemtypability*

\maybehide{\begin{proof}
    Assuming $t$ is closed, we only have to consider two cases for $t \in \cf$:
    \begin{itemize}
        \item Case $t = \lam p.u$. We can build $\Phi$ using rule (\ruleAbsStar).
        \item Case $t = \const{\ttc}{\vect{t_i}_n}$. We can build $\Phi_i\ \tr \seq{}{t_i}{\emul}$ for each $t_i$ using rule (\ruleMany) with no premises, and then $\Phi$ by using rule (\ruleConst) with premises $\Phi_i$.
    \end{itemize}
\end{proof}}

\begin{restatable}[Weighted Anti-substitution]{lemma}{lemweightedantisubs}
    \label{lem:weighted-antisubstitution}
    Let $t, u \in \calc$ and assume $\Phi_{t \subs{x}{u}} \tr \seq{\Gam}{t \subs{x}{u}}{\sig}$. Then, there exist $\Phi_t \tr \seq{\Sig; x : \M}{t}{\sig}$ and $\Phi_u \tr \seq{\Del}{u}{\M}$, such that $\Gam = \Sig \otimes \Del$ and $\sz{\Phi_{t \subs{x}{u}}} = \sz{\Phi_t} + \sz{\Phi_u} - \sizeof{\M}$.
\end{restatable}


\maybehide{\begin{proof}
    We generalize the original statement by allowing $\Phi_{t \subs{x}{u}}$ to conclude with either a term type $\sig$ or a multiset type $\M$.

    If $\Phi_{t \subs{x}{u}} \tr \seq{\Gam}{t \subs{x}{u}}{\T}$. Then, there exist derivations $\Phi_t \tr \seq{\Sig; x : \M}{t}{\T}$ and $\Phi_u \tr \seq{\Del}{u}{\M}$, such that $\Gam = \Sig \otimes \Del$ and $\sz{\Phi_{t \subs{x}{u}}} = \sz{\Phi_t} + \sz{\Phi_u} - \sizeof{\M}$.

    We start by splitting into two cases, according to the form of $t$:
    \begin{itemize}
        \item Case $t = x$. Then, $t \subs{x}{u} = u$ and either $\T = \sig$ or $\T = \M_0$. Let $\T = \sig$. Then, we can build $\Phi_t$ and $\Phi_u$ respectively as follows:
              \[ \begin{array}{c}
                      \begin{prooftree}
                          \hypo{\phantom{BUUUU}}
                          \infer1[(\ruleAx)]{\seq{x : \mul{\sig}}{x}{\sig}}
                      \end{prooftree}
                      \sep
                      \begin{prooftree}
                          \hypo{\Phi_{t \subs{x}{u}}}
                          \infer1[(\ruleMany)]{\seq{\Del}{u}{\mul{\sig}}}
                      \end{prooftree}
                  \end{array} \]
              where $\Sig = \eset$ and $\M = \mul{\sig}$. We can conclude since $\Gam = \Del = \Sig \otimes \Del$ and $\sz{\Phi_{t \subs{x}{u}}} = \sz{\Phi_u} = \sz{\Phi_u} + 1 - 1 = \sz{\Phi_u} + \sz{\Phi_t} - \len{\M}$. Now, let $\T = \M_0 = \mul{\sig_i}_{\iI}$. Then, we can build $\Phi_t$ as follows:
              \[ \begin{prooftree}
                      \infer0[(\ruleAx)]{\seq{x : \mul{\sig_i}}{x}{\sig_i}}
                      \delims{\left(}{\right)_{\iI}}
                      \infer1[(\ruleMany)]{\hspace{1cm}\seq{x : \mul{\sig_i}_{\iI}}{x}{\mul{\sig_i}_{\iI}}\hspace{1cm}}
                  \end{prooftree} \]
              where $\Sig = \eset$ and $\M = \M_0$, and pick $\Phi_u = \Phi_{t \subs{x}{u}}$, where $\Del = \Gam$. We can conclude since $\Gam = \Del = \Sig \otimes \Del$ and $\sz{\Phi_{t \subs{x}{u}}} = \sz{\Phi_u} = \sz{\Phi_u} + \len{I} - \len{I} = \sz{\Phi_u} + \sz{\Phi_t} - \len{\M}$.
        \item Case $t \neq x$. The proof follows by induction over $\Phi_{t \subs{x}{u}}$. We reason according to the last rule:
              \begin{itemize}
                  \item Rule (\ruleAx). Then, $t = y$ and $t \subs{x}{u} = y$. Therefore, we can pick $\Phi_t = \Phi_{t \subs{x}{u}}$, where $\Sig; x : \M = \Gam$ and thus $\Sig = \Gam$ and $\M = \emul$, and build $\Phi_u$ as follows:
                        \[ \begin{prooftree}
                                \infer0[(\ruleMany)]{\seq{}{u}{\emul}}
                            \end{prooftree} \]
                        where $\Del = \eset$. We can conclude since $\Gam = \Sig = \Sig \otimes \Del$ and $\sz{\Phi_{t \subs{x}{u}}} = \sz{\Phi_t} = \sz{\Phi_t} + \sz{\Phi_u} - \len{\M}$.
                  \item Rule (\ruleMany). Then, $\Phi_{t \subs{x}{u}}$ must be of the following form:
                        \[ \begin{prooftree}
                                \hypo{\Phi_i \tr \seq{\Gam_i}{t \subs{x}{u}}{\sig_i}}
                                \delims{\left(}{\right)_{\iI}}
                                \infer1[(\ruleMany)]{\hspace{.6cm}\seq{\otimes_{\iI} \Gam_i}{t \subs{x}{u}}{\mul{\sig_i}_{\iI}}\hspace{.6cm}}
                            \end{prooftree} \]
                        where $\Gam = \otimes_{\iI} \Gam_i$ and $\T = \mul{\sig_i}_{\iI}$. By the induction hypothesis over each $\Phi_i$, there exist $\Phi'_i \tr \seq{\Sig_i; x : \M_i}{t}{\sig_i}$ and $\Phi^i_u \tr \seq{\Del_i}{u}{\M_i}$, such that $\Gam_i = \Sig_i \otimes \Del_i$ and $\sz{\Phi_i} = \sz{\Phi'_i} + \sz{\Phi^i_u} - \len{\M_i}$. By~\cref{lem:split-and-merge-for-multi-types} over the $\Phi^i_u$, we know there exists $\Phi_u \tr \seq{\Del}{u}{\M}$, such that $\Del = \otimes_{\iI} \Del_i$, $\M = \sqcup_{\iI} \M_i$, and $\sz{\Phi_u} = +_{\iI} \sz{\Phi^i_u}$. We can build $\Phi_t$ as follows:
                        \[ \begin{prooftree}
                                \hypo{\Phi^i_t}
                                \delims{\left(}{\right)_{\iI}}
                                \infer1[(\ruleMany)]{\seq{\otimes_{\iI} (\Sig_i; x : \M_i)}{t}{\mul{\sig_i}_{\iI}}}
                            \end{prooftree} \]
                        where $\Sig = \otimes_{\iI} \Sig_i$ and $\M = \sqcup_{\iI} \M_i$. We can conclude since $\Gam = \otimes_{\iI} \Gam_i = \otimes_{\iI} (\Sig_i \otimes \Del_i) = (\otimes_{\iI} \Sig_i) \otimes (\otimes_{\iI} \Del_i) = \Sig \otimes \Del$ and $\sz{\Phi_{t \subs{x}{u}}} = +_{\iI} \sz{\Phi_i} = +_{\iI} (\sz{\Phi'_i} + \sz{\Phi^i_u} - \len{\M_i}) = +_{\iI} \sz{\Phi'_i} +_{\iI} \sz{\Phi^i_u} - \len{\M} = \sz{\Phi_t} + \sz{\Phi_u} - \len{\M}$.
                  \item Rule (\ruleAbs). Then, $t \subs{x}{u} = \lam p.s$, such that $s = s' \subs{x}{u}$ and $t = \lam p.s'$. Therefore, $\var{p} \cap \fv{u} = \eset$ and $\Phi_{t \subs{x}{u}}$ must be of the following form:
                        \[ \begin{prooftree}
                                \hypo{\Phi_s \tr \seq{\Gam_s}{s}{\sig}}
                                \hypo{\Pi \tr \seqp{\Gam_s \restto{p}}{p}{\M_0}}
                                \infer2[(\ruleAbs)]{\seq{\Gam_s \mminus \var{p}}{\lam p.s}{\M_0 \ta \sig}}
                            \end{prooftree} \]
                        where $\Gam = \Gam_s \mminus \var{p}$, and $\T = \M_0 \ta \sig$. By the induction hypothesis over $\Phi_s$, there exist $\Phi_{s'} \tr \seq{\Sig_{s'}; x : \M}{s'}{\sig}$ and $\Phi_u \tr \seq{\Del}{u}{\M}$, such that $\Gam_s = \Sig_{s'} \otimes \Del$ and $\sz{\Phi_s} = \sz{\Phi_{s'}} + \sz{\Phi_u} - \len{\M}$. Since $\var{p} \cap \fv{u} = \eset$, we know $\var{p} \cap \dom{\Del} = \eset$ (by~\cref{lem:relevance}). Therefore, $\Gam_s \restto{p} = (\Sig_{s'} \otimes \Del) \restto{p} = \Sig_{s'} \restto{p}$, $\Gam \mminus \var{p} = (\Sig_{s'} \otimes \Del) \mminus \var{p} = (\Sig_{s'} \mminus \var{p}) \otimes \Del$, and we can build $\Phi_t$ as follows:
                        \[ \begin{prooftree}
                                \hypo{\Phi_{s'}}
                                \hypo{\Pi}
                                \infer2[(\ruleAbs)]{\seq{(\Sig_{s'}; x : \M) \mminus \var{p}}{\lam p.s'}{\M_0 \ta \sig}}
                            \end{prooftree} \]
                        where $(\Sig_{s'}; x : \M) \mminus \var{p} = (\Sig_{s'} \mminus \var{p}); x : \M$ (since we may assume $x \not\in \var{p}$ by $\alpha$-conversion). Thus, $\Sig = \Sig_{s'} \mminus \var{p}$, and we can conclude since $\Gam = \Gam_s \mminus \var{p} = (\Sig_{s'} \otimes \Del) \mminus \var{p} = (\Sig_{s'} \mminus \var{p}) \otimes \Del = \Sig \otimes \Del$ and $\sz{\Phi_{t \subs{x}{u}}} = \sz{\Phi_s} + \sz{\Pi} + 1 = \sz{\Phi_{s'}} + \sz{\Phi_u} - \len{\M} + \sz{\Pi} + 1 = \sz{\Phi_t} + \sz{\Phi_u} - \len{\M}$.
                  \item Rule (\ruleAbsStar). Then, $t \subs{x}{u} = \lam p.s$, such that $s = s' \subs{x}{u}$ and $t = \lam p.s'$. Therefore, $\var{p} \cap \fv{u} = \eset$ and $\Phi_{t \subs{x}{u}}$ must be of the following form:
                        \[ \begin{prooftree}
                                \hypo{\phantom{BUUUUU}}
                                \infer1[(\ruleAbsStar)]{\seq{}{\lam p.s}{\atom}}
                            \end{prooftree} \]
                        where $\Gam = \eset$ and $\T = \atom$. Therefore, we can build $\Phi_t$ as follows:
                        \[ \begin{prooftree}
                                \hypo{\phantom{BUUUUU}}
                                \infer1[(\ruleAbsStar)]{\seq{}{\lam p.s'}{\atom}}
                            \end{prooftree} \]
                        where $\Sig = \eset$ and $\M = \emul$, and $\Phi_u$ as follows:
                        \[ \begin{prooftree}
                                \hypo{\phantom{BUUUUUU}}
                                \infer1[(\ruleMany)]{\seq{}{u}{\emul}}
                            \end{prooftree} \]
                        where $\Del = \eset$. We can conclude since $\Gam = \eset = \Sig \otimes \Del$ and $\sz{\Phi_{t \subs{x}{u}}} = 1 = \sz{\Phi_t} + \sz{\Phi_u} - \len{\M}$.
                  \item Rule (\ruleApp). Then, $t \subs{x}{u} = sr$, such that $s = s' \subs{x}{u}$, $r' \subs{x}{u}$, and $t = s'r'$. Therefore, $\Phi_{t \subs{x}{u}}$ must be of the following form:
                        \[ \begin{prooftree}
                                \hypo{\Phi_s \tr \seq{\Gam_s}{s}{\M_0 \ta \sig}}
                                \hypo{\Phi_r \tr \seq{\Gam_r}{r}{\M_0}}
                                \infer2[(\ruleApp)]{\seq{\Gam_s \otimes \Gam_r}{sr}{\sig}}
                            \end{prooftree} \]
                        where $\Gam = \Gam_s \otimes \Gam_r$ and $\T = \sig$. By the induction hypothesis over $\Phi_s$ (resp. $\Phi_r$), we know there exist derivations $\Phi_{s'} \tr \seq{\Sig_{s'}; x : \M_1}{s'}{\M_0 \ta \sig}$ (resp. $\Phi_{r'} \tr \seq{\Sig_{r'}; x : \M_2}{r'}{\sig}$) and $\Phi^1_u \tr \seq{\Del_1}{u}{\M_1}$ (resp. $\Phi^2_u \tr \seq{\Del_2}{u}{\M_2}$), such that $\Gam_s = \Sig_{s'} \otimes \Del_1$ (resp. $\Gam_r = \Sig_{r'} \otimes \Del_2$) and $\sz{\Phi_{s \subs{x}{u}}} = \sz{\Phi_s} + \sz{\Phi^1_u} - \len{\M_1}$ (resp. $\sz{\Phi_{r \subs{x}{u}}} = \sz{\Phi_r} + \sz{\Phi^2_u} - \len{\M_2}$). By~\cref{lem:split-and-merge-for-multi-types} over $\Phi^1_u$ and $\Phi^2_u$, we know there exists a derivation $\Phi_u \tr \seq{\Del}{u}{\M}$, such that $\Del = \Del_1 \otimes \Del_2$, $\M = \M_1 \sqcup \M_2$ and $\sz{\Phi_u} = \sz{\Phi^1_u} + \sz{\Phi^2_u}$. We can build $\Phi_t$ as follows:
                        \[ \begin{prooftree}
                                \hypo{\Phi_{s'}}
                                \hypo{\Phi_{r'}}
                                \infer2[(\ruleApp)]{\seq{(\Gam_{s'}; x : \M_1) \otimes (\Gam_{r'}; x : \M_2)}{s'r'}{\sig}}
                            \end{prooftree} \]
                        where $(\Gam_{s'}; x : \M_1) \otimes (\Gam_{r'}; x : \M_2) = \Gam_{s'} \otimes \Gam_{r'}; x : \M$. Thus, $\Sig = \Gam_{s'} \otimes \Gam_{r'}$. We can conclude since $\Gam = \Gam_s \otimes \Gam_r = \Sig_{s'} \otimes \Del_1 \otimes \Sig_{r'} \otimes \Del_2 = \Sig \otimes \Del$ and $\sz{\Phi_{t \subs{x}{u}}} = \sz{\Phi_{s \subs{x}{u}}} + \sz{\Phi_{r \subs{x}{u}}} + 1 = \sz{\Phi_s} + \sz{\Phi^1_u} - \len{\M_1} + \sz{\Phi_r} + \sz{\Phi^2_u} - \len{\M_2} + 1 = \sz{\Phi_t} + \sz{\Phi_u} - \len{\M}$.
                  \item Rule (\ruleConst). Then, $t \subs{x}{u} = \const{\ttc}{\vect{s_i}_n}$, such that each $s_i = s'_i \subs{x}{u}$ and $t = \const{\ttc}{\vect{s'_i}_n}$. Therefore, $\Phi_{t \subs{x}{u}}$ must be of the following form:
                        \[ \begin{prooftree}
                                \hypo{\Phi_{s_i} \tr \seq{\Gam_i}{s_i}{\M^0_i}}
                                \delims{\left(}{\right)_{i \in \interval{1}{n}}}
                                \infer1[(\ruleConst)]{\hspace{.5cm}\seq{\otimes_{i \in \interval{1}{n}} \Gam_{s_i}}{\const{\ttc}{\vect{s_i}_n}}{\const{\ttc}{\vect{\M^0_i}_n}}\hspace{.5cm}}
                            \end{prooftree} \]
                        where $\Gam = \otimes_{i \in \interval{1}{n}} \Gam_{s_i}$, and $\T = \const{\ttc}{\vect{\M^0_i}_n}$. By the induction hypothesis over each $\Phi_{s_i}$, we know there exist $\Phi_{s'_i} \tr \seq{\Sig_{s'_i}; x : \M_i}{s'_i}{\M^0_i}$ and $\Phi^i_u \tr \seq{\Del_i}{u}{\M_i}$, such that $\Gam_{s_i} = \Sig_{s'_i} \otimes \Del_i$, and $\sz{\Phi_{s_i}} = \sz{\Phi_{s'_i}} + \sz{\Phi^i_u} - \len{\M_i}$. By~\cref{lem:split-and-merge-for-multi-types} over $\Phi^i_u$, we know there exists $\Phi_u \tr \seq{\Del}{u}{\M}$, such that $\Del = \otimes_{\interval{1}{n}} \Del_i$, $\M = \sqcup_{i \in \interval{1}{n}} \M^0_i$ and $\sz{\Phi_u} = +_{i \in \interval{1}{n}} \sz{\Phi^i_u}$. We can build $\Phi_t$ as follows:
                        \[ \begin{prooftree}
                                \hypo{\Phi_{s'_i}}
                                \delims{\left(}{\right)_{i \in \interval{1}{n}}}
                                \infer1[(\ruleConst)]{\seq{\otimes_{i \in \interval{1}{n}} (\Sig_{s'_i}; x : \M_i)}{s'_i}{\const{\ttc}{\vect{\M^0_i}_n}}}
                            \end{prooftree} \]
                        where $\otimes_{i \in \interval{1}{n}} (\Sig_{s'_i}; x : \M_i) = (\otimes_{i \in \interval{1}{n}} \Sig_{s'_i}); x : \M$. Thus, $\Sig = \otimes_{i \in \interval{1}{n}} \Sig_i$, and we can conclude since $\Gam = \otimes_{i \in \interval{1}{n}} \Gam_{s_i} = \otimes_{i \in \interval{1}{n}} (\Sig_{s'_i} \otimes \Del_i) = (\otimes_{i \in \interval{1}{n}} \Sig_{s'_i}) \otimes (\otimes_{i \in \interval{1}{n}} \Del_i) = \Sig \otimes \Del$ and $\sz{\Phi_{t \subs{x}{u}}} = 1 +_{i \in \interval{1}{n}} \sz{\Phi_{s_i}} = 1 +_{i \in \interval{1}{n}} (\sz{\Phi_{s'_i}} + \sz{\Phi^i_u} - \len{\M_i}) = 1 +_{i \in \interval{1}{n}} \sz{\Phi_{s'_i}} +_{i \in \interval{1}{n}} \sz{\Phi^i_u} - \len{\M} = \sz{\Phi_t} + \sz{\Phi_u} - \len{\M}$.
                  \item Rule (\ruleMatch). Then, $t \subs{x}{u} = s \match{p}{r}$, such that $s = s' \subs{x}{u}$, $r = r' \subs{x}{u}$ and $t = s' \match{p}{r'}$. Therefore, $\var{p} \cap \fv{u} = \eset$ and $\Phi_{t \subs{x}{u}}$ must be of the following form:
                        \[ \begin{prooftree}
                                \hypo{\Phi_s \tr \seq{\Gam_s}{s}{\sig}}
                                \hypo{\Pi \tr \seq{\Gam_s \restto{p}}{p}{\M_0}}
                                \hypo{\Phi_r \tr \seq{\Gam_r}{r}{\M_0}}
                                \infer3[(\ruleMatch)]{\seq{(\Gam_s \mminus \var{p}) \otimes \Gam_r}{s \match{p}{r}}{\sig}}
                            \end{prooftree} \]
                        where $\Gam = (\Gam_s \mminus \var{p}) \otimes \Gam_r$, and $\T = \sig$. By the induction hypothesis over $\Phi_s$ (resp. $\Phi_r$), we know there exist $\Phi_{s'} \tr \seq{\Sig_{s'}; x : \M_1}{s}{\sig}$ (resp. $\Phi_{r'} \tr \seq{\Sig_{r'}; x : \M_2}{r}{\M_0}$) and $\Phi^1_u \tr \seq{\Del_1}{u}{\M_1}$ (resp. $\Phi^2_u \tr \seq{\Del_2}{u}{\M_2}$), such that $\Gam_s = \Sig_{s'} \otimes \Del_1$ (resp. $\Gam_r = \Sig_{r'} \otimes \Del_2$), and $\sz{\Phi_s} = \sz{\Phi_{s'}} + \sz{\Phi^1_u} - \len{\M_1}$ (resp. $\sz{\Phi_r} = \sz{\Phi_{r'}} + \sz{\Phi^2_u} - \len{\M_2}$). By~\cref{lem:split-and-merge-for-multi-types} over $\Phi^1_u$ and $\Phi^2_u$, we know there exists $\Phi_u \tr \seq{\Del}{u}{\M}$, such that $\Del = \Del_1 \otimes \Del_2$, $\M = \M_1 \sqcup \M_2$, and $\sz{\Phi_u} = \sz{\Phi^1_u} + \sz{\Phi^2_u}$. Since $\var{p} \cap \fv{u} = \eset$, we know $\var{p} \cap \dom{\Del} = \eset$ (by~\cref{lem:relevance}). Therefore, $\Gam_s \restto{p} = (\Sig_{s'} \otimes \Del_1) \restto{p} = \Sig_{s'} \restto{p}$, $\Gam_s \mminus \var{p} = (\Sig_{s'} \otimes \Del) \mminus \var{p} = (\Sig_{s'} \mminus \var{p}) \otimes \Del$, and we can build $\Phi_t$ as follows:
                        \[ \begin{prooftree}
                                \hypo{\Phi_{s'}}
                                \hypo{\Pi}
                                \hypo{\Phi_{r'}}
                                \infer3[(\ruleMatch)]{\seq{((\Sig_{s'}; x : \M_1) \mminus \var{p}) \otimes \Sig_{r'}; x : \M_2}{s \match{p}{r}}{\sig}}
                            \end{prooftree} \]
                        where $(\Sig_{s'}; x : \M_1) \mminus \var{p} = \Sig_{s'} \mminus \var{p}; x : \M_1$ (since we may assume $x \not\in \var{p}$ by $\alpha$-conversion). Thus, $((\Sig_{s'}; x : \M_1) \mminus \var{p}) \otimes \Sig_{r'}; x : \M_2 = (\Sig_{s'} \mminus \var{p}) \otimes \Sig_{r'}; x : \M$, $\Sig = (\Sig_{s'} \mminus \var{p}) \otimes \Sig_{r'}$, and we can conclude since $\Gam = (\Gam_s \mminus \var{p}) \otimes \Gam_r = ((\Sig_{s'} \otimes \Del_1) \mminus \var{p}) \otimes \Sig_{r'} \otimes \Del_2 = \Sig \otimes \Del$ and $\sz{\Phi_{t \subs{x}{u}}} = \sz{\Phi_s} + \sz{\Pi} + \sz{\Phi_r} = \sz{\Phi_{s'}} + \sz{\Phi^1_u} - \len{\M_1} + \sz{\Pi} + \sz{\Phi_{r'}} + \sz{\Phi^2_u} - \len{\M_2} = \sz{\Phi_t} + \sz{\Phi_u} - \len{\M}$.
                  \item Rule (\ruleCase). Then, $t \subs{x}{u} = \case{r}{\vect{\branch{\const{\hat{p}_i}}{s_i}}_n}$, such that $r = r' \subs{x}{u}$, $s_i = s'_i \subs{x}{u}$ for each $\iI$, and $t = \case{r'}{\vect{\branch{\const{\hat{p}_i}}{s'_i}}_n}$. Therefore, $\var{\hat{p}_i} \cap \fv{u} = \eset$ for all $\iI$, and $\Phi_{t \subs{x}{u}}$ must be of the following form:
                        \[ \begin{prooftree}
                                \hypo{\Phi_r \tr \seq{\Gam_r}{r}{\M_0}}
                                \hypo{\Pi \tr \seq{\Gam_{s_k} \restto{\hat{p}_k}}{\hat{p}_k}{\M_0}}
                                \hypo{\Phi_{s_k} \tr \seq{\Gam_{s_k}}{s_k}{\sig}}
                                \infer3[(\ruleCase)]{\seq{(\Gam_{s_k} \mminus \var{\hat{p}_k}) \otimes \Gam_{r}}{\case{r}{\vect{\branch{\const{\hat{p}_i}}{s_i}}_n}}{\sig}}
                            \end{prooftree} \]
                        where $\Gam = (\Gam_{s_k} \mminus \var{\hat{p}_k}) \otimes \Gam_r$, and $\T = \sig$. By the induction hypothesis over $\Phi_{s_k}$ (resp. $\Phi_r$), we know there exist $\Phi_{s'_k} \tr \seq{\Sig_{s'_k}; x : \M_1}{s'_k}{\sig}$ (resp. $\Phi_{r'} \tr \seq{\Sig_{r'}; x : \M_2}{r'}{\M_0}$) and $\Phi^1_u \tr \seq{\Del_1}{u}{\M_1}$ (resp. $\Phi^2_u \tr \seq{\Del_2}{u}{\M_2}$), such that $\Gam_{s_k} = \Sig_{s'_k} \otimes \Del_1$ (resp. $\Gam_r = \Sig_{r'} \otimes \Del_2$) and $\sz{\Phi_{s_k}} = \sz{\Phi_{s'_k}} + \sz{\Phi^1_u} - \len{\M_1}$ (resp. $\sz{\Phi_r} = \sz{\Phi_{r'}} + \sz{\Phi^2_u} - \len{\M_2}$). By~\cref{lem:split-and-merge-for-multi-types} over $\Phi^1_u$ and $\Phi^2_u$, we know there exists $\Phi_u \tr \seq{\Del}{u}{\M}$, such that $\Del = \Del_1 \otimes \Del_2$, $\M = \M_1 \sqcup \M_2$, and $\sz{\Phi_u} = \sz{\Phi^1_u} + \sz{\Phi^2_u}$. Since $\var{\hat{p}_k} \cap \fv{u} = \eset$, we know $\var{\hat{p}_k} \cap \dom{\Del} = \eset$ (by~\cref{lem:relevance}). Therefore, $\Gam_{s_k} \restto{\hat{p}_k} = (\Sig_{s'_k} \otimes \Del_1) \restto{\hat{p}_k} = \Sig_{s'_k} \restto{\hat{p}_k}$, $\Gam_{s_k} \mminus \var{\hat{p}_k} = (\Sig_{s'_k} \otimes \Del) \mminus \var{\hat{p}_k} = (\Sig_{s'_k} \mminus \var{\hat{p}_k}) \otimes \Del$, and we can build $\Phi_t$ as follows:
                        \[ \begin{prooftree}
                                \hypo{\Phi_{r'}}
                                \hypo{\Pi}
                                \hypo{\Phi_{s'_k}}
                                \infer3[(\ruleCase)]{\seq{((\Sig_{s'_k}; x : \M_1) \mminus \var{\hat{p}_k}) \otimes \Sig_{r'}; x : \M_2}{\case{r}{\vect{\branch{\const{\hat{p}_i}}{s_i}}_n}}{\sig}}
                            \end{prooftree} \]
                        where $(\Sig_{s'_k}; x : \M_1) \mminus \var{\hat{p}_k} = \Sig_{s'_k} \mminus \var{\hat{p}_k}; x : \M_1$ (since we may assume $x \not\in \var{\hat{p}_k}$ by $\alpha$-conversion). Thus, $((\Sig_{s'}; x : \M_1) \mminus \var{\hat{p}_k}) \otimes \Sig_{r'}; x : \M_2 = (\Sig_{s'_k} \mminus \var{\hat{p}_k}) \otimes \Sig_{r'}; x : \M$, $\Sig = (\Sig_{s'} \mminus \var{\hat{p}_k}) \otimes \Sig_{r'}$, and we can conclude since $\Gam = (\Gam_{s_k} \mminus \var{\hat{p}_k}) \otimes \Gam_r = ((\Sig_{s'_k} \otimes \Del_1) \mminus \var{p}) \otimes \Sig_{r'} \otimes \Del_2 = \Sig \otimes \Del$ and $\sz{\Phi_{t \subs{x}{u}}} = \sz{\Phi_{s_k}} + \sz{\Pi} + \sz{\Phi_r}  + 1 = \sz{\Phi_{s'_k}} + \sz{\Phi^1_u} - \len{\M_1} + \sz{\Pi} + \sz{\Phi_{r'}} + \sz{\Phi^2_u} - \len{\M_2} + 1 = \sz{\Phi_t} + \sz{\Phi_u} - \len{\M}$.
              \end{itemize}
    \end{itemize}
\end{proof}}

\lemweightedsubjexp*

\maybehide{\begin{proof}
    The proof follows by induction over $t \redd t'$:
    \begin{itemize}
        \item Case $t = \MC_0 \lhole{\lam p.u} s \redd[(\ruleBeta)] \MC_0 \lhole{u \match{p}{s}} = t'$, such that $\bv{\MC_0} \cap \fv{s} = \eset$. We proceed by induction over the list matching context $\MC_0$:
              \begin{itemize}
                  \item Let $\MC_0 = \ehole$. Then, $\Phi_{t'}$ must be of the following form:
                        \[ \begin{prooftree}
                                \hypo{\Phi_u \tr \seq{\Gam_u}{u}{\sig}}
                                \hypo{\Pi \tr \seqp{\Gam_u \restto{p}}{p}{\M}}
                                \hypo{\Phi_s \tr \seq{\Gam_s}{s}{\M}}
                                \infer3[(\ruleMatch)]{\seq{(\Gam_u \mminus \var{p}) \otimes \Gam_s}{u \match{p}{s}}{\sig}}
                            \end{prooftree} \]
                        where $\Gam = (\Gam_u \mminus \var{p}) \otimes \Gam_s$. Therefore, we can build $\Phi_t$ as follows:
                        \[ \begin{prooftree}
                                \hypo{\Phi_u}
                                \hypo{\Pi}
                                \infer2[(\ruleAbs)]{\seq{\Gam_u \mminus \var{p}}{\lam p.u}{\M \ta \sig}}
                                \hypo{\Phi_s}
                                \infer2[(\ruleApp)]{\seq{(\Gam_u \mminus \var{p}) \otimes \Gam_s}{(\lam p.u) s}{\sig}}
                            \end{prooftree} \]
                        We can conclude with $\sz{\Phi_{t'}} = \sz{\Phi_u} + \sz{\Pi} + \sz{\Phi_s} < \sz{\Phi_u} + \sz{\Pi} + 1 + \sz{\Phi_s} + 1 = \sz{\Phi_t} $
                  \item Let $\MC_0 = \MC_1 \match{q}{r}$. Then, $\Phi_{t'}$ must be of the following form:
                        \[ \begin{prooftree}
                                \hypo{\Phi_{\MC_1 \lhole{u \match{p}{s}}} \tr \seq{\Gam_{\MC_1 \lhole{u \match{p}{s}}}}{\MC_1 \lhole{u \match{p}{s}}}{\sig}}
                                \hypo{\Pi \tr \seqp{\Gam_{\MC_1 \lhole{u \match{p}{s}}} \restto{q}}{q}{\M_q}}
                                \hypo{\Phi_r \tr \seq{\Gam_r}{r}{\M_q}}
                                \infer3[(\ruleMatch)]{\seq{(\Gam_{\MC_1 \lhole{u \match{p}{s}}} \mminus \var{q}) \otimes \Gam_r}{\MC_1 \lhole{u \match{p}{s}} \match{q}{r}}{\sig}}
                            \end{prooftree} \]
                        where $\Gam = (\Gam_{\MC_1 \lhole{u \match{p}{s}}} \mminus \var{q}) \otimes \Gam_r$. Since $\bv{\MC_1} \subseteq \bv{\MC_0}$, we can do the step $\MC_1 \lhole{\lam p.u} s \redd \MC_1 \lhole{u \match{p}{s}}$.
                        By the \ih over $\MC_1 \lhole{\lam p.u} s \redd \MC_1 \lhole{u \match{p}{s}}$, we know there exists a derivation $\Psi_{\MC_1 \lhole{\lam p.u} s}$ for $\MC_1 \lhole{\lam p.u} s$, which must be of the following form:
                        \[ \begin{prooftree}
                                \hypo{\Phi_{\MC_1 \lhole{\lam p.u}} \tr \seq{\Gam_{\MC_1 \lhole{\lam p.u}}}{\MC_1 \lhole{\lam p.u}}{\M_0 \ta \sig}}
                                \hypo{\Phi_s \tr \seq{\Gam_s}{s}{\M_0}}
                                \infer2[(\ruleApp)]{\seq{\Gam_{\MC_1 \lhole{u \match{p}{s}}}}{\MC_1 \lhole{\lam p.u} s}{\sig}}
                            \end{prooftree} \]
                        where $\Gam_{\MC_1 \lhole{u \match{p}{s}}} = \Gam_{\MC_1 \lhole{\lam p.u}} \otimes \Gam_s$ and $\sz{\Psi_{\MC_1 \lhole{\lam p.u} s}} > \sz{\Phi_{\MC_1 \lhole{u \match{p}{s}}}}$. Since $\bv{\MC_0} \cap \fv{s} = \eset$, we know that $\fv{s} \cap \var{q} = \eset$ and thus $\var{q} \cap \dom{\Gam_s} = \eset$ (\cref{lem:relevance}). Therefore, we can build $\Phi_t$ as follows:
                        \[ \begin{prooftree}
                                \hypo{\Phi_{\MC_1 \lhole{\lam p.u}}}
                                \hypo{\Pi}
                                \hypo{\Phi_r}
                                \infer3[(\ruleMatch)]{\seq{(\Gam_{\MC_1 \lhole{\lam p.u}} \mminus \var{q}) \otimes \Gam_r}{\MC_1 \lhole{\lam p.u} \match{q}{r}}{\M_0 \ta \sig}}
                                \hypo{\Phi_s}
                                \infer2[(\ruleApp)]{\seq{(\Gam_{\MC_1 \lhole{\lam p.u}} \mminus \var{q}) \otimes \Gam_r \otimes \Gam_s}{\MC_1 \lhole{\lam p.u} \match{q}{r} s}{\sig}}
                            \end{prooftree} \]
                        We can conclude since $(\Gam_{\MC_1 \lhole{\lam p.u}} \mminus \var{q}) \otimes \Gam_r \otimes \Gam_s = (\Gam_{\MC_1 \lhole{u \match{p}{s}}} \mminus \var{q}) \otimes \Gam_r$, and $\sz{\Phi_{t'}} = \sz{\Phi_{\MC_1 \lhole{u \match{p}{s}}}} + \sz{\Pi} + \sz{\Phi_r} < \sz{\Psi_{\MC_1 \lhole{\lam p.u} s}} + \sz{\Pi} + \sz{\Phi_r} = \sz{\Phi_{\MC_1 \lhole{\lam p.u}}} + \sz{\Phi_s} + 1 + \sz{\Pi} + \sz{\Phi_r} = \sz{\Phi_t}$.
              \end{itemize}
        \item Case $t = u \match{\const{\ttc}{\vect{p_i}_n}}{\MC_0 \lhole{\const{\ttc}{\vect{s_i}_n}}} \redd[(\ruleM)] \MC_0 \lhole{u \match{p_1}{s_1} \cdots \match{p_n}{s_n}} = t'$, such that $\bv{\MC_0} \cap \fv{u} = \eset$. We proceed by induction over the list matching context $\MC_0$:
              \begin{itemize}
                  \item Case $\MC_0 = \chole$. Then, $\Phi_{t'}$ must be of the following form:
                        \[ \begin{prooftree}
                                \hypo{\Phi_u \tr \seq{\Gam_u}{u}{\sig}}
                                \infer[no rule]1{\Pi_{p_1} \tr \Gam_u \restto{p_1} \Vdash p_1 : \M_1}
                                \infer[no rule]1{\Phi_{s_1} \tr \seq{\Gam_{s_1}}{s_1}{\M_1}}
                                \infer1[(\ruleMatch)]{\seq{\Gam_u \mminus \var{p_1} \otimes \Gam_{s_1}}{u \match{p_1}{s_1}}{\sig}}
                                \infer[no rule]1[]{\vdots}
                                \hypo{\hspace{-2cm} \Pi_{p_n} \tr \Gam_u \restto{p_n} \Vdash p_1 : \M_n}
                                \hypo{\Phi_{s_n} \tr \seq{\Gam_{s_n}}{s_n}{\M_n}}
                                \infer3[(\ruleMatch)]{\seq{((((\Gam_u \mminus \var{p_1}) \otimes \Gam_{s_1}) \cdots) \mminus \var{p_n}) \otimes \Gam_{s_n}}{u \match{p_1}{s_1} \cdots \match{p_n}{s_n}}{\sig}}
                            \end{prooftree} \]
                        where $\Gam = ((((\Gam_u \mminus \var{p_1}) \otimes \Gam_{s_1}) \cdots) \mminus \var{p_n}) \otimes \Gam_{s_n}$. By $\alpha$-conversion, we can assume that $\fv{\const{\ttc}{\vect{s_i}_n}} \cap \var{\const{\ttc}{\vect{p_i}_n}} = \eset$ and thus $\dom{\otimes_{\interval{1}{n}} \Gam_{s_i}} \cap \var{\const{\ttc}{\vect{p_i}_n}} = \eset$ (\cref{lem:relevance}). Therefore, $(\otimes_{\interval{1}{n}} \Gam_{s_i}) \mminus \var{\const{\ttc}{\vect{p_i}_n}} = \otimes_{\interval{1}{n}} \Gam_{s_i}$, $(\otimes_{\interval{1}{n}} \Gam_{s_i}) \restto{\const{\ttc}{\vect{p_i}_n}} = \eset$, and we can build $\Phi_t$ as follows:
                        \[ \begin{prooftree}
                                \hypo{\Phi_u}
                                \hypo{\Pi_{p_i}}
                                \delims{\left(}{\right)_{i \in \interval{1}{n}}}
                                \infer1[(\rulePatC)]{\seqp{\Gam_u \restto{\const{\ttc}{\vect{p_i}_n}}}{\const{\ttc}{\vect{p_i}_n}}{\mul{\const{\ttc}{\vect{\M_i}_n}}}}
                                \hypo{\Phi_{s_i}}
                                \delims{\left(}{\right)_{i \in \interval{1}{n}}}
                                \infer1[(\ruleConst)]{\seq{\otimes_{i \in \interval{1}{n}} \Gam_{s_i}}{\const{\ttc}{\vect{s_i}_n}}{\const{\ttc}{\vect{\M_i}_n}}}
                                \infer1[(\ruleMany)]{\seq{\otimes_{i \in \interval{1}{n}} \Gam_{s_i}}{\const{\ttc}{\vect{s_i}_n}}{\mul{\const{\ttc}{\vect{\M_i}_n}}}}
                                \infer3[(\ruleMatch)]{\seq{(\Gam_u \mminus \var{\const{\ttc}{\vect{p_i}_n}}) \otimes_{i \in \interval{1}{n}} \Gam_{s_i}}{u \match{\const{\ttc}{\vect{p_i}_n}}{\const{\ttc}{\vect{s_i}_n}}}{\sig}}
                            \end{prooftree} \]
                        We can conclude since $\Gam = ((((\Gam_u \mminus \var{p_1}) \otimes \Gam_{s_1}) \cdots) \mminus \var{p_n}) \otimes \Gam_{s_n} = (\Gam_u \mminus \var{\const{\ttc}{\vect{p_i}_n}}) \otimes_{i \in \interval{1}{n}} \Gam_{s_i}$ and $\sz{\Phi_{t'}} = \sz{\Phi_u} +_{i \in \interval{1}{n}} \sz{\Pi_{p_i}} +_{i \in \interval{1}{n}} \sz{\Phi_{s_i}} < \sz{\Phi_u} + 1 +_{i \in \interval{1}{n}} \sz{\Pi_{p_i}} + 1 +_{i \in \interval{1}{n}} \sz{\Phi_{s_i}} = \sz{\Phi_t}$.
                  \item Case $\MC_0 = \MC_1 \match{q}{r}$. Then, $\Phi_{t'}$ must be of the following form:
                        \[ \begin{prooftree}
                                \hypo{\Phi_{\MC_1 \lhole{u \match{p_1}{s_1} \cdots \match{p_n}{s_n}}} \tr \seq{\Gam_{\MC_1 \lhole{u \match{p_1}{s_1} \cdots \match{p_n}{s_n}}}}{\MC_1 \lhole{u \match{p_1}{s_1} \cdots \match{p_n}{s_n}}}{\sig}}
                                \infer[no rule]1{\Pi_q \tr \seq{\Gam_{\MC_1 \lhole{u \match{p_1}{s_1} \cdots \match{p_n}{s_n}}} \restto{q}}{q}{\M_q}}
                                \infer[no rule]1{\Phi_r \tr \seq{\Gam_r}{r}{\M_q}}
                                \infer1[(\ruleMatch)]{\seq{(\Gam_{\MC_1 \lhole{u \match{p_1}{s_1} \cdots \match{p_n}{s_n}}} \mminus \var{q}) \otimes \Gam_r}{\MC_1 \lhole{u \match{p_1}{s_1} \cdots \match{p_n}{s_n}} \match{q}{r}}{\sig}}
                            \end{prooftree} \]
                        where $\Gam = (\Gam_{\MC_1 \lhole{u \match{p_1}{s_1} \cdots \match{p_n}{s_n}}} \mminus \var{q}) \otimes \Gam_r$. Since $\bv{\MC_0} \cap \fv{u} = \eset$, we can do the step $u \match{\const{\ttc}{\vect{p_i}_n}}{\MC_1 \lhole{\const{\ttc}{\vect{s_i}_n}}} \redd[(\ruleM)] \MC_1 \lhole{u \match{p_1}{s_1} \cdots \match{p_n}{s_n}}$. By the \ih over $u \match{\const{\ttc}{\vect{p_i}_n}}{\MC_1 \lhole{\const{\ttc}{\vect{s_i}_n}}} \redd[(\ruleM)] \MC_1 \lhole{u \match{p_1}{s_1} \cdots \match{p_n}{s_n}}$, we know there exists a derivation $\Psi_{u \match{\const{\ttc}{\vect{p_i}_n}}{\MC_1 \lhole{\const{\ttc}{\vect{s_i}_n}}}}$ for $u \match{\const{\ttc}{\vect{p_i}_n}}{\MC_1 \lhole{\const{\ttc}{\vect{s_i}_n}}}$, which must be of the following form:
                        \[ \begin{prooftree}
                                \hypo{\Phi_u \tr \seq{\Gam_u}{u}{\sig}}
                                \hypo{\Pi \tr \seqp{\Gam_u \restto{\const{\ttc}{\vect{p_i}_n}}}{\const{\ttc}{\vect{p_i}_n}}{\M}}
                                \hypo{\Phi_{\MC_1 \lhole{\const{\ttc}{\vect{s_i}_n}}} \tr \seq{\Gam_{\MC_1 \lhole{\const{\ttc}{\vect{s_i}_n}}}}{\MC_1 \lhole{\const{\ttc}{\vect{s_i}_n}}}{\M}}
                                \infer3[(\ruleMatch)]{\seq{\Gam_{\MC_1 \lhole{u \match{p_1}{s_1} \cdots \match{p_n}{s_n}}}}{u \match{\const{\ttc}{\vect{p_i}_n}}{\MC_1 \lhole{\const{\ttc}{\vect{s_i}_n}}}}{\sig}}
                            \end{prooftree} \]
                        where $\Gam_{\MC_1 \lhole{u \match{p_1}{s_1} \cdots \match{p_n}{s_n}}} = (\Gam_u \mminus \var{\const{\ttc}{\vect{p_i}_n}}) \otimes \Gam_{\MC_1 \lhole{\const{\ttc}{\vect{s_i}_n}}}$ and $\Psi_{u \match{\const{\ttc}{\vect{p_i}_n}}{\MC_1 \lhole{\const{\ttc}{\vect{s_i}_n}}}} > \Phi_{\MC_1 \lhole{u \match{p_1}{s_1} \cdots \match{p_n}{s_n}}}$. Since $\bv{\MC_0} \cap \fv{u}$, we know that $\fv{u} \cap \var{q} = \eset$ and thus $\var{q} \cap \dom{\Gam_u} = \eset$ (\cref{lem:relevance}). Therefore, we can build $\Phi_t$ as follows:
                        \[ \begin{prooftree}
                                \hypo{\Phi_u}
                                \hypo{\Pi}
                                \hypo{\Phi_{\MC_1 \lhole{\const{\ttc}{\vect{s_i}_n}}}}
                                \hypo{\Pi_q}
                                \hypo{\Phi_r}
                                \infer3[(\ruleMatch)]{\seq{(\Gam_{\MC_1 \lhole{\const{\ttc}{\vect{s_i}_n}}} \mminus \var{q}) \otimes \Gam_r}{\MC_1 \lhole{\const{\ttc}{\vect{s_i}_n}} \match{q}{r}}{\M}}
                                \infer3[(\ruleMatch)]{\seq{(\Gam_u \mminus \var{\const{\ttc}{\vect{p_i}_n}}) \otimes (\Gam_{\MC_1 \lhole{\const{\ttc}{\vect{s_i}_n}}} \mminus \var{q}) \otimes \Gam_r}{u \match{\const{\ttc}{\vect{p_i}_n}}{\MC_1 \lhole{\const{\ttc}{\vect{s_i}_n}} \match{q}{r}}}{\sig}}
                            \end{prooftree} \]
                        We can conclude since $\var{\const{\ttc}{\vect{p_i}_n}} = \var{p_1} \cup \cdots \cup \var{p_n}$, $\Gam = (\Gam_{\MC_1 \lhole{u \match{p_1}{s_1} \cdots \match{p_n}{s_n}}} \mminus \var{q}) \otimes \Gam_r = (\Gam_u \mminus \var{\const{\ttc}{\vect{p_i}_n}}) \otimes (\Gam_{\MC_1 \lhole{\const{\ttc}{\vect{s_i}_n}}} \mminus \var{q}) \otimes \Gam_r$, and $\sz{\Phi_{t'}} = \Phi_{\MC_1 \lhole{u \match{p_1}{s_1} \cdots \match{p_n}{s_n}}} + \sz{\Pi_q} + \sz{\Phi_r} < \Psi_{u \match{\const{\ttc}{\vect{p_i}_n}}{\MC_1 \lhole{\const{\ttc}{\vect{s_i}_n}}}} + \sz{\Pi_q} + \sz{\Phi_r} = \sz{\Phi_u} + \sz{\Pi} + \sz{\Phi_{\MC \lhole{\const{\ttc}{\vect{s_i}_n}}}} + \sz{\Pi_q} + \sz{\Phi_r} = \sz{\Phi_t}$.
              \end{itemize}
        \item Case $t = u \match{x}{s} \redd[(\ruleE)] u \subs{x}{s} = t'$. By~\cref{lem:weighted-antisubstitution} over $\Phi_{t'}$, we know there exist $\Phi_u \tr \seq{\Gam_u; x : \M}{u}{\sig}$ and $\Phi_s \tr \seq{\Gam_s}{s}{\M}$, such that $\Gam = \Gam_u \otimes \Gam_s$ and $\sz{\Phi_{t'}} = \sz{\Phi_u} + \sz{\Phi_s} - \len{\M}$. Therefore, we can build $\Phi_t$ as follows:
              \[ \begin{prooftree}
                      \hypo{\Phi_u}
                      \infer0[(\rulePatV)]{\seqp{x : \M}{x}{\M}}
                      \hypo{\Phi_s}
                      \infer3[(\ruleMatch)]{\seq{\Gam_u \otimes \Gam_s}{u \match{x}{s}}{\sig}}
                  \end{prooftree} \]
              And we can conclude with $\sz{\Phi_{t'}} = \sz{\Phi_u} + \sz{\Phi_s} - \len{\M} < \sz{\Phi_u} + \sz{\Phi_s} + 1 = \sz{\Phi_t}$.
        \item Case $t = \case{\MC_0 \lhole{\const{\ttc}{\vect{s_i}_n}}}{(\ldots, \branch{\const{\ttc}{\vect{p_i}_n}}{u_i}, \ldots)} \redd[(\ruleC)] \MC_0 \lhole{u_i \match{p_1}{s_1} \cdots \match{p_n}{s_n}} = t'$, such that $\bv{\MC_0} \cap \fv{s_i} = \eset$. We proceed by induction over the list matching context $\MC_0$:
              \begin{itemize}
                  \item Case $\MC_0 = \chole$. Then, $\Phi_{t'}$ must be of the following form:
                        \[ \begin{prooftree}
                                \hypo{\Phi_{u_i} \tr \seq{\Gam_{u_i}}{u_i}{\sig}}
                                \infer[no rule]1{\Pi_{p_1} \tr \Gam_{u_i} \restto{p_1} \Vdash p_1 : \M_1}
                                \infer[no rule]1{\Phi_{s_1} \tr \seq{\Gam_{s_1}}{s_1}{\M_1}}
                                \infer1[(\ruleMatch)]{\seq{\Gam_{u_i} \mminus \var{p_1} \otimes \Gam_{s_1}}{u_i \match{p_1}{s_1}}{\sig}}
                                \infer[no rule]1[]{\vdots}
                                \hypo{\hspace{-2cm} \Pi_{p_n} \tr \Gam_{u_i} \restto{p_n} \Vdash p_n : \M_n}
                                \hypo{\Phi_{s_n} \tr \seq{\Gam_{s_n}}{s_n}{\M_n}}
                                \infer3[(\ruleMatch)]{\seq{((((\Gam_{u_i} \mminus \var{p_1}) \otimes \Gam_{s_1}) \cdots) \mminus \var{p_n}) \otimes \Gam_{s_n}}{u_i \match{p_1}{s_1} \cdots \match{p_n}{s_n}}{\sig}}
                            \end{prooftree} \]
                        where $\Gam = ((((\Gam_{u_i} \mminus \var{p_1}) \otimes \Gam_{s_1}) \cdots) \mminus \var{p_n}) \otimes \Gam_{s_n}$. By $\alpha$-conversion, we can assume that $\fv{\const{\ttc}{\vect{s_i}_n}} \cap \var{\const{\ttc}{\vect{p_i}_n}} = \eset$ and thus $\dom{\otimes_{\interval{1}{n}} \Gam_{s_i}} \cap \var{\const{\ttc}{\vect{p_i}_n}} = \eset$ (\cref{lem:relevance}). Therefore, $(\otimes_{\interval{1}{n}} \Gam_{s_i}) \mminus \var{\const{\ttc}{\vect{p_i}_n}} = \otimes_{\interval{1}{n}} \Gam_{s_i}$, $(\otimes_{\interval{1}{n}} \Gam_{s_i}) \restto{\const{\ttc}{\vect{p_i}_n}} = \eset$, and we can build $\Phi_t$ as follows:
                        \[ \begin{prooftree}
                                \hypo{\Phi_{s_i}}
                                \delims{\left(}{\right)_{i \in \interval{1}{n}}}
                                \infer1[(\ruleConst)]{\seq{\otimes_{i \in \interval{1}{n}} \Gam_{s_i}}{\const{\ttc}{\vect{s_i}_n}}{\const{\ttc}{\vect{\M_i}_n}}}
                                \infer1[(\ruleMany)]{\seq{\otimes_{i \in \interval{1}{n}} \Gam_{s_i}}{\const{\ttc}{\vect{s_i}_n}}{\mul{\const{\ttc}{\vect{\M_i}_n}}}}
                                \hypo{\Pi_{p_i}}
                                \delims{\left(}{\right)_{i \in \interval{1}{n}}}
                                \infer1[(\rulePatC)]{\seqp{\Gam_{u_i} \restto{\const{\ttc}{\vect{p_i}_n}}}{\const{\ttc}{\vect{p_i}_n}}{\mul{\const{\ttc}{\vect{\M_i}_n}}}}
                                \hypo{\Phi_{u_i}}
                                \infer3[(\ruleMatch)]{\seq{(\Gam_{u_i} \mminus \var{\const{\ttc}{\vect{p_i}_n}}) \otimes_{i \in \interval{1}{n}} \Gam_{s_i}}{\case{\const{\ttc}{\vect{s_i}_n}}{(\ldots, \branch{\const{\ttc}{\vect{p_i}_n}}{u_i}, \ldots)}}{\sig}}
                            \end{prooftree} \]
                        We can conclude since $\var{\const{\ttc}{\vect{p_i}_n}} = \var{p_1} \cup \cdots \cup \var{p_n}$, $\Gam = ((((\Gam_{u_i} \mminus \var{p_1}) \otimes \Gam_{s_1}) \cdots) \mminus \var{p_n}) \otimes \Gam_{s_n} = (\Gam_{u_i} \mminus \var{\const{\ttc}{\vect{p_i}_n}}) \otimes_{i \in \interval{1}{n}} \Gam_{s_i}$, and $\sz{\Phi_{t'}} = \sz{\Phi_{u_i}} +_{i \in \interval{1}{n}} \sz{\Pi_{p_i}} +_{i \in \interval{1}{n}} \sz{\Phi_{s_i}} < \sz{\Phi_{u_i}} + 1 +_{i \in \interval{1}{n}} \sz{\Pi_{p_i}} + 1 +_{i \in \interval{1}{n}} \sz{\Phi_{s_i}} = \sz{\Phi_t}$.
                  \item Case $\MC_0 = \MC_1 \match{q}{r}$. Then, $\Phi_{t'}$ must be of the following form:
                        \[ \begin{prooftree}
                                \hypo{\Phi_{\MC_1 \lhole{u_i \match{p_1}{s_1} \cdots \match{p_n}{s_n}}} \tr \seq{\Gam_{\MC_1 \lhole{u_i \match{p_1}{s_1} \cdots \match{p_n}{s_n}}}}{\MC_1 \lhole{u_i \match{p_1}{s_1} \cdots \match{p_n}{s_n}}}{\sig}}
                                \infer[no rule]1{\Pi_q \tr \seq{\Gam_{\MC_1 \lhole{u_i \match{p_1}{s_1} \cdots \match{p_n}{s_n}}} \restto{q}}{q}{\M_q}}
                                \infer[no rule]1{\Phi_r \tr \seq{\Gam_r}{r}{\M_q}}
                                \infer1[(\ruleMatch)]{\seq{(\Gam_{\MC_1 \lhole{u_i \match{p_1}{s_1} \cdots \match{p_n}{s_n}}} \mminus \var{q}) \otimes \Gam_r}{\MC_1 \lhole{u_i \match{p_1}{s_1} \cdots \match{p_n}{s_n}} \match{q}{r}}{\sig}}
                            \end{prooftree} \]
                        where $\Gam = (\Gam_{\MC_1 \lhole{u_i \match{p_1}{s_1} \cdots \match{p_n}{s_n}}} \mminus \var{q}) \otimes \Gam_r$. Since $\bv{\MC_0} \cap \fv{u_i} = \eset$, we can do the step $\case{\MC_1 \lhole{\const{\ttc}{\vect{s_i}_n}}}{(\ldots, \branch{\const{\ttc}{\vect{p_i}_n}}{u_i}, \ldots )} \redd[(\ruleM)] \MC_1 \lhole{u_i \match{p_1}{s_1} \cdots \match{p_n}{s_n}}$. By the \ih over $\case{\MC_1 \lhole{\const{\ttc}{\vect{s_i}_n}}}{(\ldots, \branch{\const{\ttc}{\vect{p_i}_n}}{u_i}, \ldots )} \redd[(\ruleM)] \MC_1 \lhole{u_i \match{p_1}{s_1} \cdots \match{p_n}{s_n}}$, we know there exists a derivation $\Psi_{\case{\MC_1 \lhole{\const{\ttc}{\vect{s_i}_n}}}{(\ldots, \branch{\const{\ttc}{\vect{p_i}_n}}{u_i}, \ldots)}}$ for $\case{\MC_1 \lhole{\const{\ttc}{\vect{s_i}_n}}}{(\ldots, \branch{\const{\ttc}{\vect{p_i}_n}}{u_i}, \ldots)}$, which must be of the following form:
                        \[ \begin{prooftree}
                                \hypo{\Phi_{\MC_1 \lhole{\const{\ttc}{\vect{s_i}_n}}} \tr \seq{\Gam_{\MC_1 \lhole{\const{\ttc}{\vect{s_i}_n}}}}{\MC_1 \lhole{\const{\ttc}{\vect{s_i}_n}}}{\M}}
                                \hypo{\Pi \tr \seqp{\Gam_{u_i} \restto{\const{\ttc}{\vect{p_i}_n}}}{\const{\ttc}{\vect{p_i}_n}}{\M}}
                                \hypo{\Phi_{u_i} \tr \seq{\Gam_{u_i}}{u_i}{\sig}}
                                \infer3[(\ruleMatch)]{\seq{\Gam_{\MC_1 \lhole{u_i \match{p_1}{s_1} \cdots \match{p_n}{s_n}}}}{\case{\MC_1 \lhole{\const{\ttc}{\vect{s_i}_n}}}{(\ldots, \branch{\const{\ttc}{\vect{p_i}_n}}{u_i}, \ldots)}}{\sig}}
                            \end{prooftree} \]
                        where $\Gam_{\MC_1 \lhole{u_i \match{p_1}{s_1} \cdots \match{p_n}{s_n}}} = (\Gam_{u_i} \mminus \var{\const{\ttc}{\vect{p_i}_n}}) \otimes \Gam_{\MC_1 \lhole{\const{\ttc}{\vect{s_i}_n}}}$ and $\Psi_{\case{\MC_1 \lhole{\const{\ttc}{\vect{s_i}_n}}}{(\ldots, \branch{\const{\ttc}{\vect{p_i}_n}}{u_i}, \ldots)}} > \Phi_{\MC_1 \lhole{u_i \match{p_1}{s_1} \cdots \match{p_n}{s_n}}}$. Since $\bv{\MC_0} \cap \fv{u_i}$, we know that $\fv{u_i} \cap \var{q} = \eset$ and thus $\var{q} \cap \dom{\Gam_{u_i}} = \eset$ (\cref{lem:relevance}). Therefore, we can build $\Phi_t$ as follows:
                        \[ \hspace{-1cm}\begin{prooftree}
                                \hypo{\Phi_{\MC_1 \lhole{\const{\ttc}{\vect{s_i}_n}}}}
                                \hypo{\Pi_q}
                                \hypo{\Phi_r}
                                \infer3[(\ruleMatch)]{\seq{(\Gam_{\MC_1 \lhole{\const{\ttc}{\vect{s_i}_n}}} \mminus \var{q}) \otimes \Gam_r}{\MC_1 \lhole{\const{\ttc}{\vect{s_i}_n}} \match{q}{r}}{\M}}
                                \hypo{\Pi}
                                \hypo{\Phi_{u_i}} \infer3[(\ruleMatch)]{\seq{(\Gam_{u_i} \mminus \var{\const{\ttc}{\vect{p_i}_n}}) \otimes (\Gam_{\MC_1 \lhole{\const{\ttc}{\vect{s_i}_n}}} \mminus \var{q}) \otimes \Gam_r}{\case{\MC_1 \lhole{\const{\ttc}{\vect{s_i}_n}} \match{q}{r}}{(\ldots, \branch{\const{\ttc}{\vect{p_i}_n}}{u_i}, \ldots)}}{\sig}}
                            \end{prooftree} \]
                        We can conclude since $\Gam = (\Gam_{\MC_1 \lhole{u_i \match{p_1}{s_1} \cdots \match{p_n}{s_n}}} \mminus \var{q}) \otimes \Gam_r = (\Gam_{u_i} \mminus \var{\const{\ttc}{\vect{p_i}_n}}) \otimes (\Gam_{\MC_1 \lhole{\const{\ttc}{\vect{s_i}_n}}} \mminus \var{q}) \otimes \Gam_r$ and $\sz{\Phi_{t'}} = \sz{\Phi_{\MC_1 \lhole{u_i \match{p_1}{s_1} \cdots \match{p_n}{s_n}}}} + \sz{\Pi_q} + \sz{\Phi_r} < \\ \sz{\Psi_{\case{\MC_1 \lhole{\const{\ttc}{\vect{s_i}_n}}}{(\ldots, \branch{\const{\ttc}{\vect{p_i}_n}}{u_i}, \ldots)}}} + \sz{\Pi_q} + \sz{\Phi_r} = \sz{\Phi_{\MC_1 \lhole{\const{\ttc}{\vect{s_i}_n}}}} + \sz{\Pi} + \sz{\Phi_{u_i}} + \sz{\Pi_q} + \sz{\Phi_r} = \sz{\Phi_t}$.
              \end{itemize}
        \item Case $t = us \redd[(\ruleAppL)] u's = t'$, such that $\neg\isabs{u}$ and $u \redd u'$. Then, $\Phi_{t'}$ must be of the following form:
              \[ \begin{prooftree}
                      \hypo{\Phi_{u'} \tr \seq{\Gam_{u'}}{u'}{\M \ta \sig}}
                      \hypo{\Phi_s \tr \seq{\Gam_s}{s}{\M}}
                      \infer2[(\ruleApp)]{\seq{\Gam_{u'} \otimes \Gam_s}{u' s}{\sig}}
                  \end{prooftree} \]
              where $\Gam = \Gam_{u'} \otimes \Gam_s$. By the \ih over $u \redd u'$, we know there exists $\Phi_u \tr \seq{\Gam_{u'}}{u}{\M \ta \sig}$, such that $\sz{\Phi_u} > \sz{\Phi_{u'}}$. Therefore, we can build $\Phi_t$ as follows:
              \[ \begin{prooftree}
                      \hypo{\Phi_u}
                      \hypo{\Phi_s}
                      \infer2[(\ruleApp)]{\seq{\Gam_{u'} \otimes \Gam_s}{us}{\sig}}
                  \end{prooftree} \]
              We can conclude with $\sz{\Phi_{t'}} = \sz{\Phi_{u'}} + \sz{\Phi_s} + 1 < \sz{\Phi_u} + \sz{\Phi_s} + 1 = \sz{\Phi_t}$.
        \item Case $t = u \match{\const{\ttc}{\ptuple}}{s} \redd[(\ruleEsL)] u' \match{\const{\ttc}{\ptuple}}{s} = t'$, such that $\neg\isdata[\ttc]{s}$ and $u \redd u'$. Then, $\Phi_{t'}$ must be of the following form:
              \[ \begin{prooftree}
                      \hypo{\Phi_{u'} \tr \seq{\Gam_{u'}}{u'}{\sig}}
                      \hypo{\Pi \tr \seqp{\Gam_{u'} \restto{\const{\ttc}{\ptuple}}}{\const{\ttc}{\ptuple}}{\M}}
                      \hypo{\Phi_s \tr \seq{\Gam_s}{s}{\M}}
                      \infer3[(\ruleMatch)]{\seq{(\Gam_{u'} \mminus \var{\const{\ttc}{\ptuple}}) \otimes \Gam_s}{u' \match{\const{\ttc}{\ptuple}}{s}}{\sig}}
                  \end{prooftree} \]
              where $\Gam = (\Gam_{u'} \mminus \var{\const{\ttc}{\ptuple}}) \otimes \Gam_s$. By the \ih over $u \redd u'$, we know there exists $\Phi_u \tr \seq{\Gam_{u'}}{u}{\sig}$, such that $\sz{\Phi_u} > \sz{\Phi_{u'}}$. Therefore, we can build $\Phi_t$ as follows:
              \[ \begin{prooftree}
                      \hypo{\Phi_u}
                      \hypo{\Pi}
                      \hypo{\Phi_s}
                      \infer3[(\ruleMatch)]{\seq{(\Gam_{u'} \mminus \var{\const{\ttc}{\ptuple}}) \otimes \Gam_s}{u \match{\const{\ttc}{\ptuple}}{s}}{\sig}}
                  \end{prooftree} \]
              We can conclude with $\sz{\Phi_{t'}} = \sz{\Phi_{u'}} + \sz{\Phi_s} < \sz{\Phi_u} + \sz{\Phi_s} = \sz{\Phi_t}$.
        \item Case $t = u \match{\const{\ttc}{\ptuple}}{s} \redd[(\ruleEsR)] u \match{\const{\ttc}{\ptuple}}{s'} = t'$, such that $\neg\isdata[\ttc]{u}$, $u \not\redd$, and $s \redd s'$. Then, $\Phi_{t'}$ must be of the following form:
              \[ \begin{prooftree}
                      \hypo{\Phi_u \tr \seq{\Gam_u}{u}{\sig}}
                      \hypo{\Pi \tr \seqp{\Gam_u \restto{\const{\ttc}{\ptuple}}}{\const{\ttc}{\ptuple}}{\M}}
                      \hypo{\Phi_{s'} \tr \seq{\Gam_{s'}}{s'}{\M}}
                      \infer3[(\ruleMatch)]{\seq{(\Gam_u \mminus \var{\const{\ttc}{\ptuple}}) \otimes \Gam_{s'}}{u \match{\const{\ttc}{\ptuple}}{s'}}{\sig}}
                  \end{prooftree} \]
              where $\Gam = (\Gam_u \mminus \var{\const{\ttc}{\ptuple}}) \otimes \Gam_{s'}$. By the \ih over $s \redd s'$, we know there exists $\Phi_s \tr \seq{\Gam_{s'}}{s}{\M}$, such that $\sz{\Phi_s} > \sz{\Phi_{s'}}$. Therefore, we can build $\Phi_t$ as follows:
              \[ \begin{prooftree}
                      \hypo{\Phi_u}
                      \hypo{\Pi}
                      \hypo{\Phi_s}
                      \infer3[(\ruleMatch)]{\seq{(\Gam_u \mminus \var{\const{\ttc}{\ptuple}}) \otimes \Gam_s}{u \match{\const{\ttc}{\ptuple}}{s}}{\sig}}
                  \end{prooftree} \]
              We can conclude with $\sz{\Phi_{t'}} = \sz{\Phi_{u'}} + \sz{\Phi_s} < \sz{\Phi_u} + \sz{\Phi_s} = \sz{\Phi_t}$.
        \item Case $t = \case{s}{\vect{\branch{\const{\ttc_i}{\ptuple_i}}{u_i}}_n} \redd[(\ruleCaseIn)] \case{s'}{\vect{\branch{\const{\ttc_i}{\ptuple_i}}{u_i}}_n} = t'$. Then, $\Phi_{t'}$ must be of the following form:
              \[ \begin{prooftree}
                      \hypo{\Phi_{s'} \tr \seq{\Gam_{s'}}{s'}{\M}}
                      \hypo{\Pi \tr \seqp{\Gam_{u_k} \restto{\const{\ttc_k}{\ptuple_k}}}{\const{\ttc_k}{\ptuple_k}}{\M}}
                      \hypo{\Phi_{u_k} \tr \seq{\Gam_{u_k}}{u_k}{\sig}}
                      \hypo{ \text{for some $k \in \interval{1}{n}$}}
                      \infer4[(\ruleCase)]{\seq{(\Gam_{u_k} \mminus \var{\const{\ttc_k}{\ptuple_k}}) \otimes \Gam_{s'}}{\case{s}{\vect{\branch{\const{\ttc_i}{\ptuple_i}}{u_i}}_n}}{\sig}}
                  \end{prooftree} \]
              where $\Gam = (\Gam_{u_k} \mminus \var{\const{\ttc_k}{\ptuple_k}}) \otimes \Gam_{s'}$. By the \ih over $s \redd s'$, we know there exists $\Phi_s \tr \seq{\Gam_{s'}}{s}{\M}$, such that $\sz{\Phi_s} > \sz{\Phi_{s'}}$. Therefore, we can build $\Phi_{t'}$ as follows:
              \[ \begin{prooftree}
                      \hypo{\Phi_s}
                      \hypo{\Pi}
                      \hypo{\Phi_{u_k}}
                      \infer3[(\ruleCase)]{\seq{(\Gam_{u_k} \mminus \var{\const{\ttc_k}{\ptuple_k}}) \otimes \Gam_{s'}}{\case{s'}{\vect{\branch{\const{\ttc_i}{\ptuple_i}}{u_i}}_n}}{\sig}}
                  \end{prooftree} \]
              We can conclude since $\sz{\Phi_{t'}} = \sz{\Phi_{s'}} + \sz{\Phi_{u_k}} + 1 < \sz{\Phi_s} + \sz{\Phi_{u_k}} + 1 = \sz{\Phi_t}$.
    \end{itemize}
\end{proof}}




\end{document}